\documentclass[pra,showpacs,twocolumn,aps]{revtex4-1}

\usepackage{latexsym}
\usepackage{amssymb}
\usepackage{amsfonts}
\usepackage{epsfig}
\usepackage{psfrag}
\usepackage{graphicx}

\usepackage{epsfig,color}
\usepackage{psfrag}
\usepackage[utf8]{inputenc} 
\definecolor{darkgreen}{rgb}{0,0.4,0}

\newcommand{\bea}{\begin{eqnarray}}
\newcommand{\ea}{\end{eqnarray}}
\newcommand{\eea}{\end{eqnarray}}

\newcommand{\vc}[1]{\mathbf{#1}}

\usepackage{amsmath}

\usepackage{color}

\begin{document}

\title{
Quantum electrodynamics of non-demolition detection of single microwave photon by superconducting qubit array
}

\author{P. Navez, A. G. Balanov, S. E. Savel'ev, A. M. Zagoskin}

\affiliation{
Department of Physics, Loughborough University,
Loughborough LE11 3TU, United Kingdom
}

\date{\today}

\begin{abstract}
By consistently applying the formalism of quantum electrodynamics we developed a comprehensive theoretical
framework describing the interaction of single microwave photons with an array of superconducting
transmon qubits in a waveguide cavity resonator. In particular, we analyse the effects of microwave
photons on the array’s response to a weak probe signal exciting the resonator. The study reveals
that a high quality factor cavities provide better spectral resolution of the response, while cavities
with moderate quality factor allow better sensitivity for a single photon detection. Remarkably, our
analysis showed that a single-photon signal can be detected by even a sole qubit in cavity under the
realistic range of system parameters. We also discuss how quantum properties of the microwave radiation and
electrodynamical properties of resonators affect the response of qubits’ array. Our results provide
an efficient theoretical background for informing the development and design of quantum devices
consisting of arrays of qubits, especially for those using a cavity where an explicit expression for the transmission or reflection is required.
\end{abstract}


\maketitle


\section{Introduction}

Microwaves are a natural frequency range for superconducting qubits, 
which  makes them especially attractive for applications related to detecting a single microwaves photons \cite{PhysRevA.69.062320,PhysRevX.12.011026,Zagoskin2011}, where one has to contend with strong ambient noise. These important studies relates not only to pivotal technological applications, but also to  paradigm building research such as searching of galactic axions \cite{Duffy_2009,BECK20156,PhysRevLett.126.141302}, one of dark matter candidates.
Due to all these reasons, experimental and theoretical studies of the  interaction of  
a microwave radiation
with arrays of qubits attract increasing research attention
\cite{PhysRevLett.107.240501,PhysRevA.76.042319,wendin2017quantum,wallraff2004strong,
PhysRevLett.91.097906,Hoi_2013,Brehm2021}.

Despite a significant progress over the last decades in the theoretical investigation of the interaction of superconducting qubit systems with electromagnetic field (EMF) \cite{rakhmanov2008quantum,zagoskin2009quantum,savel2012two,ivic2016qubit,PhysRevX.7.031024}, the applied approaches are predominantly based on rather qualitative descriptions of the system under consideration to effective equations, which limits applicability of such models by very specific problems. 
Here we develop a consistent and comprehensive theory of a quantized electromagnetic field interacting with a transmon qubits embedded in a superconducting waveguide cavity. 

The main distinctive feature of our approach  \cite{PhysRevLett.126.141302,GU20171,Zagoskin2011,Schuster2007,Inomata2016,PhysRevA.105.033519,PhysRevB.105.104516} is that we start from a consistent non-relativistic quantum electrodynamic (QED) formalism describing both photon and Cooper pair fields,  
where decoherence is derived explicitly. 
Within this framework we distinguish the cavity mode from the external mode, as the photon field,  while the transmon field has to be built as superpositions of  two mode state describing tunnelling through Josephson junctions as the electron field. This description enables us to express the model parameters in terms of fundamental  quantities associated with the architectural geometry of a system consisting of transmons and coplanar waveguide lines. 

Usually, when using a resonator waveguide, the qubits are probed theoretically by making use of the power spectral function or correlation function providing the spectral properties \cite{Zagoskin_2013,PhysRevA.74.042318}. However this analysis is not explicitly  related to the transmission or reflection factor that is really measured. Many missing informations  occur, as for instance the intensity peak, that may be crucial for a full experimental characterization. 

In this work, we apply the developed theoretical framework to a microwave photon detection scheme based on the Stark shift response to the probe beam to calculate  explicitly the  transmission signal, depending on the parameters of resonant cavity and the properties of the photons to detect. We also discuss some conditions where the system under study is capable of working as a single photon detector.

Other proposals on the single photon detection \cite{PhysRevLett.102.173602}, in particular based on the itinerant photon \cite{PhysRevX.10.021038,PhysRevA.79.052115,PhysRevLett.120.203602},  exist in the literature but they do not employ explicitely a probe beam used for a subsequent transmission analysis of the Stark shift profile. In contrast in our work, the performance of a detection  relies on the ability of this macroscopic transmitted beam to probe a single photon signal, which can be interpreted as an amplification process.

If the actual trends in single photon detection is to separate the storage cavity from the measurement cavity to read out the qubit \cite{Johnson_2010}, 
it is at the cost of using a time protocol of different pulses and adiabatic coupling and decoupling processes by means of a magnetic field, preventing the possibility of a continuous measurement. Our approach aims at benefiting from this last property in order to characterize the nature of the beam of single photon entering into the cavity. 
It could be viewed as a more involved theoretical  description of the early experiment 
\cite{Schuster2007} with an emphasis on the quantum 
state discrimination (coherent, incoherent, thermal) under  various parameter conditions, including the possibility of using many qubits.      


The paper is organized as follows. In Section \ref{sec:mod} we 
 provide theoretical approaches
 leading to a practical model that allows to 
 describe the transmission signal.
In Section \ref{sec:Res}, we discuss the results obtained for various parameters conditions before ending with the conclusion in Section \ref{sec:conc}. All accounts explained in those sections are explicitly detailed in the appendices.

\begin{figure}
\begin{center}
\includegraphics[width=9cm]{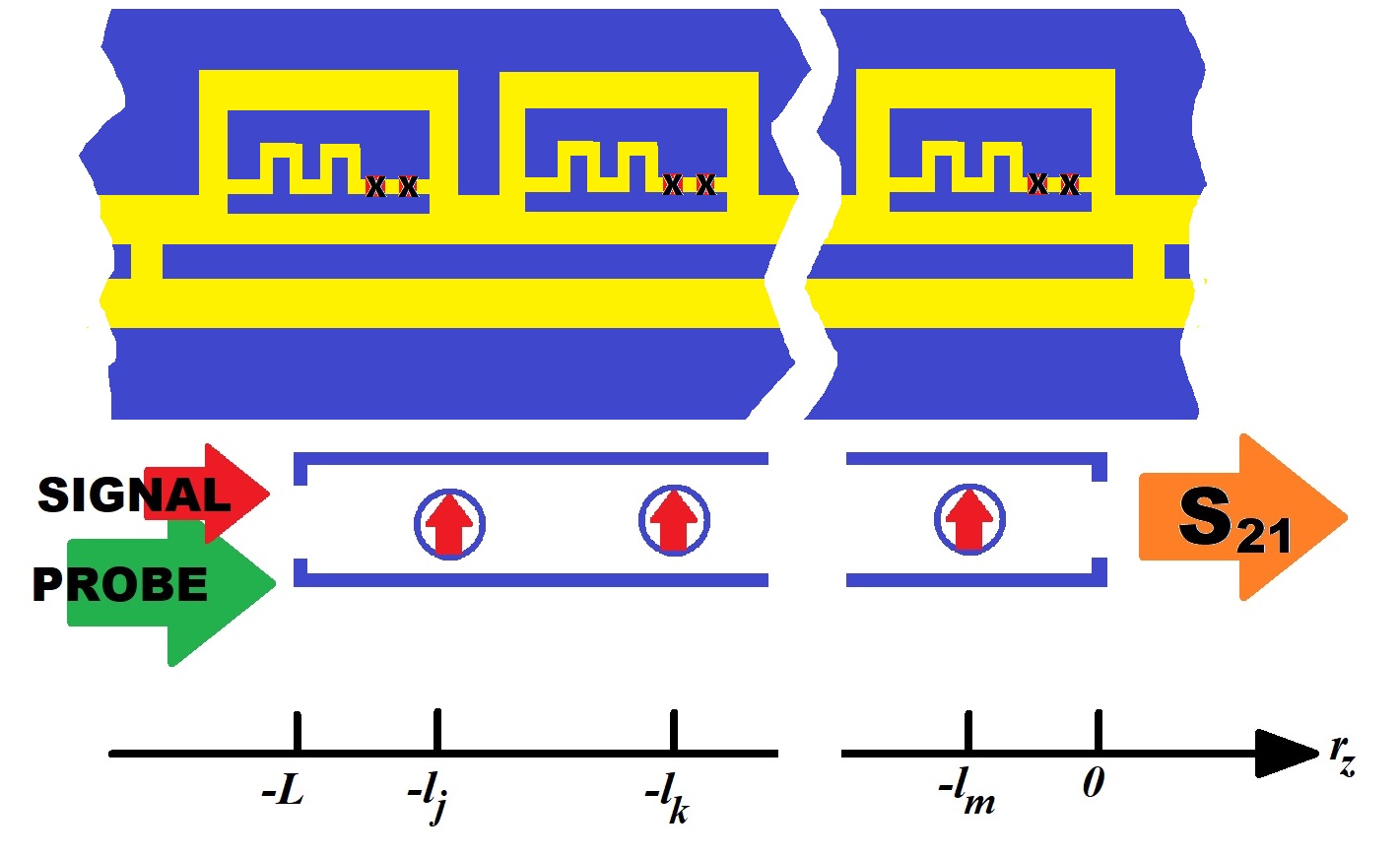}
\end{center}
\caption{\label{fig0} 
(Color online) {\bf Upper panel}: Artistic rendering of the QED device with an array of transmon qubits; yellow color corresponds to insulator region, blue shows the superconductor region; crosses indicate the position of the Josephson junctions.
{\bf Lower panel}: Schematic representation of the QED circuit associated with the device depicted in the upper panel; blue box with open edges is a cavity resonator; ovals with vertical arrow are the qubits; horizontal arrows represent the input signal, probe field and measured transmission ${\bf S}_{21}$; axis is used to indicate the positions  $-l_j$ of the $j^{th}$ transmon along the direction $r_z$ (the  origin is placed after the device as the usual convention in transmission line theory \cite{Pozar:882338}). 
}
\end{figure}

\section{Modelling}
\label{sec:mod}

A generic quantum electrodynamics (QED) description of a qubit array embedded in a microwave waveguide is discussed in appendices \ref{A}, \ref{B} and \ref{C}.
Here we consider a QED system, which represents an array of transmon qubits coupled to a coplanar waveguide resonator, see Fig.\ref{fig0}.  The upper part of the figure shows an artistic image of the system, while the lower panel provides with its schematic representation. In the schematic representation, the probe (green arrow online) and  signal (red arrow online) fields form the input and transmission $S_{21}$ (orange arrow online) is the output for the measurement. 

In appendix \ref{A}, we define the 3D Hamiltonian for EMF and quantum Cooper pair field and transform it into the more convenient dipole representation. In appendix \ref{B},  
we use a restricted set of basis for the Cooper field that describes the tunneling at the Josephson junction  while,  in appendix \ref{C}, we restrict the basis to the TME modes.    
For the transverse electromagnetic components propagating within the waveguide in the direction  $r_z$  as time $t$ changes, 
the two canonically conjugated field variables are the integrated nonlocal flux $\hat \Phi_B(r_z,t)$ through the insulator region surrounding perpendicularly   the signal (blue line in the middle) and an infinitesimal charge element $dQ(r_z,t)$.
They obey the nontrivial commutation relations:
\begin{eqnarray}\label{commwave}
[\hat \Phi_B(r'_z,t),\partial_{r_z} \hat Q(r_z,t)]=i\hbar \delta(r_z -r'_z) \, .
\end{eqnarray}
For freely propagating electromagnetic waves,  these operators are related to the local potential $V(r_z,t)$ through the Lenz law $\hat V(r_z,t)=-\partial_t \hat \Phi_B(r_z,t)$ and the local current $I(r_z,t)$ through charge conservation $\partial_{r_z}\hat I(r_z,t)=-\partial_t \partial_{r_z} \hat Q(r_z,t)$.

The transmons  denoted by $j=1 \dots N$ are arranged in an array near the  waveguide. 
They are formed of two superconducting islands shaped as interlocking combs (to increase mutual capacitance) and connected by a pair of Josephson junctions.
The electrodynamics of transmon qubits can be described in terms of
the real charge operators $\hat q_{j}(t)$ of the Cooper pair having charge $2e$ on each side of the islands and their 
conjugated flux operators $\hbar\hat \Phi_j(t)$ that correspond to the phase difference between the wave functions of the two macroscopic quantum states, associated to each island, multiplied by $(2e)/\hbar$.

Similarly, for the longitudinal modes related to $j^{th}$ transmon the nontrivial commutation relations read
\begin{eqnarray}\label{commplasmon}
[\hat \Phi_j(t),\hat q_{j'}(t)]=i\hbar \delta_{j,j'}.
\end{eqnarray}
Although for the description of both the waveguide and the qubit we use similar type of variables, they have some differences.
Namely, the charge on the surface of the waveguide is determined by the Gauss theorem  for the electric field. It is induced  by a transverse mode of the electromagnetic field and differs from the  one $ \hat q_{j'}$ resulting from a longitudinal Coulombian field commonly  referred to plasmon excitations.

Unlike the case of bulk superconductors, the (plasmon) frequencies of our Josephson structure lie within the microwave range. Therefore these excitations have their own dynamics, which needs to be taken into account. 
More details are given in  appendices \ref{B} and \ref{C}.
Despite the above fundamental difference between the introduced charge components, the defined set of canonically conjugated variables allows to describe a quantum device 
as a circuit of lump elements (quantum circuit). Such a description greatly simplifies the understanding of the underlying physics since the proposed derivation allows a  unambiguous well-defined physical representation of these variables in Fig.\ref{fig0}.

Using the outcomes of appendices \ref{A}-\ref{C} with the addition of the description of the resonator in 
appendix \ref{D} and a unitary transformation on transmon variables in   appendix \ref{E}, we derive the Hamiltonian 
 in terms of these quantum circuit variables:
\begin{eqnarray}\label{HA}
\hat H
&=& -\sum_{j=1}^N E_{J,j}
\cos\left[\frac{2e}{\hbar}(\hat \Phi_j- f_j \hat \Phi_B (-l_j))\right] 
+ \frac{\hat q^2_j}{2C_j}+
\nonumber 
\\
\int_{-\infty}^\infty \!\!\!\!\!&dr_z&
\left[\frac{\partial_{r_z}\hat{Q}^2(r_z)}{2C'} + \frac{\partial_{r_z}{\hat \Phi_B}^2(r_z)}{2L'}\right]
+ \frac{\hat{Q}^2_-(-L)+\hat{Q}^2_+(0)}{2C} \, , 
\nonumber \\
\end{eqnarray}
where $E_{J,j}$ is the Josephson junction energy responsible for an inductance-like coupling with 
 the EMF  and $C_j$ the capacitance of the transmon. 
The geometric  prefactor $f_j$ controls the effective flux inside the transmon and $-l_j$ is the position of the qubit inside the cavity (see Fig. \ref{fig0}).
$L'$ and $C'$ are the line inductance and capacitance respectively and $C$ is the capacitance  at the ends of the cavity having length $L$ 
(see Fig.\ref{fig0}). 
Since the capacitances $C$ are placed longitudinally, their charge corresponds precisely to the effective charge accumulated 
through the waveguide line. The integration has to be taken over the whole length of the waveguide which is assumed infinite from both sides.
 The integrated charge $\hat{Q}_\pm(r_z,t)=\pm \int_{\pm \infty}^{r_z} \partial_{r_z} \hat Q(r_z,t)dr_z$ is a nonlocal operator,  which is opposite to the charge accumulated  throughout the waveguide. The integration is from the left side of the waveguide $-$ or from the right side $+$. 
 
Note, the presence of a
cavity  suppresses the superradiative relaxation rate that scales like $N^3$ as observed experimentally in \cite{Brehm2021} in order to  remain mainly with the dephasing rate $\Gamma_{\phi,j}$ and some other additional rates $\Gamma_j$ coming from other decay into other electric modes (TM or TE). 

After restricting consideration by two first transmon energy levels,  the Hamiltonian (\ref{HA}) is rewritten in appendix \ref{E} in a nonlocal form using new field operator and a qubit variable as:
\begin{eqnarray}\label{HB}
\hat H
&=&  
\int_{-\infty}^\infty \!\!\!dr_z
\frac{
\hat{\dot{\alpha}}^2(r_z)+ c^2 
\partial_{r_z} \!\hat{\alpha}^2(r_z)}{2} +
\frac{\hat{Q}^2_-(-L)+\hat{Q}^2_+(0)}{2C} 
\nonumber \\
&+&\sum_{j=1}^N\frac{\hbar \omega_j}{2}\hat \sigma^z_j-\sqrt{\hbar c}\kappa_j\hat \sigma^x_j\hat{\alpha}(-l_j) \, ,
\end{eqnarray}
where the flux is now identified as $\hat \Phi_B(r_z)= \hat{\alpha}(r_z)/\sqrt{C'}$ and the charge has now a non local expression: $\hat{Q}_\mp(r_z) = \pm \int_{\mp \infty}^{r_z} dr'_z\sqrt{C'}\hat{\dot{\alpha}}(r'_z)$. The new parameters of the Hamiltonian (\ref{HB}) are
$c=1/\sqrt{L'C'}$, 
$\hbar\omega_j=\sqrt{4e^2 E_{J,j}/C_j} - e^2/( 2C_j)$ and $\kappa_j=\frac{2e}{\hbar}f_j\sqrt{E_{J,j}\omega_j/(2cC')}$. 
Then the nontrivial commutation relation between the conjugate variables becomes:
\begin{eqnarray}
[\hat{{\alpha}}(r_z,t),\hat{\dot{\alpha}}(r'_z,t)]=i\hbar \delta(r_z -r'_z) \, .
\end{eqnarray}
Note that the Heisenberg equations imply that the conjugated momentum is not a time derivative $\hat{\dot{\alpha}}(r_z,t)\not= \partial_t\hat{\alpha}(r_z,t)$.
 Depending on the waveguide architecture, the propagation speed $c$ can be significantly smaller that the speed of light in the free space.  For example, in the specific case of coplanar waveguide depicted in Fig. \ref{cpw} of appendix \ref{C},  $C'$ should be substituted by an efficient capacitance $C'_{eff}$, which accounts also the interface current to the  dielectric materials composing the waveguide. Such consideration corrects the estimated speed of propagation $c$ and allows to improve previous theoretical description  \cite{Watanabe_1994,375223}.

As shown in appendix \ref{E}, the quantities of interest available for  measurement in experiment are the forward ($+$) and backward wave ($-$) quantum operators propagating throughout the waveguide:
\begin{eqnarray}
\hat A_\pm (r_z,t)&=&\frac{\hat V(r_z,t) \pm Z \hat I(r_z,t)}{2\sqrt{Z}}
\nonumber \\
&=&\pm c^{3/2} \partial_{r_z}\hat \alpha(r_z,t)- c^{1/2} \hat {\dot \alpha}(r_z,t) \, ,
\end{eqnarray}
where the impedance is $Z=\sqrt{L'/C'}$. 

Taking the temporal Fourier transform, the transmission of the signal is ${ S}_{21}=\langle \hat A_{\omega,+}(0)\rangle/
\langle \hat A^{in}_{\omega,+}(0)\rangle$, where $\hat A^{in}_{\omega,+}(r_z)$ is the input field operator, while the full operator $\hat A_{\omega,+}(r_z)=\hat A^{in}_{\omega,+}(r_z)+\hat A^{sc}_{\omega,+}(r_z)$ contains the scattering contribution $\hat A^{sc}_{\omega,+}(r_z)$ resulting from the interaction  of EMF  with the cavity and the qubits. 

Despite the simplification in terms of qubits, the Hamiltonian (\ref{HB}) cannot still be solved exactly in the continuous limit. For a high quality cavity however, it is possible to separate the continuous field of the waveguide and the other degree of freedom inside the cavity. 
In appendix \ref{F}, these cavity modes are determined in terms of the two end capacitances  $C$  assumed to be small in comparison to the line capacitance  $C'$, i.e. $C \ll C'L$. 
In the leading order, we obtain the harmonics 
$\omega_{n,0}=n\pi c/L$ for any natural integer. 
To the next orders, the  frequency shift and the 
half-width are:
\begin{eqnarray}\label{spec}
\omega_n-\omega_{n,0} =-\frac{2C}{C'L}\omega_{n,0} \, ,  \quad \quad 
\gamma_n=\frac{4c}{L}\left(\frac{C \omega_{n,0} }{cC'}\right)^2 \, ,
\end{eqnarray}
in agreement with the reference \cite{doi:10.1063/1.3010859}.
The quality factor  of $n^{th}$ mode is defined as:
\begin{eqnarray}\label{Qn}
{\cal Q}_n=\frac{\omega_n}{\gamma_n}\cong 
\frac{C'L}{2\omega_{n,0} Z C^2} \, .
\end{eqnarray}
It  differs from the definition in 
 \cite{doi:10.1063/1.3010859}, where 
${\cal Q}_n=C'L/(2\omega_{n,0} R_L C^2)$ 
with $R_L$  being the resistance outside the resonator, usually chosen to match the line impedance $Z$. 
Within our consideration, we don't use any resistance 
since, in the situation where the qubits are absent, a lossless  cavity does not dissipate any energy but rather scatters part of its energy. 
Therefore, the finite linewidth previously associated with dissipation through the effective external load resistance is understood  in our case in terms of impedances of the EMF  without any dissipation.
The expression (\ref{Qn}) does not account for the quality factor saturation
\cite{doi:10.1063/1.3010859}. The latter can be explained by the additional currents at the dielectric interfaces.  

In appendix \ref{H}, we restrict the cavity dynamics to the first mode $n=1$ described in terms  of the annihilation operator $\hat a_1$ with a frequency close to the qubit frequencies and apply the rotating wave approximation.
We arrive eventually at the Jaynes-Cumming Hamiltonian for many qubits \cite{fink2008climbing}
which additionally contains the external backward and forward fields:
\begin{eqnarray}\label{Stoch}
\frac{\hat H(t)}{\hbar}&=&\sum_{j=1}^N
-\frac{\delta \omega_i}{2}\hat \sigma^z_i
+g_j(\hat a^\dagger_1 \hat \sigma^-_j+\hat a_1 \hat \sigma^+_j)
+(\omega_{c} -\omega)\hat a_1^\dagger \hat a_1
\nonumber \\&+&i\sqrt{\frac{\gamma_c }{\hbar \omega_c}}\sum_\pm
\hat A^{in}_{\pm}(0,t)(\hat a_1 e^{-i\omega t}- \hat a^\dagger_1 e^{i\omega t}) \, .
\end{eqnarray}
Here, the coupling  $g_j=\kappa_j \cos(\pi l_j/L) /\sqrt{\pi}$ depends on the position of the qubits, while   the cavity frequency $\omega_c=\omega_1$ and the damping $\gamma_c=\gamma_1$ are adjusted to the first resonance mode. The latter induces a decay rate of the qubits  as a result of the Purcell effect.
At the difference of a similar scheme described in 
\cite{PhysRevA.76.042319}, the input field $\hat A^{in}_{\pm}(0,t)= \langle \hat A^{in}_{\pm}(0,t)\rangle +\delta \hat A^{in}_{\pm}(0,t)$ not only accounts for the classical coherent part of the field but also for the quantum part of the field which either corresponds to the fluctuating vacuum mode viewed as an input quantum noise \cite{gardiner00} or the incoherent mode to be detected. This is an interesting fundamental feature that these two apparently unrelated contributions originate from  one unique quantum field. A similar expression appears in absence of the cavity in the appendix 
\ref{G} and leads to similar developments in \cite{PhysRevA.88.043806,doi:10.1126/science.1181918}.

The Hamiltonian (\ref{Stoch}) with such a time dependent quantum operator is generally treated using the master equation  approach or stochastic techniques \cite{gardiner00}. The former is used in our work in order to get analytical expressions derived in Appendices \ref{G} and \ref{H}.  
In complement to this approach,  the line transmission ${ S}_{21}$ is determined for a coherent forward field passing through the cavity in appendices \ref{F} and \ref{H}.
For a frequency $\omega_p$ close to the first resonance mode of the cavity, it is a sum 
of two contributions from the cavity and from the qubits:
\begin{eqnarray}\label{S21m}
&&{ S}_{21}
 =
{S}^{cv}_{21}+ {S}^{qb}_{21}=
\nonumber \\
&&
\sum_{\pm}
\frac{-i\gamma_c/2}{\omega_p \mp \omega_c + i\gamma_c/2}
\left[1 \pm
\sqrt{\frac{\hbar \omega_c}{\gamma_c}}\sum_{j=1}^N \frac{g_j\langle \hat \sigma_{\omega_p,j}^\mp \rangle}{\langle  \hat A^{in}_{\omega_p,+}(0) \rangle} \right]\, .
\end{eqnarray}
Comparing with earlier works \cite{Zagoskin_2013,PhysRevB.103.064503}, this last expression suggests that, since the transmission factor 
is only sensitive to the lowering and raising operators, the qubit state does not have to be  prepared in a superposition of ground and excited states for an effective action. 
As a consequence, the discrete photon set inside a cavity  is monitored  continuously using a coherent probe beam tuned at the qubit frequency without any additional protocol required.

\section{Results and discussions}
\label{sec:Res}
\subsection{Setup}

As detailed in appendix \ref{I} and illustrated in Fig.\ref{fig0}, the setup involves two continuous beams: the weak photon signal beams with a flux ${\cal J}$ is  tuned at the cavity resonance $\omega=\omega_c$; a second coherent probe beam of frequency  $\omega_p$ close to the qubit frequencies. 

For the purpose of quantum non-demolition measurements (QND) \cite{Zagoskin_2013},
the cavity and qubit frequencies, 
$\omega_c$  and $\omega_j$  respectively, must be strongly detuned 
such that their difference much larger compared to the coupling 
$|\omega_j-\omega_c| \gg g_j$.
Then, as shown in appendix \ref{H}, the inelastic processes are suppressed, and the effective qubit-cavity interaction has  the form photon number-dependent qubit energy shift with the Hamiltonian terms $\chi_j \hat \sigma^z_j \hat a_1^\dagger \hat a_1$. The  energy shift $\chi_j= g_j^2/(\omega_j-\omega_c)$ 
is usually referred as the Stark shift and must be very low in comparison to the microwave frequencies. The latter  discriminates the peaks of cavity associated to different photon population in the probe transmission spectrum. 

Using the formula for the relaxation rate $\Gamma_j = \kappa^2_j/\omega_j$ in appendix \ref{F} in absence of a cavity, the constant $\chi_j$ is estimated using the data available in \cite{Brehm2021}. Namely the  radiative relaxation rate $\Gamma_j \sim  2\pi \, 6.4 {\rm MHz}$ and the qubit frequency $\omega_j \sim 2\pi \,8 {\rm GHz}$, we estimate  the coupling $g_j \sim \kappa_j/2 \sim \sqrt{\omega_j \Gamma }/2 \sim 2\pi \, 100 {\rm MHz}$ and deduce $\chi_j \sim 2\pi 10 {\rm MHz}$ for $|\omega_j-\omega_c|\sim 2\pi \, 1 {\rm GHz}$. 

In presence  of a cavity with the relaxation rate $\gamma_c$, we add phenomenologically two other damping mechanisms associated to the qubits \cite{doi:10.1126/science.1181918,PhysRevLett.107.240501}. 
The first accounts for the qubit radiative time $T_{1,j}=1/\Gamma_j$ decaying in other modes than in the TEM mode, while the second for the 
dephasing time (lifetime of  coherent superposition of qubits state  not described explicitly in our formalism) $T_{2,j}=1/\Gamma_{\phi,j}$. We shall choose the best value $\Gamma_{\phi,j}={\rm 2\pi\,250 kHz}$ obtained in \cite{Brehm2021} keeping in mind that these are difficult to control in the fabrication process.
Although the origin of the time $T_{2,j}$ has been thoroughly discussed in \cite{PhysRevA.76.042319}, the possibility of other decay electromagnetic modes (TE or TM) included in  $T_{1,j}=1/\Gamma_j$ has not been mentioned and could be negligible in an optimal designed circuit \cite{Brehm2021}. In contrast, according to our estimation of the experimental data \cite{doi:10.1126/science.1181918} in appendix \ref{F}, these other modes contribute to 90\% of the relaxation in comparison to the TEM mode.

The quantum input states under consideration are detailed in appendix \ref{I} and are described as follows:
\begin{enumerate}
\item{\bf Coherent state:} This state has non zero amplitude average $\langle \hat a_1 \rangle \not= 0$ with a well defined phase with infinite decoherence time and is the most common input radiation produced by a microwave source. 
\item{\bf Incoherent state:}
This state is monochromatic and  has a decoherence time $\tau_c$ much larger than the cavity lifetime $2/\gamma_c$ with 
the zero amplitude average and thus no defined phase. These input radiations may be  produced by a fluorescent source resulting from 
a spontaneous emission of artificial atoms. 
{\item \bf Thermal state:}
A quantum  state with a  much smaller decoherence time $\tau_c$ such that  
$\tau_c \chi_j  \ll 1$ has  a much larger broadening of its frequency spectrum. 
Once it passes through the cavity, it cannot be distinguished from a thermal state resulting from a black body radiation. In comparison with the coherent and incoherent cases, for the thermal case, we note  a much more important increase of line broadening, as well as a reduction of the photon flux once 
$\tau_c < 2/\gamma_c$.
\end{enumerate}

Quite generally, the calculation of the transmission $S_{21}^{\nu}$ is done in appendix \ref{I} for the three cases of photon signal above with the labelings $\nu={\rm coh,inc,th}$ as a function of the probe frequency $\omega_p$.
We obtain the general analytical formulas (\ref{sigcoh}) (\ref{siginc}) and (\ref{sigth}) combined with (\ref{S21p}) that allow to deal with imperfect situations of absence of line resolution.

\subsection{Well-resolved spectrum:  $\gamma_c \ll \chi_j$}

For the positive  frequency $\omega_p$ close to the qubit frequencies and the signal frequency $\omega$  with a photon flux per unit of time ${\cal J}$ exactly tuned at the cavity resonance $\omega=\omega_c$ and $\gamma_c \ll \chi_j$, the transmission can be approximated as: 
\begin{eqnarray}\label{S21}
&&{ S}^{\nu}_{21} \cong-\frac{i\gamma_c}{2(\omega_p-\omega_c)
}+ 
\nonumber \\ 
&&\!\! \!\!\!\! \sum_{j=1}^{N}
\sum_{n=0}^\infty
\frac{i P_\nu(n)\gamma_c\chi_j/[2(\omega_j-\omega_c)]}{\omega_p -[\omega_j
+2\chi_jn -
i(\Gamma_{\nu}(n) +\Gamma_j/2+\Gamma_{\phi,j})]} \, ,
\end{eqnarray}
where $P_\nu(n)$ is the probability distribution and $\Gamma_{\nu}(n)$ is the decay rate due to the cavity with explicit expressions given in table \ref{table:1} for various input quantum states of the signal field $\nu$.
Note that the formula for $\Gamma_{\nu}(n)$ agrees for coherent state with the one derived in \cite{PhysRevA.74.042318} using the spectral power for the spin. In contrast, a more fundamental approach allows to determine explicitly the intensity peak for the transmission.

For the validity of formula Eq.(\ref{S21}) in the incoherent case, we need also to satisfy the high quality factor condition   $\gamma_c \ll (1+\overline{n})(\Gamma_j/2+\Gamma_{\phi,j})/\overline{n}$. 

\begin{table}[h!]
\begin{tabular}{|c|c|c|c|c|}
\hline 
State  &$\nu$ & $P_{\nu}(n)$ & $\Gamma_{\nu}(n)$ &  $\overline{n}$ \\
\hline
Vacuum & $0$ & $\delta_{n,0}$ & $0$  & $0$  \\
Coherent & coh & $e^{-\overline{n}}
{\overline{n}}^{n}/n!$ & $(n+\overline{n})\gamma_c/2$ & $2{{\cal J}}/\gamma_c$ \\
Incoherent & inc & $\overline{n}^n/(\overline{n}+1)^{n+1}$ & $n\gamma_c/2$ & $2{{\cal J}}/\gamma_c$ \\
Thermal & th & $\overline{n}^n/(\overline{n}+1)^{n+1}$ &  $[(2\overline{n}+1)n+\overline{n}]\gamma_c$ &  $
\tau_c {\cal J}$ \\
\hline 
\end{tabular}
\caption{Representation of the probability distribution and cavity decay rate 
for a given average photon number inside the cavity $\overline{n}$. The index $\nu$ labels the various analyzed quantum states while $\nu=0$ corresponds to the background response without any signal $\overline{n}=0$.  The last column shows the  proportionality relation between the cavity photon average and the photon flux ${\cal J}$.}
\label{table:1}
\end{table}

\begin{figure}
\begin{center}
 \includegraphics[width=9cm]{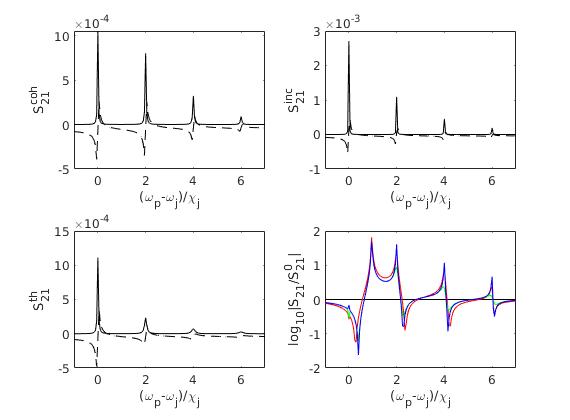}
\end{center}
\caption{\label{fig1} 
Transmission $S^\nu_{21}$  through the cavity containing one qubit and the average signal photon number $\overline{n}=1$.  System parameters: Values: 
$\omega_j=2\pi\, 10 {\rm GHz}$, $\omega_c=2\pi\, 9 {\rm GHz}$, $\chi_j=2\pi \,10  {\rm MHz}$,  $\gamma_c= 2\pi\, 100 {\rm 
kHz}$, $\Gamma_j + 2\Gamma_{\phi,j}= 2\pi\, 250 {\rm kHz}$   \\
{\bf Panels 1-3:} 
Real (solid line) and imaginary (dashed line) of $S^\nu_{21}$ for the coherent (panel 1), incoherent (2) or thermal (3) photon field. 
Peaks are labeled by the actual photon number $n$. Peak heights are proportional to $P_\nu(n)$ with linewidths:
$\Gamma_{\nu}(n) +\Gamma_j/2+\Gamma_{\phi,j}$. \\
{\bf Panel 4:} 
Transmission relative to the case of no signal, $|S^\nu_{21}|/|S^0_{21}|$, for coherent (red), incoherent
(blue) and thermal (green) photon field. 
Transmission  $S^\nu_{21}$  for a cavity containing one  qubit with an average signal  photon number $\overline{n}=1$. } 
\end{figure}

\begin{figure}
\begin{center}
 \includegraphics[width=9cm]{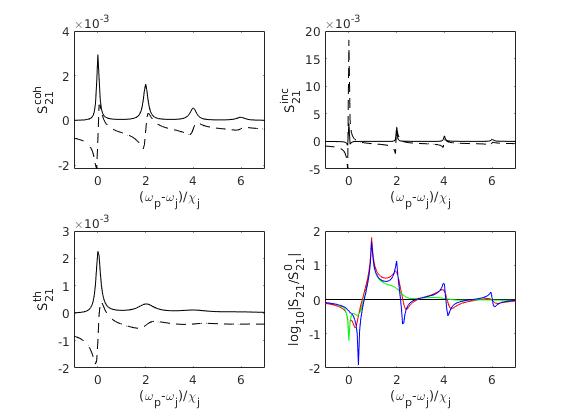}
\end{center}
\caption{\label{fig2} Same as Fig.\ref{fig1} for a lower cavity quality factor. Parameters: $\omega_j=2\pi\, 10 {\rm GHz}$, $\omega_c=2\pi\, 9 {\rm GHz}$, $\chi_j=2\pi \,10  {\rm MHz}$, $\gamma_c= 2\pi\, 1 {\rm 
MHz}$, $\Gamma_j + 2\Gamma_{\phi,j}= 2\pi\, 250 {\rm 
kHz}$.}
\end{figure}

\begin{figure}
\begin{center}
 \includegraphics[width=9cm]
 {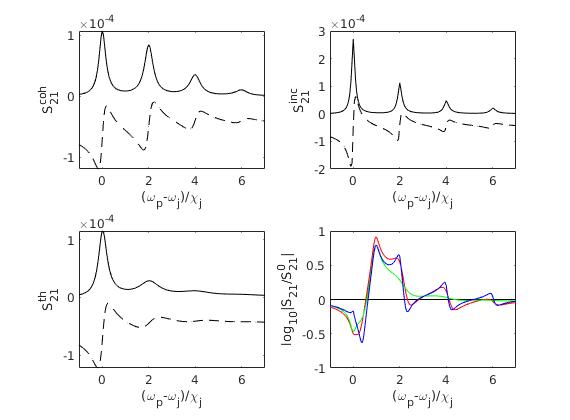}
\end{center}
\caption{\label{fig2bis} Same as Fig.\ref{fig1} for lower Stark shift.
Parameters: $\omega_j=2\pi\, 10 {\rm GHz}$, $\omega_c=2\pi\, 9 {\rm GHz}$, $\chi_j=2\pi \,1 {\rm MHz}$, $\gamma_c= 2\pi\, 100 {\rm 
kHz}$, $\Gamma_j + 2\Gamma_{\phi,j}= 2\pi\, 250 {\rm 
kHz}$.}
\end{figure}

\begin{figure}
\begin{center}
 \includegraphics[width=9cm]{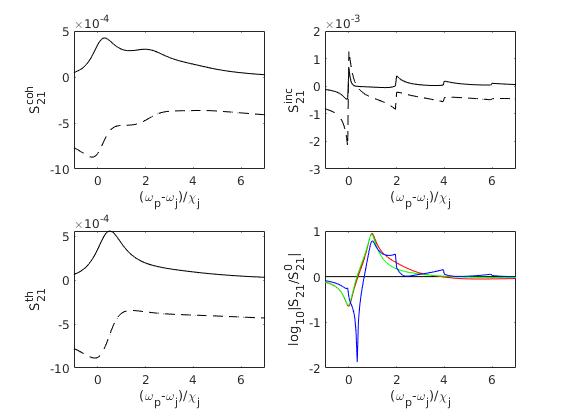}
\end{center}
\caption{\label{fig3}
Same as Fig.\ref{fig1} for lower Stark shift and quality factor.
Parameters: $\omega_j=2\pi\, 10 {\rm GHz}$, $\omega_c=2\pi\, 9 {\rm GHz}$, $\chi_j=2\pi \,1 {\rm MHz}$, $\gamma_c= 2\pi\, 1 {\rm 
MHz}$, $\Gamma_j + 2\Gamma_{\phi,j}= 2\pi\, 250 {\rm 
kHz}$.}
\end{figure}

\begin{figure}
\begin{center}
\includegraphics[width=9cm]{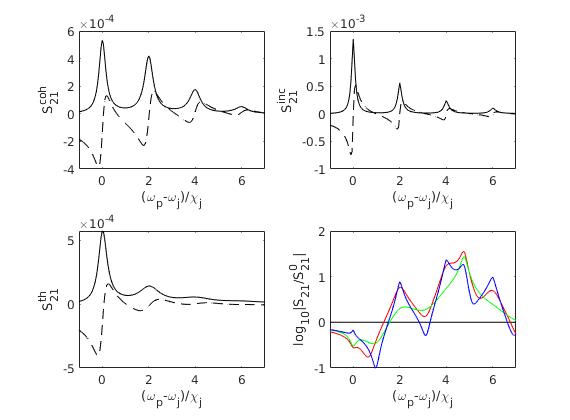}
\end{center}
\caption{\label{fig35q}
Same as Fig.\ref{fig2bis} 
for 5 qubits.
Parameters: $\omega_j=2\pi\, 10 {\rm GHz}$, $\omega_c=2\pi\, 9 {\rm GHz}$, $\chi_j=2\pi \,1 {\rm MHz}$, $\gamma_c= 2\pi\, 1 {\rm 
MHz}$, $\Gamma_j + 2\Gamma_{\phi,j}= 2\pi\, 250 {\rm 
kHz}$.}
\end{figure}

For one qubit, the graphs for the  transmission  are displayed in Figs.\ref{fig1}, \ref{fig2}, \ref{fig2bis}, \ref{fig35q} and 
\ref{fig6} and represent a sideband comb structure whose the distribution determines the nature and the average of the signal. The other graphs in Fig.\ref{fig3} or \ref{fig5} display too much broadening to be approximated by Eq.(\ref{S21}) despite the possible line resolution for the incoherent case. 
Fig.\ref{fig1} corresponds to the observed Stark shift with the best to date values of coupling and quality factor so far obtained in current experiments, while other figures represent results for less perfect devices.

The probe beam acts in the range around the qubit frequency so that the transmission due to the cavity $S_{21}^{cv}$  in the first term of Eq.(\ref{S21}) is far off resonance and therefore can be neglected as shown in Fig.\ref{fig1} where  the imaginary part for large $\omega_p$ is close to zero. In contrast in Figs.\ref{fig2},\ref{fig2bis} and \ref{fig6}, the imaginary part becomes asymptotically negative
and may introduce a significant background noise if not well distinguished from the signal in the real part. 

As shown in Fig.\ref{fig35q}, the increase of the qubit number contributes to suppress the background from the cavity and  increase the sensitivity to the signal detection. 

\begin{figure}
\begin{center}
 \includegraphics[width=9cm]{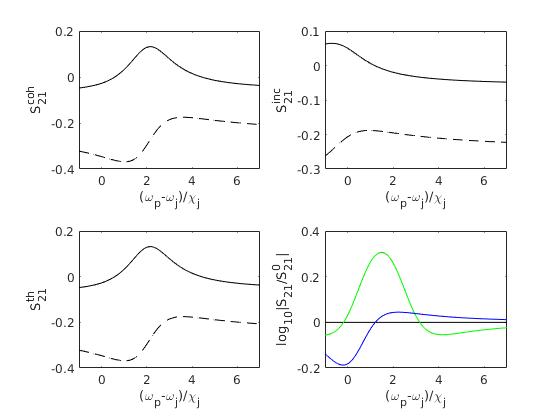}
\end{center}
\caption{\label{fig4}
Same as Fig.\ref{fig1} in the limit of very low quality factor. The thermal case becomes indistinguishable from the coherent case.
Parameters: $\omega_j=2\pi\, 10 {\rm GHz}$, $\omega_c=2\pi\, 9 {\rm GHz}$, $\chi_j=2\pi \,100 {\rm kHz}$, $\gamma_c= 2\pi\, 500 {\rm 
MHz}$, $\Gamma_j + 2\Gamma_{\phi,j}= 2\pi\, 250 {\rm 
kHz}$.}
\end{figure}

\begin{figure}
\begin{center}
 \includegraphics[width=9cm]
 {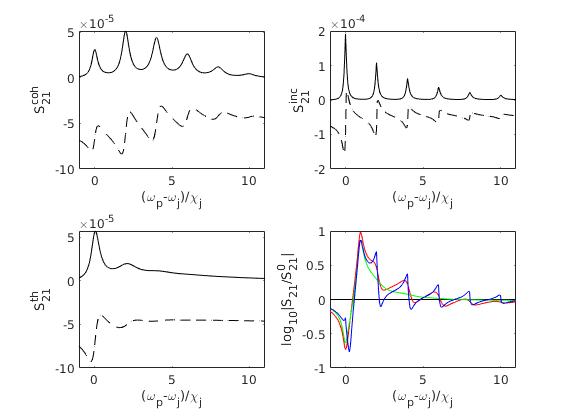}
\end{center}
\caption{\label{fig6}
Same as Fig.\ref{fig2bis} for $\overline{n}=2$ for high quality factor. 
Parameters: $\omega_j=2\pi\, 10 {\rm GHz}$, $\omega_c=2\pi\, 9 {\rm GHz}$, $\chi_j=2\pi \,1 {\rm MHz}$, $\gamma_c= 2\pi\, 100 {\rm 
kHz}$, $\Gamma_j + 2\Gamma_{\phi,j}= 2\pi\, 250 {\rm 
kHz}$.}
\end{figure}

\begin{figure}
\begin{center}
 \includegraphics[width=9cm]{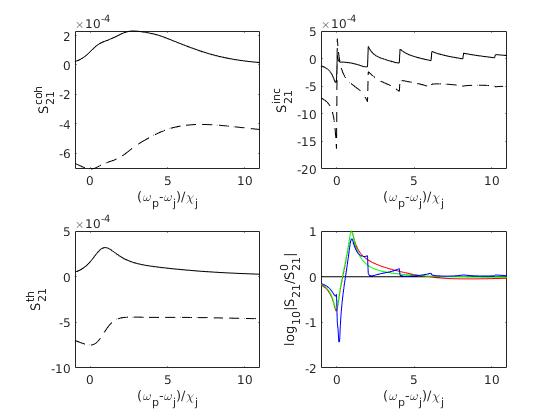}
\end{center}
\caption{\label{fig5}
Same as Fig.\ref{fig3} for $\overline{n}=2$ in the limit of very low quality factor. 
Parameters: $\omega_j=2\pi\, 10 {\rm GHz}$, $\omega_c=2\pi\, 9 {\rm GHz}$, $\chi_j=2\pi \,1 {\rm MHz}$, $\gamma_c= 2\pi\, 1 {\rm 
MHz}$, $\Gamma_j + 2\Gamma_{\phi,j}= 2\pi\, 250 {\rm 
kHz}$.}
\end{figure}

\begin{figure}
\begin{center}
 \includegraphics[width=9cm]{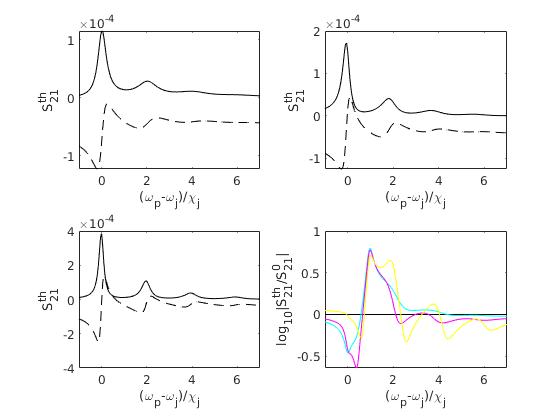}
\end{center}
\caption{\label{fig7} Same as Fig.\ref{fig2bis} for three values of the coherence time 
$\tau_c = (10^{-12},10^{-9},10^{-8}) {\rm sec}/2\pi$ with their respective colors: cyan, magenta and yellow for the panel 4.
Parameters: $\omega_j=2\pi\, 10 {\rm GHz}$, $\omega_c=2\pi\, 9 {\rm GHz}$, $\chi_j=2\pi \,1 {\rm MHz}$,  $\gamma_c= 2\pi\, 100 {\rm 
kHz}$, $\Gamma_j + 2\Gamma_{\phi,j}= 2\pi\, 250 {\rm 
kHz}$.}
\end{figure}



\begin{figure}
\begin{center}
 \includegraphics[width=9cm]{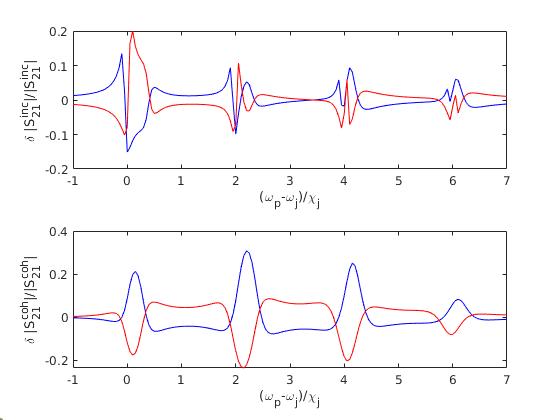}
\end{center}
\caption{\label{fig10} Relative error for values of  Fig.\ref{fig2bis} for the coherent case in panel 1 and incoherent case in panel 2 and  two  values of the detuning  
$\omega -\omega_c = -\gamma_c/3,\gamma_c/3$    corresponding to the blue and red curves respectively.
Parameters: $\omega_j=2\pi\, 10 {\rm GHz}$, $\omega_c=2\pi\, 9 {\rm GHz}$, $\chi_j=2\pi \,1 {\rm MHz}$,  $\gamma_c= 2\pi\, 100 {\rm 
kHz}$, $\Gamma_j + 2\Gamma_{\phi,j}= 2\pi\, 250 {\rm 
kHz}$.}
\end{figure}

As far as the sensitivity of the probe is concerned,  the transmission of qubit for the spectrum scales at resonance like $g_j^2/(\omega_j -\omega_c)^2$ which is always a very small quantity and this imposes that $\gamma_c \gg \Gamma_j , \Gamma_{\phi,j}$ in order 
to have a high response. Therefore a high quality factor is not desired unless the qubit decay rate and dephasing are negligible. Since dephasing is about ${\rm 250 kHz}$, a similar value has to be chosen for the cavity linewidth for an optimum transmission.

\subsection{Lower quality factor regime $\gamma_c \gg \chi_j$}

The case of low broadening of the spectral line may be difficult to achieve experimentally since it requires  high quality factor
and extreme high sensitivity of the response signal. 

In the weak coupling regime and for $\chi_j \ll \sqrt{\gamma_c (\Gamma_j/2+\Gamma_{\phi,j})}$, the line amplitude vanishes except for the main line, but the signal amplitude increases due to the lower quality factor allowing for 
a better transmission of the probe beam. 
We find in appendix \ref{I} the same results for the coherent and thermal cases:
\begin{eqnarray}
&S_{21}^{\rm coh}& = S_{21}^{\rm th}
\nonumber \\
&=&S_{21}^{cv} +
\sum_{j=1}^N
\frac{i\gamma_c\chi_j/[2(\omega_j-\omega_c)]}{\omega_p -[\omega_j
+2\chi_j \overline{n} -
i(\Gamma_j/2+\Gamma_{\phi,j})]} \, .
\end{eqnarray}
The position of the unique peak is displaced by the offset $2\chi \overline{n}$ in comparison to the situation of vacuum signal. 
For the incoherent case the transmission is strongly suppressed due to the large uncertainty of the phase leading to additional broadening. Using the Kummer special function (detailed in Appendix \ref{I}), we find: 
\begin{eqnarray}
\!\!\!{ S}^{\rm inc}_{21} =  S^{cv}_{21}+\sum_{j=1}^N
\frac{i\gamma_c
U\left(1,1,\frac{\omega_p - \omega_j +i(\Gamma_j/2+\Gamma_{\phi,j})}{2\overline{n}\chi_j}\right)
}{(\omega_j-\omega_c)4\overline{n}} \, .
\end{eqnarray}
In this context, the transmission of the pure vacuum state is higher than the one of the mixed incoherent state which acts as an inhibitor of the probe beam propagation.  

Even in presence of large broadening, it is nevertheless still possible 
to discriminate in the probe response between coherent or thermal and  incoherent states in the weak Stark coupling regime as seen in Fig.\ref{fig4}.


\subsection{Influence of a finite coherence time for $\gamma_c \tau_c \ll 1$}

The infinite coherence time limit corresponds to the ideal case of the thermal state case. 
Once the coherence time $\tau_c$ becomes smaller, both the probe response and the line resolution deteriorate. Figure \ref{fig7} illustrates these features for realistic experimental values.

We did not address however the crossover case:  $\gamma_c \tau_c \geq 1$ for which a more elaborated theory is required. However, by interpolation with the incoherent case, we could obtain a rough idea of the outcome in this regime. 
The validity  of such an interpolation has not yet been proved  from our present approach and would require more theoretical investigation.

\subsection{Detuned case: $\omega \not= \omega_c$}

The proposed setup implies an exact knowledge of the single photon frequency 
to be detected. Deviations from this frequency with respect to the cavity one deteriorate the quality of the probe response.  Fig.\ref{fig10} illustrates how 
the quality of the response is affected significantly by this indetermination for coherent state and incoherent state respectively, once the detuning is of order of $\gamma_c$.

\subsection{Non-identical and interacting qubits}

The probe response to the signal increases proportionally to  the number of qubits, but only if all the essential qubit parameters ($\omega_j$,$\Gamma_j$, $\Gamma_{\phi,j}$,$\chi_j$) are identical.  
In reality, due to the difficulty of manufacturing the qubit with high precision \cite{Brehm2021}, these parameters will likely demonstrate a dispersion of the order of at least 10\%. 

In principle, a properly tuned interqubit coupling (e.g., due to mutual capacitance) can mitigate the effects of dispersion and optimize the signal-to-noise ratio and achieve the theoretical Heisenberg limit of sensitivity \cite{PhysRevB.103.064503}. A rigorous analysis of this situation in a realistic experimental setup requires a special consideration.

\section{Conclusions and perspectives}

\label{sec:conc}

Using the consistent QED approach, we have developed a general formalism for treating the system of transmon qubits in a waveguide cavity interacting with low-intensity electromagnetic field. In limiting cases, our results are in accordance with earlier calculations based on the phenomenological descriptions of the system. As an example, we have obtained solutions for non-interacting qubits in the presence of three realistic versions of the photon field: coherent, incoherent and thermal.  Our approach is particularly relevant for the quantitative treatment of transmon-based detectors of microwave photons that requires an explicit expression for the transmission.


The formalism can be further applied to the more interesting case of strongly interacting qubits, which should achieve the over-SQL Heisenberg limit of sensitivity.   


{\bf Acknowledgments:}
This work was performed in the framework of the EU project SUPERGALAX (Grant agreement ID: 863313). PN thanks Artur Sowa for his hospitality and fruitful discussions during his visit at the University of Saskatchewan.

\bibliographystyle{apsrev4-1}
\bibliography{paperrefs2}

\appendix
\begin{widetext}

\section{Quantum electrodynamics formulation}\label{A}

\subsection{Quantum field Hamiltonian}

Here we develop an effective Hamiltonian description of quantum dynamics of the electromagnetic field interacting with transmon qubits in a waveguide, improving on the earlier work \cite{GU20171}.

Our goal is, starting from the full quantum electrodynamics (QED) of the system, to derive a reduced description valid at low temperatures and at low signal intensity, i.e., in the expected regime of its operation.


We consider the structure depicted in Fig.\ref{fig0} and assume for simplicity that its conducting parts are made of the same Type I, s-wave superconductor. At temperatures much less than the superconducting gap, the effects of quasiparticle excitations can be neglected 
($k_B T \ll \Delta$). The superconductor can be described by the bosonic quantum spatiotemporal field $\hat \psi(\vc{r},t)$ of  Cooper pairs  of charge $2e$ and mass $m$ interacting with the vector potential  
$\hat {\mathbf{A}}(\vc{r},t)$ of the quantized electromagnetic field in the Coulomb gauge $\nabla \cdot \hat {\mathbf{A}}(\vc{r},t)=0$. 
As point like particle, the Cooper pair gas must be seen as a whole in the bulk and not as having a momentum distribution but rather as condensed composite objects of electrons. Only their spatial distribution can change so that the field operator is written in terms of the electron operator $\hat \psi_{e,\sigma}(\vc{r},t)$ with spin $\sigma=\uparrow, \downarrow$ as
\begin{eqnarray}\label{pair}
\hat \psi(\vc{r},t)
=\int d^3\vc{r'} F(\vc{r'}) \hat \psi_{e,\uparrow}(\vc{r}+\vc{r'}/2,t)\hat \psi_{e,\downarrow}(\vc{r}-\vc{r'}/2,t) \, ,
\end{eqnarray}
where $F(\vc{r'})$ is  the pairing function.  
We assume the nontrivial commutation relations  \cite{Ryder1996Quantum}:
\begin{eqnarray}\label{comm}
[\hat \psi(\vc{r},t),\hat \psi^\dagger(\vc{r}',t)]=\delta^{(3)}(\vc{r} -\vc{r'}) \, ,
\quad \quad \quad
[\hat{{A}}^p  (\vc{r},t),\partial_t \hat{{A}}^q (\vc{r'},t)]=i\frac{\hbar}{\epsilon}(\delta_{p,q}- \frac{\nabla^p_{\vc{r}} \nabla^q_{\vc{r}}}{(\nabla_{\vc{r}})^2})
\delta^{(3)}(\vc{r} -\vc{r'}) \, ,
\end{eqnarray}
with $p,q=r_x,r_y,r_z$ and
where the inverse operator defines for any function $f(\vc{r})$ the nonlocal transformation: $(1/\nabla_{\vc{r}})^2f(\vc{r})= \int d^3\vc{r'} (4\pi|\vc{r} -\vc{r'}|)^{-1} f(\vc{r'})$. 
The electric and magnetic fields are expressed through the vector potential in the standard way $\hat {\vc{E}}(\vc{r},t)=-\partial_t \hat {\vc{A}}(\vc{r},t)$ and  $\hat {\vc{B}}(\vc{r},t)=\nabla \times {\vc{A}}(\vc{r},t)$. 

In this regime, we can write the low energy quantum Hamiltonian in terms of the time dependent operators as: 
\begin{eqnarray}\label{H0}
\hat H&=& \int d^3 \vc{r} \frac{1}{2m} |\left(\frac{\hbar}{i} \nabla_{\vc{r}} -2e\hat {\vc{A}}(\vc{r},t)\right)\hat \psi(\vc{r},t)|^2 
+ V_s(\vc{r})|\hat \psi(\vc{r},t)|^2 +
\frac{1}{2}\int d^3\vc{r'}  \frac{(2e)^2 (|\hat \psi(\vc{r'},t)|^2-n_s(\vc{r'})))(|\hat \psi(\vc{r},t)|^2-n_s(\vc{r})))}{4\pi\epsilon|\vc{r} -\vc{r'}|} 
\nonumber \\
&+& \frac{\epsilon}{2}\left[(\partial_t \hat {\vc{A}}(\vc{r},t))^2 - c^2 {\hat {\vc{A}}(\vc{r},t)}. \nabla^2_{\vc{r}} {\hat {\vc{A}}(\vc{r},t)} \right] \, .
\end{eqnarray}
The first term corresponds to the kinetic energy, the second is the lattice potential $V_s(\vc{r})$ which confines it inside
the superconductors. It contains the contribution of the  anomalous term due to  pairing \cite{DeGennes:566105} and is assumed to be static and constant inside the superconductors and zero outside them.
The third term describes the energy cost of local Cooper pair density exceeding the average local density of superconducting electrons, $n_s$. 
The fourth term is the energy of the tranverse electromagnetic field energy.
Replacing the operator by their expectation value  $\langle \hat \psi(\vc{r},t) \rangle$ and $\langle \hat A(\vc{r},t) \rangle$ in Eq.(\ref{H0}), we recover a form analog to the Landau-Ginsburg Hamiltonian \cite{DeGennes:566105}.


\begin{figure}
\begin{center}
 \includegraphics[width=9cm]{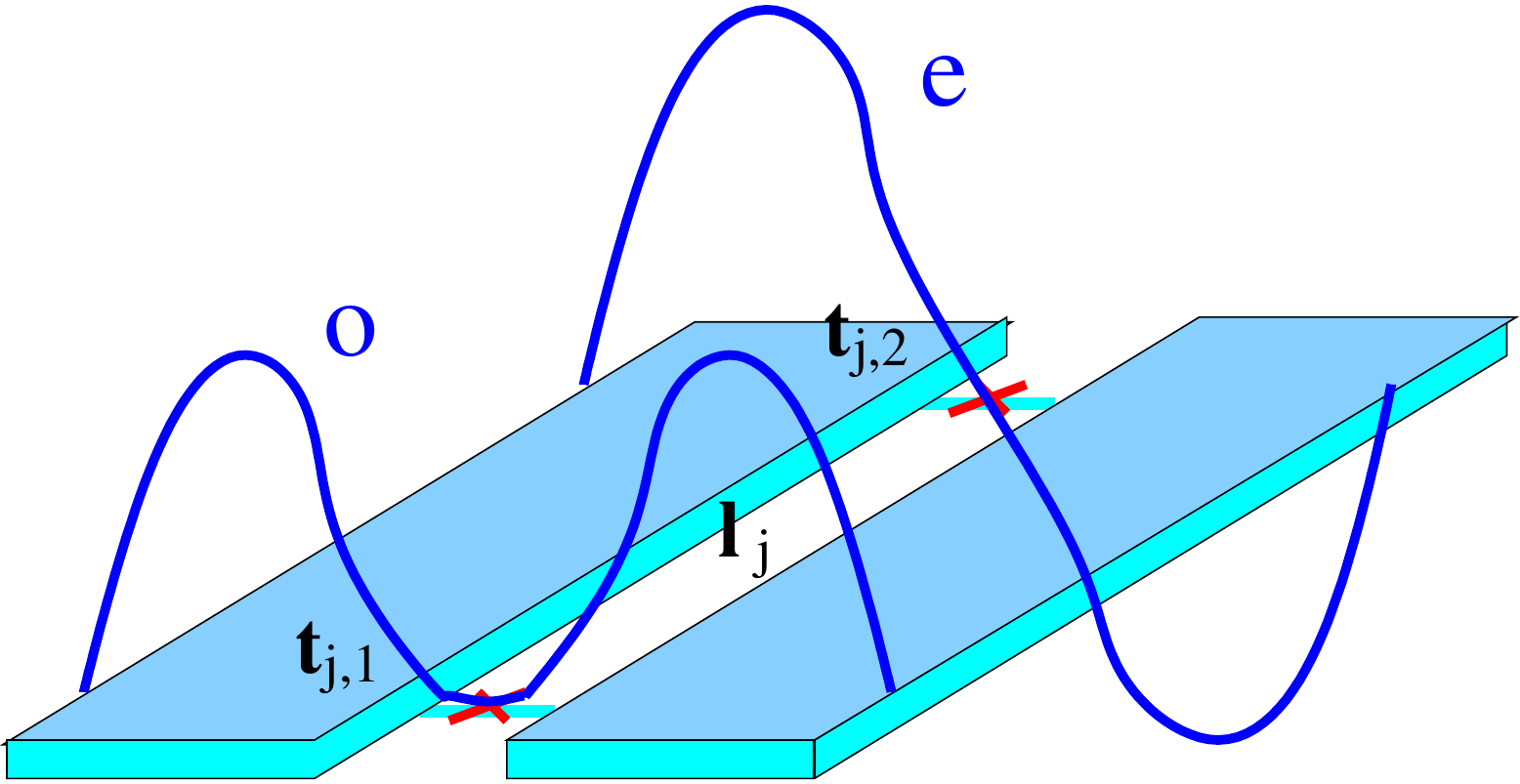}
\end{center}
    \caption{\label{A1} Schematic representation of the ground (o) and excited (e) pair wavefunctions inside the transmon $\vc{l}_j$. The red  crosses represents the Josephson junctions placed at $\vc{t}_{j,1}$ and $\vc{t}_{j,2}$ respectively.}
 \end{figure}

\subsection{The dipole representation for the transmons}

The transmons depicted in Fig.\ref{fig0} are a couple of superconducting islands with large mutual capacitance connected by
two  Josephson junctions 
which allows a control of the qubit frequency by a magnetic field. 
The superconductor volume in the circuit $\Omega$ is the sum of the  waveguide volume $\Omega_0$ 
and the  transmon volumes $\Omega_j$ i.e. $\Omega=\Omega_0 + \sum_{j=1}^N \Omega_j$.
 
The transfer of charge at the junction creates an electric dipole that is sensitive to the electromagnetic field. 
In order to reflect this observation, the Hamiltonian is expressed in terms of polarisation operator $\hat P(\vc{r},t)$ associated to each transmon.
To this end, we use the Power-Zienau-Wolley  transformation  acting on each transmon 
\cite{Cohen-Tannoudji:101367}:
\begin{eqnarray}\label{setem}
\hat \psi(\vc{r},t)&=& 
\exp\left[i\frac{2e}{\hbar}\sum_{j=1}^N \xi_j(\vc{r})
\int_0^1 du \,
\vc{r}\cdot
\hat {\vc{A}}(\vc{l_j}+u(\vc{r}-\vc{l_j}),t)
\right]\hat \psi'(\vc{r},t) \, ,
\end{eqnarray}
where $\vc{l_j}$ is the coordinate of the transmon $j$ between the two Josephson junctions, see Fig.\ref{fig0}, and  the window function 
$\xi_j(\vc{r})=1$ for $\vc{r} \in \Omega_j$ and is zero otherwise. 
The expression (\ref{setem}) takes into account of the phase accumulation within the transmon along the trajectory between the two junctions. 

The new  field $\hat \psi'(\vc{r},t)$ no longer commutes with  $\partial_t \hat{\vc{A}} (\vc{r},t)$. We derive following commutation relation:
\begin{eqnarray}
[\hat \psi'(\vc{r},t),\partial_t \hat{\vc{A}} (\vc{r''},t)]
&=&
\left[\exp\left(-
i\frac{2e}{\hbar}\sum_{j=1}^N \xi_j(\vc{r})
\int_0^1 du 
\vc{r}\cdot
\hat {\vc{A}}(\vc{l_j}+u(\vc{r}-\vc{l_j}),t)\right)
\hat \psi(\vc{r},t),\partial_t \hat{\vc{A}} (\vc{r''},t)\right]
\nonumber \\
&=&
\left[\exp\left(-i\frac{2e}{\hbar}\sum_{j=1}^N \xi_j(\vc{r})
\int_0^1\!\!\!  du\, 
\vc{r}\cdot
\hat {\vc{A}}(\vc{l_j}+u(\vc{r}-\vc{l_j}),t)
\right),\partial_t \hat{\vc{A}} (\vc{r''},t)\right]
\hat \psi(\vc{r},t)
\nonumber \\
&=&
2e\left(\vc{1}- \frac{\nabla_{\vc{r''}} \nabla_{\vc{r''}}}{(\nabla_{\vc{r''}})^2}\right)
\sum_{j=1}^N \xi_j(\vc{r})\vc{r}\int_0^1 du\,
\delta^{(3)}(\vc{l_j}+u(\vc{r}-\vc{l_j})-\vc{r''})
\nonumber \\
&&
\times
\exp\left(-
i\frac{2e}{\hbar}\sum_{j=1}^N \xi_j(\vc{r})\int_0^1 du 
\vc{r}\cdot
\hat {\vc{A}}(\vc{l_j}+u(\vc{r}-\vc{l_j}),t)\right)
\hat \psi(\vc{r},t)
\nonumber \\
&=&
\frac{2e}{\epsilon}
\sum_{j=1}^N \xi_j(\vc{r})
\left(\vc{1}- \frac{\nabla_{\vc{r''}} \nabla_{\vc{r''}}}{(\nabla_{\vc{r''}})^2}\right)
\vc{r}\int_0^1 \!\!\! du\,
\delta^{(3)}(\vc{l_j}+u(\vc{r}-\vc{l_j})-\vc{r''})
\hat \psi'(\vc{r},t) \, .
\end{eqnarray}
In order to preserve the canonicity of the commutation relation for the creation/annihilation operators, 
we use the displacement current operator $\hat {\vc{D}}(\vc{r},t)$ resulting from an unitary transformation on the vector potential time derivative. More explicitly, we define quantum constitutive relation: 
$ \hat {\vc{D}}(\vc{r},t)=
-\epsilon \hat {\dot{\vc{A}}}'(\vc{r},t) = 
 -\epsilon\partial_t \hat {\vc{A}}(\vc{r},t) + \hat {\vc{P}}(\vc{r},t)$
 with the new vector potential such that $[\hat {\dot{\vc{A}}}'(\vc{r},t),\hat \psi'(\vc{r},t)]=0$.
To satisfy this property, we use the transverse polarisation operator $\hat {\vc{P}}(\vc{r},t)$ 
defined as:
\begin{eqnarray}
\hat {\vc{P}}(\vc{r},t)&=&
\frac{2e}{\epsilon}
\left(\vc{1}- \frac{\nabla_{\vc{r}} \nabla_{\vc{r}}}{(\nabla_{\vc{r}})^2}\right)\int d^3 \vc{r'} 
\sum_{j=1}^N \int_0^1 du \,\vc{r'}(|\hat \psi'(\vc{r'},t)|^2 -n_s(\vc{r'})) \delta^{(3)}(\vc{l_j}+u(\vc{r'}-\vc{l_j})-\vc{r})
\\
&&[\hat{{A}}^p  (\vc{r},t),\hat{{D}}^q (\vc{r'},t)]=-i\hbar(\delta_{p,q}- \frac{\nabla^p_{\vc{r}} \nabla^q_{\vc{r}}}{(\nabla_{\vc{r}})^2})
\delta^{(3)}(\vc{r} -\vc{r'}) \, .
\end{eqnarray}
This expression is written in terms of the charge distribution  excess $2e\left(|\hat \psi(\vc{r'},t)|^2 -n_s(\vc{r'})\right)$ for each transmon $j$ which acts now as a electric dipole.
Under this transformation, the new vector potential itself remains unchanged $\hat {\vc{A}}'(\vc{r},t)=\hat {\vc{A}}(\vc{r},t)$ 
but it is important to note the difference between its time derivative and its conjugate momentum.
$\partial_t \hat {\vc{A}}'(\vc{r},t)\not= \hat {\dot {\vc{A}}}'(\vc{r},t)$. 
Applied to the Hamiltonian, this transformation will change the differential operator as:
\begin{eqnarray}
\left(\frac{\hbar}{i} \nabla_{\vc{r}} -2e\hat {\vc{A}}(\vc{r},t)\right)^2
\exp\left[
i\frac{2e}{\hbar}\sum_{j=1}^N \xi_j(\vc{r}) 
\int_0^1 du 
\vc{r}\cdot
\hat {\vc{A}}(\vc{l_j}+u(\vc{r}-\vc{l_j}),t)\right]\hat \psi'(\vc{r},t)=
\nonumber \\
\exp\left[
i\frac{2e}{\hbar}\sum_{j=1}^N \xi_j(\vc{r}) 
\int_0^1 du 
\vc{r}\cdot
\hat {\vc{A}}(\vc{l_j}+u(\vc{r}-\vc{l_j}),t)\right]
\left(\frac{\hbar}{i} \nabla_{\vc{r}} 
-(2e)\sum_{j=1}^N \xi_j(\vc{r})\int_0^1 du u \hat{\vc{B}}(\vc{l_j}+u(\vc{r}-\vc{l_j}),t) \times \vc{r} 
\right)^2
\hat \psi'(\vc{r},t) \, .
\nonumber \\
\end{eqnarray}
Therefore, the Hamiltonian becomes:
\begin{eqnarray}\label{H1}
\hat H&=& \sum_{j=1}^N  \int_{\Omega_i} d^3 \vc{r} \frac{1}{2m} |\left(\frac{\hbar}{i}\nabla_{\vc{r}}-(2e)\int_0^1 du u \hat{\vc{B}}(\vc{l_j}+u(\vc{r}-\vc{l_j}),t) \times \vc{r} \right)\hat \psi'(\vc{r},t)|^2
+\int_{\Omega_0} d^3 \vc{r} 
\frac{1}{2m}|\left(\frac{\hbar}{i} \nabla_{\vc{r}}-(2e)\hat {\vc{A}}(\vc{r},t) \right) \hat \psi'(\vc{r},t)|^2 
\nonumber \\
 &+&\int_{\Omega_0} d^3 \vc{r}\,V_s(\vc{r})|\hat \psi'(\vc{r},t)|^2 + 
 \frac{1}{2}\int d^3\vc{r'}  \frac{(2e)^2 (|\hat \psi'(\vc{r'},t)|^2-n_s(\vc{r'}))(|\hat \psi'(\vc{r},t)|^2-n_s(\vc{r}))}{4\pi\epsilon|\vc{r} -\vc{r'}|} 
+\frac{\epsilon}{2}
\left(\hat {\dot{\vc{A}}}'(\vc{r},t)-\frac{\hat {\vc{P}}(\vc{r},t)}{\epsilon}\right)^2 \, .
\end{eqnarray}  
The effect of the magnetic field can be neglected inside a superconductor due to the 
Meissner effect so  that the vector potential is zero  in the bulk of a superconductor. 

\section{The transmon quantum state description} \label{B}

\subsection{The angular momentum representation}

The quantum description of the  transmon in terms of the flux charge variables at the Josephson junctions, starting from the consistent QED picture.
Very often in literature, any quantum approach is developed by replacing phenomenologically these classical variables into canonically conjugated quantum operators. On the contrary in this section, we develop a consistent framework based on a fundamental quantum field  theory from which we deduce flux charge  operators, that are subsequently  
approximated as real variables in the classical limit.

In addition, as we shall see, the phase operator can be only defined formally as a discrete variables over the interval $[0,2\pi)$. The continuum limit appears usually not well defined unless we artificially prolongate their domain beyond this interval to the whole set of real number \cite{doi:10.1080/09500349708241868}.  

We approximate the quantum field operator of the Cooper pair $\hat \psi(\vc{r},t)$  considering only the two lowest-energy modes when describing the dynamics of the junctions in Fig.\ref{A1}.
It allows a consistent description of a transmon using the angular momentum representation, that is transformed afterwards, into the approximate flux charge representation for the purpose of a qubit description.

Let us consider the case in the absence of the transverse electromagnetic field. Since every
transmon subsystem is electrically neutral, we neglect the dipolar and multipolar  Coulomb interaction between them and with the elements of the waveguide so that each transmon $j$ is described by the one-mode approximation. Here the  wave function  $\psi_{j,\alpha}(\vc{r})$ describing the state with $N_{s,j}$ of pairs is a solution of the mean-field Hartree equation 
(or Gross-Pitaevskii equation \cite{Pitaevskii}):
\begin{eqnarray}\label{GP}
\left[-\frac{\hbar^2}{2m}\nabla_\vc{r}^2 + V_s(\vc{r})+ 
\int_{\Omega_j} d^3\vc{r'}  \frac{(2e)^2 (|\psi_{j\alpha}(\vc{r'})|^2 N_{s,j}-n_{s}(\vc{r'}))}{4\pi\epsilon|\vc{r} -\vc{r'}|} \right]
\psi_{j\alpha} (\vc{r})=\mu_{j,\alpha}(N_{s,j})\psi_{j\alpha}(\vc{r})  \, .
\end{eqnarray}
Here $j=0, \dots, N$  labels the wave functions for each part of the system (waveguide and transmons), $\alpha=o,e$ labels the ground and  excited states respectively and $\mu_{j\alpha}(N_{s,j})$ is the corresponding chemical potential. The Hartree energy is obtained from the relation 
$d E_{j\alpha}(N_{s,j})/dN_{s,j} = \mu_{j\alpha}(N_{s,j})$. 
The total ground state energy is the sum of ground state energies $E_0=\sum_{j=0}^N E_{j,o}(N_{s,j})$. 

The wave function of the first excited state $\alpha=e$  has  a single node changing sign at the Josephson junction, as illustrated in Fig.(\ref{A1}). Neglecting the influence of higher energy excited state, we can write approximately the quantum field operator as: 
\begin{eqnarray}
\hat \psi'(\vc{r},t) \simeq 
\psi_{0}(\vc{r}) \hat c_{0}(t) +
\sum_{j=1}^N \psi_{j,o}(\vc{r}) \hat c_{j,o}(t)+  \psi_{j,e}(\vc{r}) \hat c_{j,e}(t)  \, .
\end{eqnarray}
Here
$\int_{\Omega_j} d^3\vc{r} \, 
\psi^*_{j\alpha}(\vc{r})\psi_{j\alpha'}(\vc{r})=\delta_{\alpha \alpha'}$, $\xi_{j}(\vc{r})\psi_{j'\alpha}(\vc{r})=\delta_{j,j'}\psi_{j\alpha}(\vc{r})$ and  $\hat c_{j\alpha}$ is the annihilation operator. 
In other words, in the absence of external electromagnetic field, the general quantum state is a combination  of the Fock states of the form $|\Psi \rangle =(\hat c^\dagger_0)^{N_{s,0}}/\sqrt{N_{s,0}!}\prod_{j=1}^{N}(\hat c^\dagger_j)^{N_{s,j,o}}(\hat c^\dagger_{j\alpha})^{N_{s,j,e}}|0 \rangle/\sqrt{N_{s,j,o}!N_{s,j,e}!}$ with the particle number constraint: $N_{s,j,e}+N_{s,j,o}=N_{s,j}$.

The energy difference between the  excited 
and the ground state is Josephson energy
$E_{J,j}(N_{s,j})= E_{j,e}(N_{s,j})-E_{j,o}(N_{s,j})$. 
Assuming for simplicity that the two island parts  of the transmon are perfectly symmetric, we can define their localized states:
\begin{eqnarray}
\psi_{j\pm}(\vc{r}) =\frac{\psi_{j,o}(\vc{r})\pm  \psi_{j,e}(\vc{r})}{\sqrt{2}}  \, ,
\quad \quad \quad
\hat c_{j\pm}(t)=\frac{\hat c_{j,o}(t)\pm\hat c_{j,e}(t)}{\sqrt{2}} \, .
\end{eqnarray}
With a good accuracy, we can consider these as orthonormal (wave functions almost do not overlap much each other) 
so that we satisfy the commutation relation
$[\hat c_{j\pm}, \hat c_{j'\pm}^\dagger]=\delta_{j,j'}\delta_{\pm,\pm}$. Also, since the superconducting waveguide part is large and electrically neutral, the Cooper pair mode can be described by a coherent state and 
we can replace its quantum operators description  with a c-number: $\hat c_{0}(t)\simeq \sqrt{N_{s,0}}$. 

Therefore, the quantum field is rewritten as:
\begin{eqnarray}
\hat \psi'(\vc{r},t)\simeq  \sqrt{N_{s,0}} \psi_0(\vc{r}) +\sum_{j=1}^N \sum_{\pm}\psi_{j,\pm}(\vc{r}) 
\hat c_{j,\pm}(t)  \, .
\end{eqnarray}
As the wave functions of two different transmons do not overlap $\psi_{j,\pm}(\vc{r})\psi_{j',\pm}(\vc{r}) \simeq 0$ for $j \not= j'$,  the density operator becomes: 
\begin{eqnarray}\label{norm}
| \hat \psi'(\vc{r},t)|^2&=&
N_{s,0} |\psi_0(\vc{r})|^2 +\sum_{j=1}^N (|\psi_{j,+}(\vc{r})|^2+|\psi_{j,-}(\vc{r})|^2)
\hat n_j(t)+(|\psi_{j,+}(\vc{r})|^2-|\psi_{j,-}(\vc{r})|^2)\hat m^z_{j}(t)
\nonumber \\
&&+
\sum_{\pm}\psi^*_{j,\pm}(\vc{r})\psi_{j,\mp}(\vc{r})
\hat m^\pm_{j}(t)
\\
\hat n_j(t) &=&\frac{\hat c^\dagger_{j,+}(t)\hat c_{j,+}(t)+\hat c^\dagger_{j,-}(t)\hat c_{j,-}(t)}{2}  \, , \quad \quad 
\hat m^{z}_j(t) =\frac{\hat c^\dagger_{j,+}(t)\hat c_{j,+}(t)-\hat c^\dagger_{j,-}(t)\hat c_{j,-}(t)}{2} \, , \quad \quad 
\hat m^{\pm}_j(t) =\hat c^\dagger_{j,\pm}(t)\hat c_{j,\mp}(t) \, ,
\nonumber \\
\end{eqnarray}
where the operators $\hat m^{z}_j(t)$ and $\hat m^{\pm}_j(t)$ are subject to the algebra of the angular momentum operator and $\hat n_j(t)$ is the particle number operator for the transmon $j$. 
Indeed, $[\hat m_j^+,\hat m_j^-]=\hat m_j^z$, $[\hat m_j^z,\hat m_j^\pm]=\pm 2 \hat m_j^\pm$.
Similarly for the kinetic term, we obtain:
\begin{eqnarray}\label{grad}
&&|\nabla \hat \psi'(\vc{r},t)|^2=
N_0 |\nabla \psi_0(\vc{r})|^2 +\sum_{j=1}^N (|\nabla \psi_{j,+}(\vc{r})|^2+|\nabla\psi_{j,-}(\vc{r})|^2)
\hat n_j(t)+(|\nabla \psi_{j,+}(\vc{r})|^2-|\nabla \psi_{j,-}(\vc{r})|^2)\hat m^z_{j}(t)
\nonumber \\
&&+
\sum_{\pm}\nabla \psi_{j,\pm}(\vc{r})\nabla\psi_{j,\mp}(\vc{r})\hat m^\pm_{j}(t)  \, .
\end{eqnarray}
We substitute the expressions  Eq.(\ref{norm}) and 
Eq.(\ref{grad}) into the Hamiltonian (\ref{H1}), and apply the following  simplifications:
\begin{enumerate}
\item In (\ref{H1}), the magnetic field coupling term is neglected over the linear electric field coupling term ${\hat {\dot {\vc{A}}}}(\vc{r},t)$ with the polarisation term.
\item For a small intensity of the waveguide field, the quadratic term in the transverse polarisation term is neglected.
 \item We chose eigenstates of $\hat n_j$  with $N_{s,j}$ particles, approximating $|\psi_{j,+}(\vc{r})|^2+|\psi_{j,-}(\vc{r})|^2 \cong |\psi_{j}(\vc{r})|^2$
 \item 
We use the Eq.(\ref{GP}) to express the matrix element in terms of eigenenergies
\item We neglect the product term $m^z_{j}(t)\hat m^\pm_{j'}(t)$ and $\hat m^\pm_{j}(t)\hat m^\pm_{j'}(t)$ in the Coulomb term as the overlapping integral over $\vc{r}$ and $\vc{r'}$ is very small.
\end{enumerate}
As a result, the expressions (\ref{H1}) becomes: 
\begin{eqnarray}\label{H1bis}
\hat H&=&E_0 + 
\sum_{j=1}^N \frac{E_{j,J}(N_{s,j})}{2}- 
\frac{\mu_{j,e}(N_{s,j})-\mu_{j,o}(N_{s,j})}{2}
\left(\hat m^-_{j}(t) + c.c. \right) 
\nonumber \\
&+& 
\frac{1}{2}\int d^3\vc{r'}  \frac{(2e)^2 \sum_{j,j'=1}^N (|\psi_{j,+}(\vc{r})|^2-|\psi_{j,-}(\vc{r})|^2)\hat m^z_{j}(t)(|\psi_{j',+}(\vc{r'})|^2-|\psi_{j',-}(\vc{r'})|^2)\hat m^z_{j'}(t)}{4\pi\epsilon|\vc{r} -\vc{r'}|} 
\nonumber \\
&+& \frac{\epsilon}{2}\biggl[
{\hat {\dot {\vc{A}}}}^2(\vc{r},t) - c^2 
{\hat {\vc{A}}(\vc{r},t)}. \nabla^2_{\vc{r}} {\hat {\vc{A}}(\vc{r},t)}
+ \frac{N_{s,0} |(2e)\psi_{0}(\vc{r})|^2}{2\epsilon m}
\hat{\vc{A}}^2(\vc{r},t)\biggr]
\nonumber \\
&-& 
\sum_{j=1}^N\int_0^1 du \,\int d^3 \vc{r} \hat {\dot{\vc{A}}}(\vc{l_j}+u(\vc{r}-\vc{l_j}),t).\vc{r} (2e)
(|\psi_{j,+}(\vc{r})|^2-|\psi_{j,-}(\vc{r})|^2)\hat m^z_{j}(t) \, ,
\end{eqnarray} 
where now  we omit the prime in the time derivative of the vector potential. Assuming a linear dependence for the energy difference, we can further approximate: $\mu_{j,e}(N_{s,j})-\mu_{j,o}(N_{s,j})=d E_{j,J}(N_{s,j})/dN_{s,j}\cong E_{j,J}(N_{s,j})/N_{s,j}$. 

\subsection{The presence of the external uniform magnetic field}

In the presence of a uniform magnetic field $\vc{B}_0$,
the vector potential contains in addition to the time-dependent waveguide field the classical component:
\begin{eqnarray}
\hat {\vc{A}}(\vc{r},t) \rightarrow \hat {\vc{A}}(\vc{r},t) + \vc{A}_0 (\vc{r}) \, .
\end{eqnarray} 
Therefore, we add a phase $\theta_{j,\pm}(\vc{r})$ so that the quantum field parametrisation is modified into:
\begin{eqnarray}
\hat \psi'(\vc{r},t)\simeq  \sqrt{N_{s,0}} \psi_0(\vc{r}) +\sum_{j=1}^N \sum_{\pm}\psi_{j,\pm}(\vc{r}) e^{i\theta_{j,\pm}(\vc{r})}\hat c_{j,\pm}(t) \, .
\end{eqnarray}
Substituting these expressions into Eq.(\ref{H1}) and minimizing the energy, we obtain the London equation $\nabla_{\vc{r}}^2 \vc{A}_0(\vc{r}) =
\frac{(2e)^2 n_s(\vc{r})}{\epsilon c^2 m} 
\vc{A}_0(\vc{r})$ where the Cooper pair density is approximated as $n_s(\vc{r})\cong \sum_ {j=0}^N N_{s,j} 
|\psi_{j,o}(\vc{r})|^2$. The solution of the latter is such that the asymptotic expression is $\frac{1}{2}(\vc{B}_0 \times \vc{r})$ and implies that surface current appears  to avoid the magnetic field to penetrate the bulk.  
Therefore one deduces the phases $\theta_{j,\pm}(\vc{r})=\int_{C_{j,\pm}} d\vc{r'}\cdot \vc{A}_0 (\vc{r'})$ along a contour $C_{j,\pm}$ on the surface of the conducting islands starting from one junction located in Fig.\ref{A1} at $\vc{t}_{j,1}$ and finishing at $\vc{r} \rightarrow \vc{t}_{j ,2}$ so that it encloses the transmon loop at the other junction located at $\vc{t}_{j ,2}$.  
Since the boundary condition imposes a zero  
normal component on the surface, the phases are independent on the choice of the contour path. 
We deduce  the relation to the total flux crossing the transmon loop as $\Phi_{B_0}=\oint \vc{dS} \cdot \vc{B}_0=\theta_{j,+}(\vc{t}_{j ,2}) -  \theta_{j,-}(\vc{t}_{j ,2})$. As a result, 
the presence of the field changes the Hamiltonian into:
\begin{eqnarray}
\sum_{j=1}^N
\frac{E_{J_1,j }(N_{j,s})+ E_{J_2,j}(N_{j,s})\exp(i2e\Phi_{B_0}/\hbar)}{2N_{j,s}}\hat m^+_{j}(t) + c.c.
=\sum_{j=1}^N
\frac{E_{J_,j}(N_{j,s},B_0)}{2N_{j,s}}
\left[e^{i \varphi_{j}(N_{j,s},B_0)}\hat m^+_{j}(t) + c.c.\right] \, ,
\end{eqnarray}
where the reparametrization with the effective Josephson energy $E_{J_,j}(B_0)$ and the phase $\varphi_{j}(N_{s,j},B_0)$ is found in \cite{PhysRevA.76.042319}. 
In what follows, the phase will be reabsorbed in the raising-lowering operators: $\hat m^{\pm}_{j}(t)$.
Finally:
\begin{eqnarray}\label{H1ter}
\hat H&=&E_0 + 
\sum_{j=1}^N \frac{E_{J,j}(N_{s,j})}{2}
\left(2- \hat m^-_{j}(t) - \hat m^+_{j}(t)\right) 
\nonumber \\
&+& 
\frac{1}{2}\int d^3\vc{r'}  \frac{(2e)^2 (\sum_{j,j'=1}^N (|\psi_{j,+}(\vc{r})|^2-|\psi_{j,-}(\vc{r})|^2)\hat m^z_{j}(t)(|\psi_{j',+}(\vc{r'})|^2-|\psi_{j',-}(\vc{r'})|^2)\hat m^z_{j'}(t)}{4\pi\epsilon|\vc{r} -\vc{r'}|} 
\nonumber \\
&+& \frac{\epsilon}{2}\biggl[
{\hat {\dot {\vc{A}}}}^2(\vc{r},t) - c^2 
{\hat {\vc{A}}(\vc{r},t)}. \nabla^2_{\vc{r}} {\hat {\vc{A}}(\vc{r},t)}
+ \frac{N_{s,0} |(2e)\psi_{0}(\vc{r})|^2}{2\epsilon m}
\hat{\vc{A}}^2(\vc{r},t)\biggr]
\nonumber \\
&-& 
\sum_{j=1}^N\int_0^1 du \,\int d^3 \vc{r} \hat {\dot{\vc{A}}}(\vc{l_j}+u(\vc{r}-\vc{l_j}),t).\vc{r} (2e)
(|\psi_{j,+}(\vc{r})|^2-|\psi_{j,-}(\vc{r})|^2)\hat m^z_{j}(t)
\, .
\end{eqnarray} 

\subsection{Phase and charge basis formalism: charge qubit}

We define the charge operator as:
$\hat q_j(t)= (2e)\hat m^z_j(t)$
with the quantum numbers limited to integer value $m^z_j=- M,-M+1,\dots, M-1 , M$ . We define the phase basis through
\begin{eqnarray}
|\phi_j \rangle &\equiv& | {\tilde m}_j \rangle = 
\sum_{m_j=-M}^{M} \frac{e^{i m_j \phi_j}}{\sqrt{2M+1}}|m_j \rangle  \, ,\quad \quad \quad 
\phi_j=2\pi {\tilde m}_j/(2M+1) \, , \quad \quad \quad {\tilde m}_j=0,\dots, 2M
\\
|m_j \rangle &=&
\frac{1}{\sqrt{2M+1}}\sum_{{\tilde m}_j =0}^{2M+1} e^{-i \frac{2\pi m_j {\tilde m}_j}{2M+1}}
| {\tilde m}_j \rangle \, .
\end{eqnarray}
In this restricted set, we shall assume also the basis are defined modulo $2M+1$, i.e.  $|m_j +2M+1 \rangle_j=|m_j \rangle_j$ and 
$|\tilde{m}_j +2M+1 \rangle_j=|\tilde{m}_j\rangle_j$. This assumption may 
not be correct especially for $M$ not large.
Nevertheless, once we achieve the limit $M\rightarrow \infty$ through this procedure, the phase operator goes from  a set of discrete variables to a continuous one over the range $[0,2\pi )$. From the above definition, 
we find:
\begin{eqnarray}
\sum_{m_j=-M}^{M} |m_j+ 1\rangle \langle m_j|&=&
\sum_{{\tilde m}_j=0}^{2M}e^{-i \phi_j}
\frac{|{\tilde m}_j \rangle_j \langle {\tilde m}_j |}{2M+1}
\stackrel{M \rightarrow \infty}{=}\int_0^{2\pi}\frac{d\phi_j}{2\pi} e^{-i \phi_j}|\phi_j \rangle \langle \phi_j |=
e^{-i \hat \phi_j}
\\
\hat m_j^s=
\sum_{m_j=-M}^{M} m_j^s|m_j\rangle \langle m_j|&=&
\sum_{\tilde{m}_j=0}^{2M}\sum_{\tilde{m}_j=0}^{2M}
(-i\partial_{\phi'_j})^s \frac{\sin[(2M+1)(\phi_j-\phi'_j)]}{\sin(\phi_j-\phi'_j)}
\frac{|{\tilde m}_j \rangle \langle {\tilde m}_j |}{2M+1}
\\
&\stackrel{M\rightarrow \infty}{=}&
\int_0^{2\pi}\frac{d\phi_j}{2\pi}
(-i\partial_{\phi_j})^s |\phi_j \rangle \langle \phi_j |={\dot {\hat \phi}}_j^s \, .
\end{eqnarray}
In the new basis, 
$[e^{-i \hat \phi_j},\hat m_j]= e^{-i \hat \phi_j}$ from which it follows that
$[\hat \phi_j,\hat m_{j'}]=i\delta_{j,j'}$.
Using the flux $\hat \Phi_j = (2e)\hat \phi_j /\hbar$ and charge variable, we rewrite it as $[\hat \Phi_j,\hat q_{j'}]=i\hbar\delta_{j,j'}$. 
The creation operator is approximated for large $M = N_{s,j}/2$ as:
\begin{eqnarray}
\hat m^{\pm}_j =
\sum_{m_j=-M}^{M} \sqrt{M(M+1)-m_j(m_j\pm 1)}|m_j\pm 1\rangle \langle m_j| \cong 
\frac{N_{s,j}}{2}
e^{\mp i \hat \phi_j}  \, .
\end{eqnarray}
In the language of these variables, the Hamiltonian  (\ref{H1ter}) becomes:
\begin{eqnarray}\label{chargephase}
\hat H
&=& E_0+ \sum_{j=1}^N E_{J,j}\left[1-
\cos\left(\frac{2e}{\hbar}\hat \Phi_j(t) \right) \right]
+ \frac{1}{2} \sum_{j,j'=1}^N C^{-1}_{j,j'} 
\hat q_j(t) \hat q_{j'}(t)  
\nonumber \\
&+& \frac{\epsilon}{2}\biggl[
{\hat {\dot {\vc{A}}}}^2(\vc{r},t) - c^2 
{\hat {\vc{A}}(\vc{r},t)}. \nabla^2_{\vc{r}} {\hat {\vc{A}}(\vc{r},t)}
+ \frac{N_{0} |(2e)\psi_{0}(\vc{r})|^2}{2\epsilon m}
\hat{\vc{A}}^2(\vc{r},t)\biggr]
\nonumber \\
&-& 
\sum_{j=1}^N\int_0^1 du \,\int d^3 \vc{r} \hat {\dot{\vc{A}}}(\vc{l_j}+u(\vc{r}-\vc{l_j}),t).\vc{r}
(|\psi_{j,+}(\vc{r})|^2-|\psi_{j,-}(\vc{r})|^2)\hat q_{j}(t) \, ,
\end{eqnarray}
where we define the capacitance:
\begin{eqnarray}
C^{-1}_{j,j'}&=& 
\frac{1}{2}\int d^3\vc{r'}  \frac{ (\sum_{j,j'=1}^N (|\psi_{j,+}(\vc{r})|^2-|\psi_{j,-}(\vc{r})|^2)(|\psi_{j',+}(\vc{r'})|^2-|\psi_{j',-}(\vc{r'})|^2)}{4\pi\epsilon|\vc{r} -\vc{r'}|} \, ,
\end{eqnarray}
where $E_{J,j}$ is the Josephson energy. 

\section{Quantum description of the  electromagnetic field in the  waveguide} 
\label{C}

\subsection{General treatment for a waveguide in  vacuum}

In this section, we revisit the usual approach for the waveguide description  in order to consistently account for quantum effects.  The main results are as follows: 1) The capacitance  which characterizes the electromagnetic wave propagation, differs from the static capacitance in presence of dielectric materials; 2)  Explicit formulas for the speed propagation $v$ and impedance $Z$ in a coplanar waveguide; 3) A formula for the coupling factor $f_j$ between the electromagnetic field and transmon.

Consider a quantum electromagnetic field propagating along the superconducting wire axis of a constant cross section depicted in Fig.\ref{waveguide}. 
The property of translational invariance along the z axis is required for a good quality waveguide that does not emit radiation outside the device  to eliminate the stray 
radiation.  Due to experimental constraints, it is not generally strictly valid since the waveguide has a curved geometry.
Therefore, using Eq.(\ref{chargephase}), the Heisenberg equations for  the  vector potential field 
correspond to the quantum Ampere equations: 
\begin{eqnarray}\label{amp}
(c^2\partial_t^2-\nabla_{\vc{r}}^2) \hat {\vc{A}}(\vc{r},t) 
=\mu \hat{\vc{j}}_\perp (\vc{r},t)\cong 
-\frac{N_{s,0} |(2e)\psi_{0}(\vc{r})|^2}{\epsilon c^2 m} 
\hat{\vc{A}}(\vc{r},t)\cong
\begin{cases}
-\frac{n_s (2e)^2}{\epsilon c^2 m} \hat{\vc{A}}(\vc{r},t) & \text{conductors} \\
0 & \text{insulators}
\end{cases} \, ,
\end{eqnarray}
where $n_s$ is Cooper pair density.
The right hand side corresponds to a renormalized transverse current responsible for the dynamical Meissner effect.

We restrict our analysis to the only relevant TEM mode of the electromagnetic field.  
Then the vector potential operator can be approximated  by    
$\hat{\vc{A}}(\vc{r},t) =\hat{\alpha}(r_z,t)\vc{A}_\perp(\vc{r}_\perp)$ where $\hat{\alpha}(r_z,t)$ 
is the quantum operator describing the electromagnetic  propagation along the z axis and  $\vc{A}_\perp(\vc{r}_\perp)
=(A^x_\perp(\vc{r}_\perp),A^y_\perp(\vc{r}_\perp),0)$ 
is a classical transverse vector field dependent only on the transverse coordinates $\vc{r}_\perp=
(r_x,r_y,0)$. 

In the simplest configuration of two parallel plates at $(\pm D/2,r_y,r_z)$, for a plane wave $\langle \hat \alpha (r_z,t) \rangle \sim \exp[i(k_z r_z-\omega t)]$, the solution is a constant for  ${A}_\perp^x$ normal the conductor and zero otherwise. Note the absence of a parallel component inside the bulk  with the London penetration length $\lambda_s=\sqrt{\epsilon c^2 m/(n_s (2e)^2)}$. As consequence a net electrical charge appears at the surface to compensate the discontinuity of the vector potential.
This surface charge distribution has normally its own dynamics associated the plasma frequency $\omega_{pl}=c/\lambda_s$.
Since generally the penetration depth $\lambda_s \ll 1/k$  for the wavelengths which we are concerned with (microwave), the surface charge responds instantaneously to the field and 
we can consider the limit $\lambda_s \rightarrow 0$ in the Ampere equation, so that the $z$ component of the vector potential cancels out. It corresponds to neglecting the influence of the kinetic inductance due to current compared to the geometric (magnetic inductance) \cite{doi:10.1063/1.3010859}.

In the London gauge ($\nabla.\hat{\vc{A}}(\vc{r},t)=0$), Eq.(\ref{amp}) is transformed into a two-dimensional problem  for the transverse electromagnetic component:
\begin{eqnarray}
\nabla_\perp^2  \vc{A}_\perp (\vc{r}_\perp)= 0 \, , \quad \quad \quad \vc{n}\times \vc{A}_\perp =0
{\rm \ at \ the \ surface \ boundary }\, , 
\quad \quad \nabla_\perp . \vc{A}_\perp(\vc{r}_\perp)=0   \, ,
\end{eqnarray}
where $\vc{n}=(n_x,n_y,0)$ is the normal unit vector at the boundary and is pointing from the signal  to the ground part of the waveguide.
The transverse components are normalized such that $\int d^2\vc{r}_\perp \vc{A}^2_\perp(\vc{r}_\perp)=1/\epsilon$.  
It is then possible to define a scalar potential 
$\vc{A_\perp}=\nabla_\perp \chi(\vc{r}_\perp)$ so the problem reduces to the Laplace 
equation:
\begin{eqnarray}\label{Lchi}
\nabla_\perp^2  \chi (\vc{r}_\perp)= 0 \, , \quad \quad \quad 
\chi(\vc{r}_\perp)=\chi_s
{\rm \ at \ signal \ wire   } \, , \quad \quad 
\chi(\vc{r}_\perp)= 0
{\rm \ at \ ground  \ wire  }  \, .
\end{eqnarray}
These boundary conditions define the gauge of the vector potential and tells that the  potential is set to $\chi_s$  at the z-axis ($r_x=0$ and  $r_y=0$). 
The substitution of this approximated expression into the commutation relation and the integration over the transverse coordinate produce
the non trivial commutation relation:
\begin{eqnarray}
[\hat{{\alpha}}(r_z,t),\hat{\dot{\alpha}}(r'_z,t)]=i\hbar \delta(r_z -r'_z) \, .
\end{eqnarray}
Note again that in general $\hat{\dot{\alpha}}(r_z,t)\not= \partial_t\hat{\alpha}(r_z,t)$.

\begin{figure}
\begin{center}
\includegraphics[width=9cm]{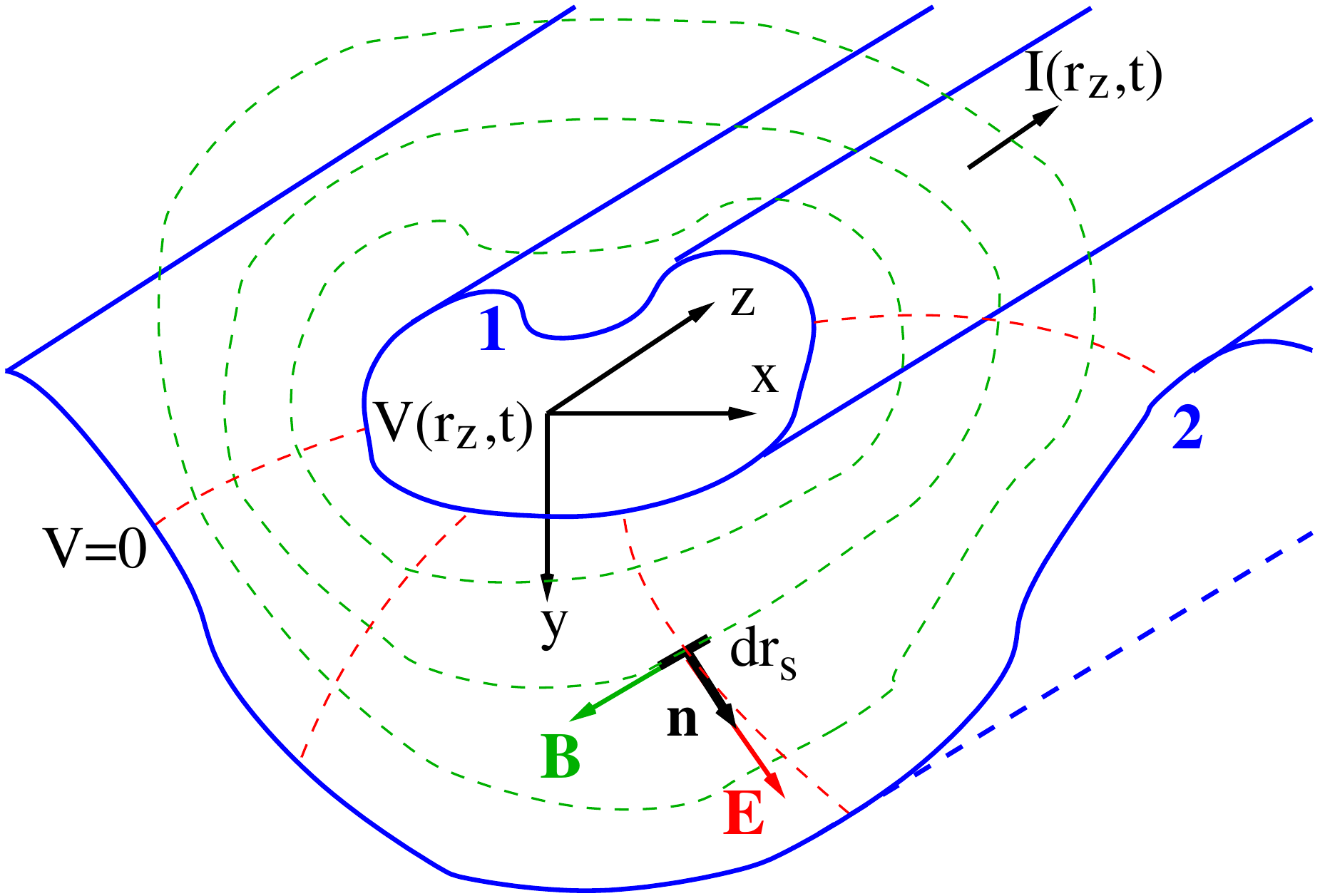}
\end{center}
\caption{\label{waveguide} 
Schematic representation of a linear waveguide with signal line (1) and ground line (2) with red electric field line and green magnetic line conformal to each other. 
}
\end{figure}

\subsection{Quantum circuit formulation of the waveguide}\label{C2}

We assume a generic waveguide  represented under the form of in  Fig.\ref{waveguide}.
In order to express the electromagnetic field  in terms of the electrical circuit notations,
we define the operator potential difference  between the signal wire 1 and ground wire 2  as:
\begin{eqnarray}
\hat V(r_z,t)= -\int^{1}_{2} d\vc{r'}_\perp. \hat {\vc{E}}  (\vc{r},t)=\int_2^1 d\vc{r}_\perp.{\vc{A}}_\perp (\vc{r}_\perp)\hat{\dot \alpha}(r_z,t)= \chi_s \hat{\dot \alpha}(r_z,t)
\, .
\end{eqnarray}
According to this definition, the potential is the same everywhere 
at any point of the conducting surface since the vector potential is parallel to the normal vector. By extension, we also determine as shown in Fig.\ref{waveguide} the equipotential in green and the force line in red between the two lines.   

For each normal vector $\vc{n}$, we define the infinitesimal element $d\vc{s}= dr_s \vc{n}$ where $dr_s=n_x dr_y-n_y dr_x$ which is the variation along the equipotential in the xy plane. 
On an infinitesimal section of size $dr_z$, the Gauss theorem  defines an effective line charge that encloses the signal wire at the surface given by   
$d\hat Q(r_z,t)=\epsilon \oint  d\vc{s}. \hat {\vc{E}} (\vc{r},t)dr_z=-\epsilon \oint  d\vc{s}.{\vc{A}}_\perp (\vc{r}_\perp) \hat{\dot \alpha}(r_z,t)dr_z$ and  opposite to the ground wire line charge.
Using these definitions, we define the differential capacitance $dC$ for this charge element as the proportionality factor between the operator of electric charge on the section of the waveguide and the local voltage operator $d\hat Q(r_z,t)=dC\, \hat V(r_z,t)$. Eliminating the operator $\hat{\dot \alpha}(r_z,t)$ in the two expressions, we find explicitly:
\begin{eqnarray}\label{C'}
dC =  -
\frac{\epsilon \oint  d\vc{s}. {\vc{A}}_\perp (\vc{r}_\perp)}{\chi_s}dr_z \, ,
\end{eqnarray}
which depends only on the waveguide geometry.

The current along the wires is related to the magnetic field according to the Ampere law Eq.(\ref{amp}) expressed in a circulation integral form along a green line over the transverse plane so that the electric field does not contribute:
\begin{eqnarray}\label{cur}
\hat I (r_z,t)= \frac{1}{\mu_0}\oint  d\vc{s}. \nabla \times \left({\vc{A}}_\perp (\vc{r}_\perp) \hat{\alpha}(r_z,t)\right) 
= \frac{1}{\mu_0}\oint  d\vc{s} .{\vc{A}}_\perp (\vc{r}_\perp)\partial_{r_z} \hat{\alpha}(r_z,t) \, .
\end{eqnarray}
The magnetic flux operator in this section is defined along a trajectory between the two wires (1 and 2) as:
\begin{eqnarray}
d\hat \Phi_B(r_z,t)&=&
\int_1^2 d\vc{r}_\perp.\hat {\vc{B}}(\vc{r},t)dr_z=
\int_1^2 d\vc{r}_\perp.\nabla \times \hat {\vc{A}}(\vc{r},t)dr_z=
\int_1^2 d\vc{r}_\perp. {\vc{A}}_\perp (\vc{r}_\perp)\partial_{r_z} \hat{\alpha}(r_z,t)dr_z \, .
\end{eqnarray}
Integrating along $r_z$, the total flux is $\hat \Phi_B(r_z,t)=
\int_1^2 d\vc{r}_\perp. {\vc{A}}_\perp (\vc{r}_\perp) \hat{\alpha}(r_z,t)$. 
Note the quantum Lenz relation linking the flux and the potential:
$\partial_t \hat \Phi_B(r_z,t) = -\hat V(r_z,t)$. 
The phase operator is related to the flux  through:
$\hat \phi_B(r_z,t)= (2e)\hat \Phi_B(r_z,t)/\hbar$.
Similarly we can define the differential inductance $dL$ through 
the proportionality relation  $d\hat \Phi_B(r_z,t)=dL \, \hat I(r_z,t)$:
\begin{eqnarray}\label{L'}
dL=
\frac{\mu_0 \int_1^2 d\vc{r}_\perp.{\vc{A}}_\perp (\vc{r}_\perp) }
{ \oint  d\vc{s}. {\vc{A}}_\perp (\vc{r}_\perp)} dr_z
=-
\frac{\chi_s}
{\epsilon c^2 \oint  d\vc{s}. {\vc{A}}_\perp (\vc{r}_\perp)} dr_z \, ,
\end{eqnarray}
that depends also only on the geometry of the waveguide. 
We deduce immediately the relation linking the line inductance and the line capacitance $(dL/dr_z)(dC/dr_z)=1/c^2$. 
These expressions can be further simplified by carry out a partial integration on the normalization condition and by using the Laplace equation in: 
\begin{eqnarray}
1/\epsilon&=& 
 \int d^2\vc{r}_\perp \left(\nabla \chi(\vc{r}_\perp) \right)^2=
 \oint  d\vc{s}. (\nabla \chi(\vc{r}_\perp)) \chi(\vc{r}_\perp)|_1^2
 - \int d\vc{r}_\perp \chi(\vc{r}_\perp) \nabla^2 \chi(\vc{r}_\perp)
\nonumber \\
 &=& -\oint  d\vc{s}.  {\vc{A}}_\perp (\vc{r}_\perp) 
 \chi_s  \, .
\end{eqnarray}
The last relation expresses  all parameters in terms of the scalar potential $\chi_s$.
We find finally: 
\begin{eqnarray}\label{defs}
\partial_{r_z} \hat \Phi_B(r_z,t)=
\chi_s \partial_{r_z} \hat{\alpha}(r_z,t) \, ,
\quad \quad 
\partial_{r_z}\hat Q(r_z,t)=\frac{1}{\chi_s} 
\hat{\dot \alpha}(r_z,t) \, ,
\quad \quad
L'=\frac{dL}{dr_z} = 
\frac{\chi_s^2}
{c^2} \, , \quad \quad
C'= \frac{dC}{dr_z} =
\frac{1}{\chi_s^2}  \, .
\end{eqnarray}
Thus the electromagnetic part of the Hamiltonian can be rewritten as:
\begin{eqnarray}\label{Hem}
\int dr_z
\frac{1}{2}\left[
(\hat{\dot{\alpha}}(r_z,t))^2- c^2 
{\hat{{\alpha}}(r_z,t)\partial_z^2 \hat{{\alpha}}(r_z,t)}\right]
=\int 
\left[ \frac{d\hat{Q}^2}{2dC} + \frac{d{\hat \Phi_B}^2}{2dL}
\right]
= \int dr_z \left[\frac{1}{2C'} \left(\frac{d\hat{Q}}{dr_z}\right)^2
+ \frac{1}{2L'} \left(\frac{d{\hat \Phi_B}}{dr_z}\right)^2 \right] \, ,
\end{eqnarray}
with the non trivial commutations relation:
\begin{eqnarray}
[\hat \Phi_B(r'_z,t),\partial_{r_z} \hat Q(r_z,t)]=i\hbar \delta(r_z -r'_z)  \, .
\end{eqnarray}
Let us note that  the  charge conservation law along the conductor expressed in differential form: $\partial_t  \partial_{r_z}\hat Q(r_z,t)=-\partial_{r_z}\hat  I (r_z,t)$ is deduced easily from the  Heisenberg equation: $\partial_t \hat{\dot \alpha}(r_z,t) =c^2 \partial^2_{r_z} \hat{\alpha}(r_z,t)$ valid for a free em field.
The self Coulombian energy associated to these surface charges  is included in the Hamiltonian Eq.(\ref{Hem}) with another capacitance $C_{coul}$  and leads  to the  renormalisation of the line capacitance $C' \rightarrow 1/(1/C' +1/C_{coul})$. 

Let us show the power of this approach on some examples. 
In the simplest case of  parallel planes of large transverse width $W$ and interdistance  $D$, so that 
${\vc{A}}_\perp (\vc{r}_\perp)=(A_\perp,0,0)$, we obtain 
\begin{eqnarray}
C'= \frac{dC}{dr_z} = \frac{\epsilon \int_{-W/2}^{W/2}  dr_y}{\int_1^2 dr_x}= \frac{\epsilon W}{D} \, .
\end{eqnarray}
In the case of a coplanar waveguide of size $w$ for the signal wire and $s$ for the interspace represented in Fig.\ref{cpw} assuming the vacuum dielectric constant everywhere ($\epsilon_0=\epsilon_1=\epsilon_2$),
we need to use two conformal transformations one on the upper and the other on lower half planes \cite{Watanabe_1994,375223}.
For each areas, this process redefines up to a normalisation constant the effective width $W_0\sim 2 K(k_0)$ and the effective interdistance $D_0 \sim K(k'_0)$ where $k_0=w/(w+2s)$ and $k'_0=\sqrt{1-k_0^2}$ are arguments of the complete elliptic integral. As a result, we obtain  $C'=\epsilon 2W_0/D_0=\epsilon 4 K(k_0)/K(k'_0)$, an extra factor $2$ is added to include both half planes \cite{Watanabe_1994,375223}. 

Using $\vc{l_j}=(-l_j,\vc{l_j}_\perp)$, the Hamiltonian Eq.(\ref{chargephase}) becomes:
\begin{eqnarray}\label{H7}
\hat H
&=& E_0+ \sum_{j=1}^N E_{J,j}\left[1-
\cos\left(\frac{2e}{\hbar}\hat \Phi_j(t) \right) \right]
+ \frac{1}{2} \sum_{j,j'=1}^N C^{-1}_{j,j'} 
\hat q_j(t) \hat q_{j'}(t)  
\nonumber \\
&+&  \int_{-\infty}^\infty dr_z
\frac{1}{2}\left[(\hat{\dot{\alpha}}(r_z,t))^2- c^2 
{\hat{{\alpha}}(r_z,t)\partial_z^2 \hat{{\alpha}}(r_z,t)}\right]
- \sum_{j=1}^N f_j
\frac{\hat{\dot{\alpha}}(-l_j,t)}{\sqrt{C'}}\hat q_{j}(t) \, ,
\end{eqnarray}
where we define the dimensionless parameter: 
\begin{eqnarray}
f_j&=&\sqrt{C'}\int_0^1 du \,\int d^3 \vc{r}\,\vc{A}_{\perp}(\vc{l_j}_\perp +u(\vc{r}-\vc{l_j}_\perp)).\vc{r}_\perp 
(|\psi_{j,+}(\vc{r})|^2-|\psi_{j,-}(\vc{r})|^2)  \, .
\end{eqnarray}
This complicated formula depends on the complicated architecture of the circuit but can nevertheless be interpreted as follows. Using the rough formula $C'\sim \epsilon \mathfrak{r}/\mathfrak{l}$ where $\mathfrak{r}$ is the average spatial range of the microwave electric field in the xy plane and $\mathfrak{l}$ is the average distance between two wires, we estimate
$\vc{A}_{\perp}(\vc{r})\sim 1/\sqrt{\epsilon\mathfrak{l} \mathfrak{r} }$ from the normalisation.
If $\mathfrak{d}$ is the distance between the two islands of the transmon,
we find the magnitude $f_j \sim  \mathfrak{d}/ \mathfrak{l}$. 

In what follows, we shall omit $E_0$ in Eq.(\ref{H7}).

\subsection{Generalisation to the dielectric-filled waveguide}

In the experimentally relevant case when the insulating substrates (silicon oxide, sapphire,  etc) are present \cite{doi:10.1063/1.3010859}, the  permittivity  parameter depends on the position  
$\epsilon (\vc{r}_\perp)$ keeping the permeability $\mu_0$  constant and affect the speed of propagation. Repeating the reasoning of previous sections with such dependence in Eq.(\ref{chargephase}), we generalize  the equations  for the transverse electromagnetic component as:
\begin{eqnarray}
\nabla_\perp \times \vc{A_\perp} (\vc{r}_\perp)= 0 \, , \quad \quad \quad \vc{n}\times \vc{A}_\perp(\vc{r}_\perp)=0
{\rm \ at \ the \ surface \ boundary } \, ,
\quad \quad \nabla_\perp . (\epsilon(\vc{r}_\perp)\vc{A}_\perp(\vc{r}_\perp))=0   \, .
\end{eqnarray}
These  components are normalized such that $\int d^2\vc{r}_\perp \epsilon(\vc{r}_\perp)\vc{A}^2_\perp(\vc{r}_\perp)=1$.  
It is still possible to define a scalar potential via
$\vc{A_\perp}(\vc{r}_\perp)=\nabla_\perp \chi(\vc{r}_\perp)$ so the problem reduces into the Laplace-like 
equation:
\begin{eqnarray}\label{chidiel}
\nabla_\perp .\left(\epsilon(\vc{r}_\perp)  \nabla_\perp\chi (\vc{r}_\perp)\right)= 0  \, ,\quad \quad \quad 
\chi(\vc{r}_\perp)=\chi_s
{\rm \ at \ signal \ wire   } \, ,\quad \quad \quad 
\chi(\vc{r}_\perp)= 0
{\rm \ at \ ground  \ wire  }  \, .
\end{eqnarray} 
We find that the propagation speed $v$ is given by 
\begin{eqnarray}
v^2=\frac{\int d^2\vc{r}_\perp \vc{A}^2_\perp(\vc{r}_\perp)}{\mu_0 \int d^2\vc{r}_\perp \epsilon(\vc{r}_\perp)\vc{A}^2_\perp(\vc{r}_\perp)}=\frac{\int d^2\vc{r}_\perp \left(\nabla_\perp \chi(\vc{r}_\perp)\right))^2
}{\mu_0} \not= c^2  \, .
\end{eqnarray}
\begin{figure}
\begin{center}
 \includegraphics[width=8cm]{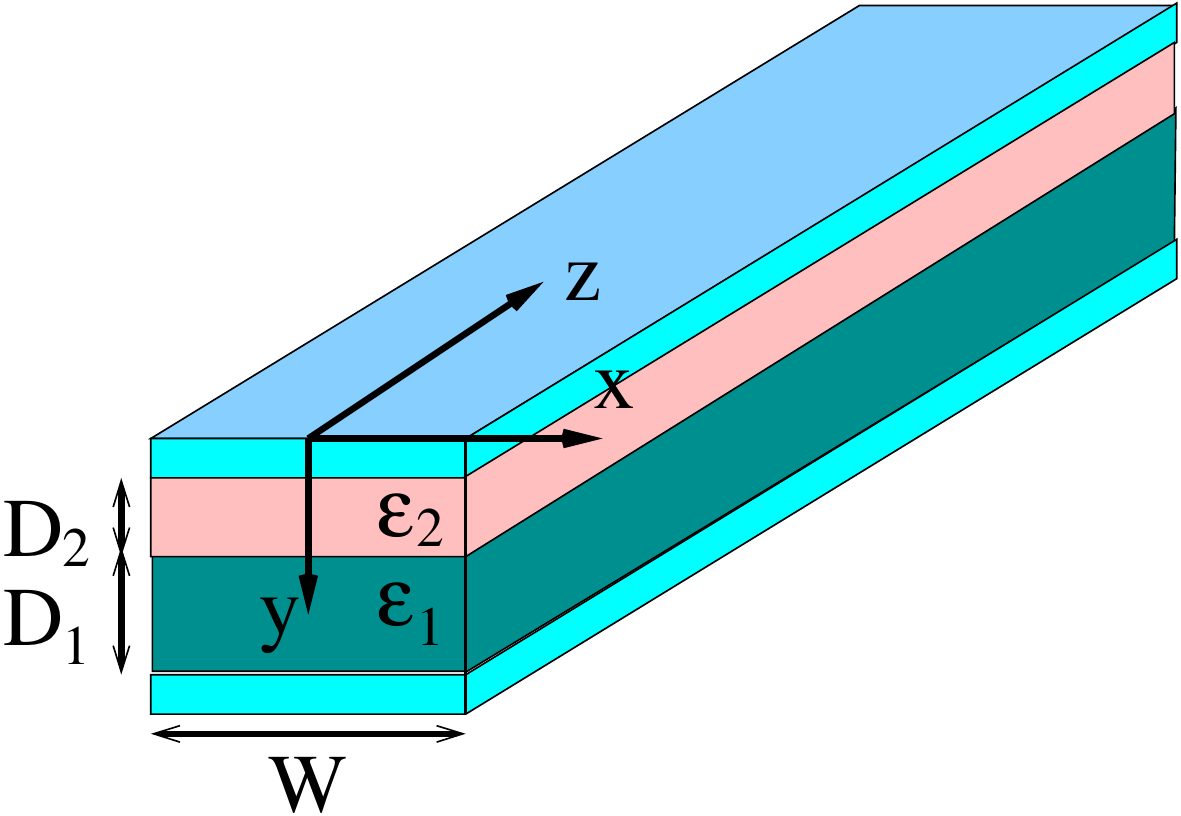}
\end{center}
\caption{\label{dielwg} Schematic representation of a waveguide made of two superconducting planes filled with two dielectrics.}
\end{figure}

In order to have a first idea of the expected outcomes, let us estimate the speed for the simple case in Fig.\ref{dielwg} of two parallel plane signal and ground wire of transverse width $W$ filled with two separate dielectrics, one  of permittivity $\epsilon_1$  of thickness $D_1$  followed  by another  of permittivity  $\epsilon_2$ of thickness $D_2$ .
Solving the equation, we obtain a different field 
${\vc{A}}_\perp (\vc{r}_\perp)=(1/[\epsilon_i \sqrt{W(D_1/\epsilon_1+ D_2/\epsilon_2})],0,0)$ for each dielectric $i=1,2$ (We neglect the edge effect). We obtain the phase velocity:
\begin{eqnarray}\label{v}
v= \sqrt{\frac{D_1/\epsilon^2_1+ D_2/\epsilon^2_2}
{\mu_0(D_1/\epsilon_1+ D_2/\epsilon_2)}}  \, .
\end{eqnarray}
This result unfortunately leads to the inequality $v\not= 1/\sqrt{L'C'}$ valid strictly for the vacuum case. The definitions Eqs.(\ref{C'}),(\ref{L'}) and (\ref{defs})  are generalized into:
\begin{eqnarray}\label{Cst}
C' = \frac{\oint \epsilon(\vc{r}_\perp){\vc{A}}_\perp (\vc{r}_\perp).d\vc{s}}{\int_1^2 d\vc{r}_\perp.{\vc{A}}_\perp (\vc{r}_\perp)}=\frac{W}
{(D_1/\epsilon_1+ D_2/\epsilon_2)} \, ,
\\ \label{Lst}
L' =
\frac{\mu_0 \int_1^2 d\vc{r}_\perp.{\vc{A}}_\perp (\vc{r}_\perp) }
{ \oint  d\vc{s}. {\vc{A}}_\perp (\vc{r}_\perp)} 
=
\frac{\mu_0 \epsilon_2(D_1/\epsilon_1+ D_2/\epsilon_2)}
{W} \, ,
\end{eqnarray}
which yields instead $1/\sqrt{L'C'}=1/\sqrt{\mu_0 \epsilon_1}$. If we would use the inductance calculated from the ground wire $\epsilon_1 L'/\epsilon_2$, another different  result would have been been found.  From this example, this result clearly differs from the results obtained using an electrostatic approach  \cite{375223} and magnetostatic approach \cite{Watanabe_1994} because no additional polarisation or ``bound'' currents and charge are included dynamically inside the dielectric at non zero frequencies. These unwanted currents not predicted in a static approach affect the magnetic field inside the dielectrics and may well create a current difference between the signal and the ground wires requiring a full dynamic treatment. 

\subsection{The coplanar waveguide}

\begin{figure}
\begin{center}
\includegraphics[width=9cm]{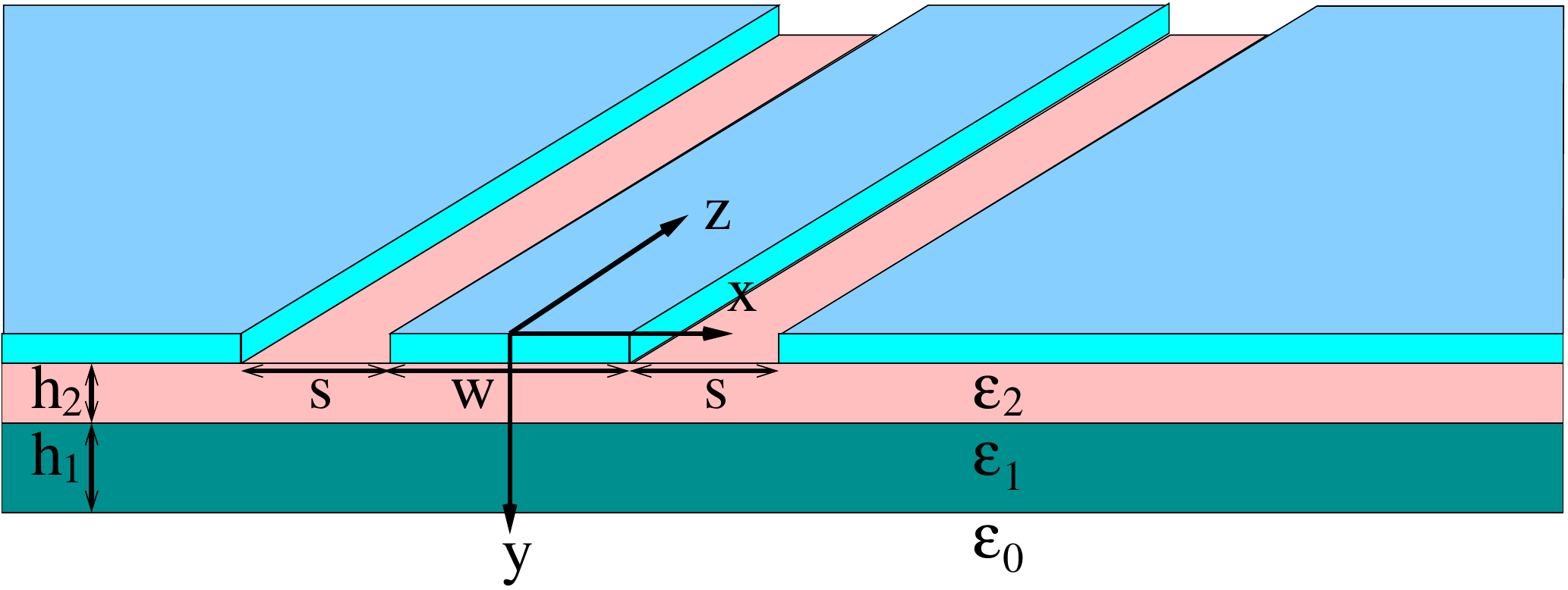}
\end{center}
\caption{\label{cpw} 
Schematic representation of the coplanair waveguide with two layers of dielectric materials.
}
\end{figure}

\begin{figure}
\begin{center}
\includegraphics[width=15cm]{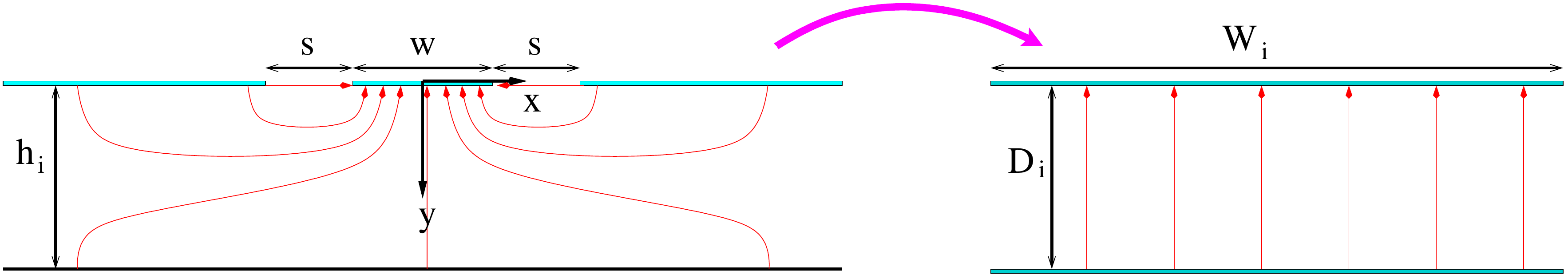}
\end{center}
\caption{\label{conformal} 
Schematic representation of the conformal transformation for the line force of the field $\chi_i(\vc{r}_\perp)$ for a CPW.
}
\end{figure}

In the case of a coplanar waveguide (CPW) \cite{doi:10.1063/1.3010859} represented in Fig.\ref{cpw}, we solve the partial differential Eqs.(\ref{chidiel}) with 
three dielectric constants: $\epsilon_i$ in two different regions $i=1,2$ of thickness $h_i$ and  $\epsilon_0$ otherwise. The shielding to the ground in the lower part is assumed at a distance $h_0+h_1+h_2$. Generally, this distance  is large and can be taken to infinity.
Typically the first region is  silicon  and the region 2 is  silicon oxide  \cite{doi:10.1063/1.3010859}. 
The thickness of the superconducting plates are negligible. In these conditions Eq.(\ref{chidiel}) reduces 
to the Laplace equation except at the surface of the dielectric where, for both interfaces 2-1 and 1-0, the boundary conditions over the normal component of the field apply, respectively $\epsilon_2 \partial_{r_y} \chi(\vc{r}_\perp) =\epsilon_1 \partial_{r_y} \chi (\vc{r}_\perp)$ and $\epsilon_1 \partial_{r_y} \chi(\vc{r}_\perp) =\epsilon_0 \partial_{r_y} \chi (\vc{r}_\perp)$.

The capacitance is a sum of two capacitances for each half plane $C'=C'_0 +C'_d$. 
In the upper half plane, we recover the result 
$C'_0=\epsilon_0 W_0/D_0$ of the previous section for the vacuum case.
In the lower half plane, the solution is of the form $\chi=\chi_0+\chi_1+\chi_2$ where $\chi_0$
is the vacuum solution of the entire half space, $\chi_1$ is the non zero solution only in the regions $1$ and $2$ and 
$\chi_2$ is the non zero solution only in the region $2$. These are solutions of the Laplace equation and obey the whole set of boundary conditions:
\begin{center}
\begin{tabular}{|l | c l|}
\hline
1 &$\chi_i(r_x,r_y=0)=\chi_{s,i}$ & for $r_x \in[-w/2,w/2]$ \\
2 & $\chi_i(r_x,r_y=0)=0$ & for $r_x \in [-\infty,-s-w/2] \cup [s+w/2,\infty]$ \\
3 & $\partial_{r_x}\chi_i(r_x,r_y=0)=0$ & for $r_x \in [-s-w/2,-w/2] \cup [w/2,s+w/2]$\\
4  & \multicolumn{2}{| l |}{$\chi_i(r_x=\pm \infty ,r_y=0)=0$}
\\
5  & 
\multicolumn{2}{| l |}{
$\chi_0(r_x,r_y=h_0+h_1+h_2)=0$} 
\\
6 & \multicolumn{2}{| l |}{
$\epsilon_0 \partial_{r_y} \chi_0(r_\perp)|_{r_y=h_1+h_2+0^+}= \epsilon_1 \partial_{r_y} (\chi_0(r_\perp)+ \chi_1(r_\perp))|_{r_y=h_1+h_2-0^+}$} 
\\
7 &
\multicolumn{2}{| l |}{
$\epsilon_1 \partial_{r_y} (\chi_0(r_\perp) + \chi_1(r_\perp))|_{r_y=h_2+ 0^+}= \epsilon_2 \partial_{r_y} (\chi_0(r_\perp)+ \chi_1(r_\perp)+ \chi_2(r_\perp))|_{r_y=h_2-0^+}$}\\
\hline
\end{tabular}
\end{center}
The  boundary conditions (1-5) are treated using the techniques 
of conformal transformation developed in  \cite{375223,Watanabe_1994} and represented in Fig.\ref{conformal}. Using the Gauss and the circulation theorems, it is not difficult to show that any region mapped by $\chi_i$ has a boundary value $\chi_{i,s}$ on the signal wire, and zero everywhere else at the other boundaries, with the exception on the line of interdistance $s$ where a zero normal derivative is imposed instead.
Therefore, each dielectric region $i=0,1,2$ associated to $\chi_i$ can be conformally mapped into a rectangle of width $W_i=2K(k_i)$ and length $D_i=K(k'_i)$ where we define 
$k_i=\tanh(\pi w/2\sum_{i'=i}^2 h_{i'})/\tanh(\pi (w+2s)/2\sum_{i'=i}^2 h_{i'})$ and $k'_i=\sqrt{1-k_i^2}$ \footnote{for a concrete calculation, we need to use the useful asymptotic expression $K(k) \sim \ln(4/\sqrt{1-k^2}) \simeq \ln(2)+\pi w_i/(2h_i)$}.
After the transformation, we deduce also the relation $W_i\chi_{s,i}/D_i=-\int_{-w/2}^{w/2}  dr_x \partial_{r_y}\chi_i (\vc{r}_\perp)$ which allows to define the vacuum capacitance for each area as $C'_i=\epsilon_0 W_i/D_i$. 

The two latest boundary conditions (6,7) correspond to the effective surface charges at the interface 01 and 12.  integrating over $r_x$ and using the Gauss theorem at the boundary for each region $1$ and $2$, 
we find the following relations between the $\chi_{s,i}$: 
\begin{eqnarray}
\epsilon_0 C'_0\chi_{s,0}=\epsilon_1(C'_0\chi_{s,0}+C'_1\chi_{s,1})=\epsilon_2(C'_0\chi_{s,0}+C'_1\chi_{s,1}+C'_2\chi_{s,2})  \, .
\end{eqnarray}
Integrating by parts the normalization condition $\int d^2\vc{r}_\perp \epsilon(\vc{r})\left(\nabla_\perp \chi(\vc{r}_\perp)\right)^2=1$, we find also that $\chi_s^2 C'=1$ similarly to the vacuum case.
With these relations,   we eliminate the variables $\chi_{s,i}$ and  we find the line capacitance of the lower plane: 
\begin{eqnarray}
C'_d&=&
-\frac{\epsilon_2}{\chi_s} \int_{-w/2}^{w/2}  dr_x \partial_{r_y}\chi (r_x,0^+)= 
\frac{1}{\displaystyle \frac{1}{C'_0}-\frac{1}{C'_1}\left(\frac{\epsilon_0}{\epsilon_1} -1\right)-\frac{1}{C'_2}
\left(\frac{\epsilon_0}{\epsilon_2} -\frac{\epsilon_0}{\epsilon_1}\right)}  \, .
\end{eqnarray}
Similarly working out $\int d^2\vc{r}_\perp \left(\nabla_\perp \chi(\vc{r}_\perp)\right)^2$ through partial integration at the  interfaces  we determine using Eq.(\ref{v}) the speed of light $v=c/\sqrt{\epsilon_{rel}}$ with an effective relative permittivity determined from:
\begin{eqnarray}
\frac{1}{\epsilon_{rel}}= \frac{2C'_0}{C'}+\left[\left(1-\frac{C'_0}{C'_1}\right)\left(\frac{\epsilon_0}{\epsilon_1}-1\right)\left[\left(\frac{\epsilon_0}{\epsilon_1}-1\right)\frac{1}{C'_1}+2\left(\frac{\epsilon_0}{\epsilon_2}-\frac{\epsilon_0}{\epsilon_1}\right)\frac{1}{C'_2}\right]+\left(1-\frac{C'_0}{C'_2}\right)\frac{1}{C'_2}\left(\frac{\epsilon_0}{\epsilon_1}-\frac{\epsilon_0}{\epsilon_2}\right)^2\right]\frac{C'^2_d}{C'} \, .
\nonumber \\
\end{eqnarray}
In order to have unambiguous definitions for the effective line capacitance and inductance, we require that their values are based on the experimental measurements of phase velocity $v$ and the impedance obtained, using a monochromatic wave of frequency $\omega$, from the ratio:
\begin{eqnarray}
Z=\frac{\langle \hat V(r_z,t) \rangle}{\langle \hat I(r_z,t) \rangle}=
\frac{\mu \int_1^2 d\vc{r}_\perp.{\vc{A}}_\perp (\vc{r}_\perp) 
\langle  \hat{\dot \alpha}(r_z,t)\rangle}
{ \oint  d\vc{s}. {\vc{A}}_\perp (\vc{r}_\perp)\langle \partial_{r_z} \hat{\alpha}(r_z,t) \rangle} 
=
\frac{\mu \int_1^2 d\vc{r}_\perp.{\vc{A}}_\perp (\vc{r}_\perp) 
\omega}
{ \oint  d\vc{s}. {\vc{A}}_\perp (\vc{r}_\perp) k} 
=
\frac{\mu v\int_1^2 d\vc{r}_\perp.{\vc{A}}_\perp (\vc{r}_\perp)}
{ \oint  d\vc{s}. {\vc{A}}_\perp (\vc{r}_\perp) } =v L' \, .
\end{eqnarray}
In the case of a CPW, we obtain $L'=\mu/(C'_u/\epsilon_0+ C'_d/\epsilon_2)$ using the definition in (\ref{Lst}).
From any experimental measurements of the speed $v$  and the impedance $Z$, we extract the static inductance using $L'=Z/v$. Normally, as for the vacuum case, the experimental extraction of the capacitance $C'_{eff}$ should be done using the formula $v=1/\sqrt{L' C'_{eff}}$ and impedance $Z=\sqrt{L'/C'_{eff}}$. 
However, in presence of dielectrics, the deduced formula $C_{eff}=1/(L' v^2)$ differs from the static formula $C'$ defined in (\ref{Cst}). Using (\ref{Lst}), we find the expression:
\begin{eqnarray}
C'_{eff} =\frac{\oint {\vc{A}}_\perp (\vc{r}_\perp).d\vc{s}}{\mu v^2\int_1^2 d\vc{r}_\perp.{\vc{A}}_\perp (\vc{r}_\perp)}\not= C'  \, .
\end{eqnarray}
As a consequence, the measured impedance $Z$  differs also  from the static one $Z_{st}=\sqrt{L'/C'}$ determined from electrostatic capacitance  and magnetostatic inductance calculations. 

The table \ref{table:2} displays the parameter values obtained for 
various model descriptions of the CPW, with comparisons between static capacitance and impedance  and effective capacitance and impedance. The two half plane model corresponds to the case where the upper half plane is the vacuum  while the lower half plane is filled with silicon; the second model corresponds to the case where the silicon oxide at the interface is not present in contrast to  the full  model. We note that the full model provides the best prediction for the effective permittivity, but with an impedance very low in comparison to the usual chosen value of $50\Omega$ as a standard in a waveguide. 
In contrast, the absence of the silicon oxide prevents a leaked polarisation current at the interface which results in a higher inductance and thus an impedance close the desired value.  Therefore  the interface current has to be taken into account and probably could also contribute to the total current at the output of the quantum circuit through ac capacitive coupling. 

The conclusion of this section is that in presence of dielectrics, we should distinguish between the real capacitance measured in static conditions and the effective capacitance. It is caused by additional polarisation currents at the dielectric  interfaces and results also in line loss caused by their intrinsic resistance but that are not studied in this work. 

For simplicity, in the main text and in what follows, we shall keep the notation of $c$ and $C'$ to actually mean  $v$ and $C'_{eff}$ instead.

\begin{table}[h!]
\centering
\begin{tabular}{|c|c|c|c|c|c|c|c|}
\hline
Cases & $C'$ & $v/c$ & $\epsilon_{eff}$ & $L'$ & $C_{eff}$ & $Z=vL'$ & $Z_{st}=\sqrt{\frac{L'}{C'}}$ \\
\hline
& $10^{-10}F$ & & & $10^{-7}H$ & $10^{-10}F$ & $\Omega$ & $\Omega$ \\
\hline
2 half planes & 1.55 & 0.398 & 6.3 & 8.32  &0.84 & 99.4 & 73 \\
$\epsilon_2=\epsilon_1$ & 1.54 & 0.409 & 5.99 & 4.54 & 1.47 & 55.6 & 54.3 \\
Full model  & 1.44 & 0.434 & 5.30 & 2.36 & 2.49 & 30.8 & 40.5\\
Simulation \cite{doi:10.1063/1.3010859} & 1.27 & 0.438 & 5.22 & 4.53 & NA & NA & 59.7 
\\
Experiment \cite{doi:10.1063/1.3010859} & & 0.455 & 5.05 & &&& \\
\hline
\end{tabular}
\caption{ \label{table:2} Comparison of various model description with experiment and simulation for the CPW waveguide. Parameter  values of \cite{doi:10.1063/1.3010859} are $s=6.6\mu m$, $w=10\mu m$, $h_1=500\mu m$, 
$h_2=550 nm$,  $\epsilon_1/\epsilon_0= 11.6$, $\epsilon_2/\epsilon_0=3.78$.}
\end{table}

\section{Accounting for the resonator capacitances}\label{D}

\subsection{Hamiltonian term derivation}

The phenomenological treatment of the resonator \cite{doi:10.1063/1.3010859} is based on the lumped-element representation and is an efficient approach for a classical  description of the microwave radiation in the system.  In order to address the quantum aspects here, we derive the
Hamiltonian term for the resonator cavity in the waveguide. In the following sections we will show how it leads to the consistent quantum derivation of the resonator's spectrum and damping.

The resonator is located between the position $r_z=-L,0$ respectively where the signal line is disrupted by a cut or an insulator gap (see Fig.\ref{fig0}). Both gaps are assumed identical with their charges located at $r_z^\pm=-L^\pm , 0^\pm$ and  the
width $w_g=|L^+ - L^-|=|0^+ - 0^-|$.  
Physically, the electric current transporting electrical charge to the gap leads to a net electrical charge at the cut of the wire. As a result, we need to modify the quantum field description of the waveguide as:
\begin{eqnarray}
\sqrt{N_{s,0}}\psi (\vc{r}) \rightarrow  \hat \psi (\vc{r},t)=
\sqrt{N_{s,0}}\psi_{0}(\vc{r}) + \delta \hat \psi (\vc{r},t) \, ,
\end{eqnarray}
with a net  local charge distribution  described by the perturbation $\delta \hat \psi (\vc{r},t)$. If the wavefunction abruptly drops to zero beyond the position $r_z=-L^-$ at the left end of the gap, 
the variation of charge is determined from the conservation law as: 
\begin{eqnarray}\label{cons}
(2e)\partial_t |\hat \psi (\vc{r},t)|^2= 
-\nabla. \hat {\vc{j}} (\vc{r},t)
\cong \delta(r_z +L^-_z) \hat {j}_z (\vc{r},t) \, .
\end{eqnarray}
From integration over the transverse plane, the net total charge accumulated at $-L^-$ is:
\begin{eqnarray}
 \hat Q_-(-L^-,t) =(2e)\int^{-L^-}_{-\infty}dr_z \int d^2\vc{r}
 (|\hat \psi (\vc{r},t)|^2-
N_{s,0}|\psi_0(\vc{r})|^2) \, .
\end{eqnarray}
Integrating over the transverse coordinates and using Eq.(\ref{cons}) and the local equation for the surface charge conservation with the current (\ref{cur}), we obtain:
\begin{eqnarray}\label{qp0}
\partial_t \hat Q_-(-L^-,t) \cong \int d^2\vc{r}_\perp \, \hat {j}_z (\vc{r}_\perp,-L^-,t)= \hat I(r_z,t)=-\int_{-\infty}^{-L^-} dr_z \partial_{t} \partial_{r_z} \hat Q(r_z,t) + \hat I(-\infty,t)
\end{eqnarray}
where we can assume that the effect of far distant current $\hat I(-\infty,t)$ is negligible for any signal localized in time and space.  The integration over time produces: 
\begin{eqnarray}\label{qp}
 \hat Q_-(-L^-,t) = -\int_{-\infty}^{-L_-} dr_z \partial_{r_z}
 \hat Q(r_z,t)  \, .
\end{eqnarray}
That is, the charge at the signal wire ends is opposite of   the total charge obtained  from integration over the line. 
A similar expression holds for a charge coming from the right:  $\hat Q_+ (0^+,t)=\int_{-\infty}^{0^+} dr_z \partial_{r_z}\hat Q(r_z,t)$. 
However, for the charge accumulated inside the resonator, 
we find instead the neutrality constraint: 
\begin{eqnarray}\label{qc}
 \hat Q_+(-L^+,t) + \hat Q_-(0^-,t)  = -\int_{-L^+}^{0^-} dr_z \partial_{r_z}
 \hat Q(r_z,t)= -\int_{-L^+}^{0^-} dr_z \epsilon \oint  d\vc{s}. \hat {\vc{E}} (\vc{r},t) \, ,
\end{eqnarray}
where the last equality results from the Gauss theorem applied to the resonator.
For each gap  of the resonator, we define the self capacitance  $C$ and the interaction capacitance $C_I$ such that the Hamiltonian term associated with these charges located at $r_z^\pm=-L^\pm , 0^\pm$ is:
\begin{eqnarray}\label{Hcb}
\hat H_c &=& \sum_{r_z=0,-L} \frac{\hat Q^2_+ (r_z^+,t)+\hat Q^2_- (r_z^-,t)}{2C}+ \frac{\hat Q_+ (r_z^+,t)\hat Q_- (r_z^-,t)}{C_I}  \, .
\end{eqnarray}
Assuming a conductor of cross
section $S$ over which the excess charges are uniformly distributed, we identify by comparison with the Coulombian term in Eq.(\ref{H0}) the self capacitance formula and the interaction capacitance: 
\begin{eqnarray}
\frac{1}{C}=\frac{1}{S^2}
\int_S d^2\vc{r}_\perp \int_S d^2\vc{r'}_\perp  \frac{ 
1}{4\pi\epsilon\sqrt{(\vc{r}_\perp -\vc{r'}_\perp)^2}} \, ,\quad \quad \quad 
\frac{1}{C_I}=\frac{1}{S^2}
\int_S d^2\vc{r}_\perp \int_S d^2\vc{r'}_\perp  \frac{ 
1}{4\pi\epsilon\sqrt{(\vc{r}_\perp -\vc{r'}_\perp)^2+
w_g^2}}   \, .
\end{eqnarray}
Usually in a classical circuit, the neutrality conditions is fulfilled so that the left and right charges are opposite $\hat Q_+(r_z^+,t)=-\hat Q_- (r_z^-,t))$.
In our case, however, we  deal 
with a sufficiently large gap so that $C_I \gg C$ and the interaction capacitance term does not contribute in (\ref{Hcb}). The validity of this assumption relies on the experimentally confirmed results Eqs.(\ref{spec}) and (\ref{Qn}), but the term $C_I$ should be generally included as well. 
Also, a too high gap  may induce the loss in the line 
leading to a saturation of the quality factor  \cite{doi:10.1063/1.3010859}.

Doing an {\it a posteriori} reasoning over the result of the appendix \ref{H}, the resonating modes of the cavity  
cancels the integral over the electric field in Eq.(\ref{qc}) and imposes neutrality of the edge charges  $\hat Q_+(-L^+,t)=-\hat Q_- (0^-,t))$. In addition, if we note that  the only remaining independent operator variable  $\hat Q_+(-L^+,t)-\hat Q_- (0^-,t)$  minimizes the Hamiltonian (\ref{Hcb}) for a zero value, we conclude that these charges inside the resonator are zero resulting in a charge asymmetry at each capacitance. 
Therefore, the Hamiltonian Eq.(\ref{Hcb}) becomes:
\begin{eqnarray}\label{Hc}
\hat H_c=\frac{\hat{Q}^2_-(-L^-,t)+\hat{Q}^2_+(0^+,t)}{2C} \, , \quad \quad \quad 
\hat{Q}_\pm(r_z,t) = \pm\int_{-\infty}^{\infty} dr'_z 1^+(\pm(r_z-r'_z)) \sqrt{C'}\hat{\dot{\alpha}}(r'_z,t)  \, .
\end{eqnarray}
where $1^+(r_z)$ is the Heaviside function.
Since the gap is very small compared to the wavelength, we will assume that $-L^-\cong -L$ and $0^+ \cong 0$ in the following.

\subsection{The quantum circuit formulation with a resonator}

As a result of the new term (\ref{Hc}), the Hamiltonian (\ref{H7}) becomes:
\begin{eqnarray}
\hat H
&=&  
\int_{-\infty}^\infty dr_z
\frac{1}{2}\left[
\hat{\dot{\alpha}}^2(r_z,t)- c^2 
{\hat{{\alpha}}(r_z,t)\partial_z^2 \hat{{\alpha}}(r_z,t)}\right] + 
\frac{C'}{2C}\left[\left(\int_{-\infty}^{L} dr'_z\hat{\dot{\alpha}}(r'_z,t)\right)^2
+\left(\int_{0}^{\infty} dr'_z\hat{\dot{\alpha}}(r'_z,t)\right) 
\right]
\nonumber \\
&+& \sum_{j=1}^N E_{J,j}\left[1-
\cos\left(\frac{2e}{\hbar}\hat \Phi_j(t)\right) \right] 
+ \frac{1}{2} \sum_{j,j'=1}^N C^{-1}_{j,j'} 
\hat q_j(t) \hat q_{j'}(t) 
-\sum_{j=1}^N \frac{\hat{\dot{\alpha}}(-l_j,t)}{\sqrt{C'}}
f_j\hat q_{j}(t) \, .
\end{eqnarray}
This Hamiltonian is nonlocal in the field $\hat{{\alpha}}(r_z,t)$ which makes the resulting quantum field theory unconventional.
In terms of the effective charge and flux of the waveguide field expressed in (\ref{Hem}), the Hamiltonian takes the more recognizable  quantum circuit form:
\begin{eqnarray}\label{Htransmon}
\hat H
&=&  
\int 
\left[ \frac{d\hat{Q}^2}{2dC} + \frac{d{\hat \Phi_B}^2}{2dL}
\right]+ \frac{\hat{Q}^2_-(-L,t)+\hat{Q}^2_+(0,t)}{2C}
\nonumber \\
&+& \sum_{j=1}^N E_{J,j}\left[1-
\cos\left(\frac{2e}{\hbar}\hat \Phi_j(t)\right) \right]  
+ \frac{1}{2} \sum_{j,j'=1}^N C^{-1}_{j,j'} 
\hat q_j(t) \hat q_{j'}(t) 
+\sum_{j=1}^N f_j\frac{d\hat{Q}(-l_j,t)}{dC}\hat q_{j}(t)
\, .
\end{eqnarray}

\section{The qubit Hamiltonian and the energy balance relation}
\label{E}

\subsection{Derivation}

For renormalisation purposes, the 
Hamiltonian (\ref{Htransmon}) is rewritten using a unitary transformation: $\hat \Phi_j(t) \rightarrow \hat \Phi_j(t) -\frac{f_j}{\sqrt{C}}
\hat{\alpha}(-l_j,t)$ and $\hat{\dot \alpha}(r_z,t) \rightarrow  \hat{\dot \alpha}(r_z,t)
+\sum_{j=1}^N \frac{f_j}{\sqrt{C'}} \hat q_j \delta(r_z+l_j)$. This result yields: 
\begin{eqnarray}
\hat H
&=&  
\int dr_z
\frac{1}{2}\left[
\hat{\dot{\alpha}}^2(r_z,t)- c^2 
{\hat{{\alpha}}(r_z,t)\partial_z^2 \hat{{\alpha}}(r_z,t)}\right] +
\frac{\hat{Q}^2_-(-L,t)+\hat{Q}^2_+(0,t)}{2C}
\nonumber \\
&+&\sum_{j=1}^N E_{J,j}\left[1-
\cos\left(\frac{2e}{\hbar}(\hat \Phi_j(t)- \frac{f_j}{\sqrt{C'}}\hat \alpha(-l_j,t))\right)\right] 
+ \frac{\hat q^2_j(t)}{2C_j} \, ,
\end{eqnarray}
where $1/C_j=1/C_{j,j}+ f_j^2 \delta(0)/C'$ in which $``\delta(0)'' \simeq 1 /\Delta r$ with $\Delta r$ is 
the size of the transmon.

From inspection of  the Hamiltonian (\ref{Htransmon}) when we omit the coupling term,
we recover the well known anharmonic oscillator dynamics of each transmon characterized by the Josephson energy much larger than  the capacitance energy  and that leads to a non equidistant spectrum \cite{PhysRevA.76.042319}. 
In these conditions and if the transmons are sufficiently far from each other to neglect their mutual capacitance interaction, we can restrict the description  to the two first states so that the transmon becomes in good approximation a qubit.   

Using the notation: 
\begin{eqnarray}
\hat \Phi_j=\frac{\hbar}{2e}\left(\frac{8E_{C,j}}{E_{J,j}}\right)^{1/4}\frac{\hat b_j +\hat b^\dagger_j}{\sqrt{2}} \, ,\quad \quad 
\hat q_j=2e\left(\frac{E_{J,j}}{8E_{C,j}}\right)^{1/4}\frac{\hat b_j -\hat b^\dagger_j}{\sqrt{2i}} \, , \quad \quad 
E_C=\frac{e^2}{2C_j}  \, ,  \quad \quad \hbar\omega_j=\sqrt{8E_{C,j}E_{J,j}} \, ,
\end{eqnarray}
we reexpress the Hamiltonian using the creation annihilation operators $\hat b^\dagger_j$ and $\hat b_j$. In the limit $E_{C,j} \ll E_{J,j}$, we can limit the expansion up to 
the quadratic terms:
\begin{eqnarray}
\hat H
&=&  
\int dr_z
\frac{1}{2}\left[
\hat{\dot{\alpha}}^2(r_z,t)- c^2 
{\hat{{\alpha}}(r_z,t)\partial_z^2 \hat{{\alpha}}(r_z,t)}\right] +
\frac{\hat{Q}^2_-(-L,t)+\hat{Q}^2_+(0,t)}{2C} +\sum_{j=1}^N \frac{c\kappa^2_j}{\omega_j}\hat{\alpha}^2(-l_j,t)
\nonumber \\
&+&\hbar \omega_j \hat b_j^\dagger(t) \hat b_j(t) -
\sqrt{\hbar c}\kappa_j\left(\hat b_j(t) +\hat b^\dagger_j(t)\right)\hat{\alpha}(-l,t) \, ,
\end{eqnarray}
where we define the coupling with electromagnetic field $\kappa_j=\frac{2e}{\hbar}\frac{f_jE_{J,j}}{\sqrt{2cC'}}\left(\frac{8E_{C,j}}{E_{J,j}}\right)^{1/4}=\frac{2e}{\hbar}\frac{f_j\sqrt{E_{J,j}\omega_j}}{\sqrt{2cC'}}$.
We shall omit the third quadratic term proportional to $\kappa^2_j/\omega_j$
in the field as it gives a negligible contribution  to the detuning frequency much small than the microwave frequency. 
Since we are interested in the two first levels, we replace the transmon operators by the 
qubit Pauli operators and obtain finally:
\begin{eqnarray}\label{Hqbit}
\hat H
&=&  
\int dr_z
\frac{1}{2}\left[
\hat{\dot{\alpha}}^2(r_z,t)- c^2 
{\hat{{\alpha}}(r_z,t)\partial_z^2 \hat{{\alpha}}(r_z,t)}\right] +
\frac{\hat{Q}^2_-(-L,t)+\hat{Q}^2_+(0,t)}{2C} 
+\sum_{j=1}^N\frac{\hbar \omega_j}{2}\hat \sigma^z_j(t)-\sqrt{\hbar c}\kappa_j\hat \sigma^x_j(t)\hat{\alpha}(-l_j,t)
\, .
\nonumber \\
\end{eqnarray} 
Corrections due to the anharmonicity  renormalize this energy into $\omega_j=\sqrt{8E_{C,j}E_{J,j}}-E_{C,j}$  as established in \cite{PhysRevA.76.042319}. 
We set $\hbar=1$ in what follows to simplify the formulas.

\subsection{Heisenberg and scattering equations}

Using $\partial_t \hat A= i[\hat H, \hat A]$ and the Pauli matrix algebra 
$[\hat \sigma^+,\hat \sigma^-]=\hat \sigma^z$, $[\hat \sigma^z,\hat \sigma^\pm]=\pm 2 \hat \sigma^\pm$, the Heisenberg equations are:
\begin{eqnarray} \label{He1}
\partial_t\hat{{\alpha}}(r_z,t)&=&\hat{\dot{\alpha}}(r_z,t)
+\frac{C'}{C}\left[1^+(L-r_z)\int_{-\infty}^{L} dr'_z\hat{\dot{\alpha}}(r'_z,t)
+1^+(r_z)\int^{\infty}_{0} dr'_z\hat{\dot{\alpha}}(r'_z,t)\right]
\\ \label{He2}
\partial_t\hat{\dot{\alpha}}(r_z,t)&= &c^2 
\partial_z^2 \hat{{\alpha}}(r_z,t)
+\sum_{j=1}^N
\sqrt{c}\kappa_j\hat \sigma^x_j(t)\delta(r_z+l_j) 
\\ \label{He3}
\partial_t\hat{\sigma}^z_j(t)&= &-
\sqrt{c}\kappa_j[\hat \sigma^y_j(t),\hat{{\alpha}}(-l_j,t)]_+
\\ \label{He4}
\partial_t\hat{\sigma}^\pm_j(t)&= &
{\pm i \omega_j}\hat{\sigma}^\pm_j(t)\pm
i \sqrt{c}\kappa_j[\hat \sigma^z_j(t),\hat{{\alpha}}(-l_j,t)]_+ /2  \, ,
\end{eqnarray}
where $[.,. ]_+$ represents the anticommutator.
For any quantum field operator $\hat O(r_z,t)$, we define its Fourier transform in space and time with various associated notations as:  
\begin{eqnarray}\label{ft}
\hat{O}(r_z,t)= 
\int_{-\infty}^\infty
\frac{d\omega}{2\pi}e^{-i(\omega+i\eta) t}
\hat{O}_{\omega}(r_z)
=\int_{-\infty}^\infty
\frac{dk}{2\pi}e^{ikr_z}\hat{O}_{k}(t)
=
\int_{-\infty}^\infty
\frac{d\omega}{2\pi}
\int_{-\infty}^\infty
\frac{dk}{2\pi}e^{i[kr_z-(\omega+i\eta)t]}\hat{O}_{k,\omega} \, ,
\end{eqnarray}
where we have introduced  the positive and infinitesimal parameter $\eta \rightarrow 0$ to ensure an adiabatic switching essential to study the retarded response to any input wave \cite{Ryder1996Quantum}. 
Using the Heaviside function expressed in the integral form: 
\begin{eqnarray}
1^+(\pm r_z)=\mp \int_{-\infty}^\infty\frac{dk}{2\pi i}\frac{e^{-ikr_z}}{k^\pm} \, , \quad \quad \quad k^\pm =k\pm i\varepsilon \, ,
\end{eqnarray}
with the infinitesimal positive parameter $\varepsilon \rightarrow 0$. 
they become: 
\begin{eqnarray}\label{He5}
\partial_t\hat{{\alpha}}_k(t)&= &
\hat{\dot{\alpha}}_k(t)+
\int_{-\infty}^\infty\frac{dk'}{2\pi}\frac{C'}{C}\left[\frac{1}{k^- k'^+}+
\frac{e^{i(k-k')L}}{k^+ k'^-}\right]\hat{\dot{\alpha}}_{k'}(t)
\\ \label{He6}
\partial_t\hat{\dot{\alpha}}_k(t)&= &-c^2 
k^2\hat{{\alpha}}_k(t)
+\sum_{j=1}^N e^{ikl_j}
\sqrt{c}\kappa_j\hat \sigma^x_j(t) \, .
\end{eqnarray}
We solve the electromagnetic  wave equation for the field operator by decomposing the field operator into an input and scattered component 
\begin{eqnarray}
\hat \alpha_k(t) 
= \hat \alpha^{in}_k(t) + \hat \alpha^{sc}_k(t) \, ,
\quad \quad \quad 
\hat {\dot \alpha}_k(t)
=\hat {\dot \alpha}^{in}_k(t) +
\hat {\dot \alpha}^{sc}_k(t) \, .
\end{eqnarray}
The input field operators have a simple noninteracting dynamics and are decomposed into  creation-annihilation operators as
\begin{eqnarray}\label{inp}
\hat \alpha^{in}_k(t)
=e^{\eta t}\frac{e^{-ic|k|t}\hat a^{in}_k +e^{ic|k|t}\hat a^{in\dagger}_{-k}}{\sqrt{2c|k|}} \, ,
\quad \quad \quad 
\hat {\dot \alpha}^{in}_k(t)
=\sqrt{c|k|}e^{\eta t}\frac{e^{-ic|k|t}\hat a^{in}_k -e^{ic|k|t}\hat a^{in\dagger}_{-k}}{\sqrt{2}i} \, ,
\end{eqnarray}
with the nontrivial commutation relation 
$[\hat  a^{in}_k, \hat a^{in\dagger}_{k'}]=
2\pi \delta(k-k')$. These obey the noninteracting 
part of Eq.(\ref{He5},\ref{He6}).
By eliminating them and using the time Fourier transform (\ref{ft}), we obtain the following equations for the scattering field part:
\begin{eqnarray} \label{sc}
-i(\omega +i\eta)\hat{{\alpha}}^{sc}_{k,\omega}&= &
\hat{\dot{\alpha}}^{sc}_{k,\omega}+
\int_{-\infty}^\infty\frac{dk'}{2\pi}\frac{C'}{C}\left[\frac{1}{k^- k'^+}+
\frac{e^{i(k-k')L}}{k^+ k'^-}\right]\hat{\dot{\alpha}}_{k',\omega}
\\ \label{sc2}
-i(\omega +i\eta)\hat{\dot{\alpha}}^{sc}_{k,\omega}&= &-c^2 
k^2\hat{{\alpha}}^{sc}_{k,\omega}
+\sum_{j=1}^N e^{ikl_j}
\sqrt{ c}\kappa_j\hat \sigma^x_{\omega,j}
\, .
\end{eqnarray}

\subsection{Energy balance equation}

In the absence of dissipation, the energy conservation requires that the net power received within the cavity is equal to the input power radiation minus the output power radiation in a steady state regime. The Noether theorem usually ensures that the variation of the energy in a volume is equal to the energy incoming energy flux  minus the outgoing flux. More specifically, assuming 
a waveguide beginning at $r_z=-L$ before the capacitance and finishing at $r_z=0$ after the capacitance,
we define the energy within the waveguide volume as:
\begin{eqnarray}\label{bal1}
 \left. \hat H(t)\right|^{0}_{-L} 
= 
\int_{-L}^0 dr_z
\frac{1}{2}\left[
\hat{\dot{\alpha}}^2(r_z,t)- c^2 
{\hat{{\alpha}}(r_z,t)\partial_z^2 \hat{{\alpha}}(r_z,t)}\right] +
\frac{\hat{Q}^2_+(-L,t)+\hat{Q}^2_-(0,t)}{2C} 
+\sum_{j=1}^N\frac{ \omega_j}{2}\hat \sigma^z_j(t)-\sqrt{ c}\kappa_j\hat \sigma^x_j(t)\hat{\alpha}(-l_j,t)
\, .
\nonumber \\
\end{eqnarray}
The time variation of this energy is found using the Heisenberg equation (\ref{He1}-\ref{He4}). We obtain 
\begin{eqnarray}
\left.\frac{d \hat H(t)}{dt}\right|^{0}_{-L}=\frac{c^2}{2} 
\left.[ \partial_{r_z} \hat \alpha(r_z,t), \hat {\dot \alpha}(r_z,t)]_+\right|^{r_z=0}_{r_z=-L}=
\hat A^2_+(-L,t) - \hat A^2_-(-L,t)
+\hat A^2_-(0,t) - \hat A^2_+(0,t)
\, .
\end{eqnarray}
The last operators are easy to determine using the results of subsection \ref{C2} in terms of the voltage and current operators and impedance by:
\begin{eqnarray}\label{Apm}
\hat A_\pm (r_z,t)&=&\frac{\hat V(r_z,t) \pm Z \hat I(r_z,t)}{2\sqrt{Z}}=\frac{\pm c^{3/2} \partial_{r_z}\hat \alpha(r_z,t)- c^{1/2} \hat {\dot \alpha}(r_z,t)}{2} \, ,
\quad \quad \quad Z=\sqrt{L'/C'}  \, .
\end{eqnarray}
These are Hermitian and correspond to the quantum analog of forward $+$ and backward  $-$ wave
respectively, while their square are the forward and backward energy fluxes. 
Similarly the operators can be decomposed into an input and scattered fields:   
\begin{eqnarray}
\hat A_{\pm} (r_z,t)= \hat A^{in}_{\pm} (r_z,t)+\hat A^{sc}_{\pm} (r_z,t) \, .
\end{eqnarray}
Using (\ref{inp}), the left and right input wave operator in the spatial and Fourier modes is expressed in terms of creation annihilation field operator $\hat a_k$ as:
\begin{eqnarray}\label{Ainp}
\hat A^{in}_{k,\omega,\pm}&=& 
\frac{\pm ic k\hat \alpha^{in}_{k,\omega} - \hat{\dot \alpha}^{in}_{k,\omega}}{2}=
2\pi \delta(\omega\mp ck)\sqrt{c|\omega|/2}[1^+(\omega) \hat a_{\pm \omega/c}-1^+(-\omega) \hat a^\dagger_{\mp \omega/c}]
\\\label{Ainpr}
\hat A^{in}_{\omega,\pm} (r_z)&=& e^{\pm i\omega r_z/c}i\sqrt{c|\omega|/2}[1^+(\omega) \hat a_{\pm \omega/c}-1^+(-\omega) \hat a^\dagger_{\mp \omega/c}] \, ,
\end{eqnarray}
with the property that $\hat A^{in*}_{\omega,\pm} (r_z)=\hat A^{in}_{-\omega,\pm} (r_z)$. From the absence of scattering coming outside the resonator, 
we deduce  that $\hat A^{sc}_{\omega,+} (-L)=0$
and $\hat A^{sc}_{\omega,-} (0)=0$.

In a steady state regime, the energy balance equation  (\ref{bal1}) after averaging is rewritten as a sum of uncorrelated and correlated (fluorescent) part:
\begin{eqnarray}\label{bal2}
\langle \hat A^{in}_+(-L,t) \rangle^2 - \langle \hat A_-(-L,t)\rangle^2
+\langle \hat A^{in}_-(0,t) \rangle^2- \langle \hat A_+(0,t)\rangle^2 
=-
\langle 
\delta^2 \hat A^{in}_+(-L,t) - \delta^2\hat A_-(-L,t)
+\delta^2 \hat A^{in}_-(0,t) - \delta^2 \hat A_+(0,t) \rangle
\, .
\nonumber \\
\end{eqnarray}
The left hand side corresponds to signal reflection and transmission of the resonator while the right hand side 
appears when some relaxation occurs such as spontaneous emission or fluorescence of the qubits \cite{doi:10.1126/science.1181918}. Thus, the Hamiltonian Eq.(\ref{Hqbit})  models the relaxation rate $T_1$ characterizing  the decay of a qubit in its ground state.

Without the correlation, the balance equation is simplified and, after integration over infinite interval of time and the use of time Fourier transform, reads:
\begin{eqnarray}\label{bal3}
\int_{-\infty}^\infty\frac{d\omega}{2\pi}
|\langle \hat A^{in}_{\omega,+}(-L) \rangle|^2 - 
|\langle \hat A_{\omega,-}(-L) \rangle|^2
+|\langle \hat A^{in}_{\omega,-}(0) \rangle|^2
-|\langle \hat A_{\omega,+}(0) \rangle|^2=0  \, .
\end{eqnarray}

\section{The transmission and reflection formalism}\label{F}

\subsection{Reflection and Transmission: definitions}

Since the transmission and reflection are usually concepts defined for classical circuit, this section aims at establishing their link with quantum field theory. 
Within the classical consideration (for instance any standard microwave engineering handbook like \cite{Pozar:882338}), one often introduces input and output for forward and backward waves, 
 $a_1=\langle \hat A^{in}_{\omega,+} (-L)\rangle$ 
$a_2=\langle \hat A^{in}_{\omega,-} (0)\rangle$ 
$b_1=\langle \hat A_{\omega,-} (-L)\rangle$ 
$b_2=\langle \hat A_{\omega,+} (0)\rangle$
where $a_{1,2}$ ($b_{1,2}$) represents the input (output), while the index 1(2) corresponds to forward (backward) wave.

The transmission and reflection are defined from the matrix relation:
\begin{eqnarray}\label{Smatrix}
{\displaystyle {\begin{pmatrix}b_{1}\\b_{2}\end{pmatrix}}
={\begin{pmatrix}\langle \hat A_{\omega,-} (-L)\rangle \\  \langle \hat A_{\omega,+} (0)\rangle\end{pmatrix}}
=
{\begin{pmatrix}{ S}_{11}&{ S}_{12}\\{ S}_{21}&{ S}_{22}\end{pmatrix}}{\begin{pmatrix}a_{1}\\a_{2}\end{pmatrix}}
=
{\begin{pmatrix}{ S}_{11}&{ S}_{12}\\{ S}_{21}&{ S}_{22}\end{pmatrix}}{\begin{pmatrix}\langle \hat A^{in}_{\omega,+} (0)\rangle \\ \langle \hat A^{in}_{\omega,-} (0)\rangle \end{pmatrix}}\,}   \,  .
\end{eqnarray}
In what follows we consider only the case where 
$\langle \hat A^{in}_{\omega,-} (0)\rangle=0$.
In absence of scattering or fluorescence, any signal frequency remains unchanged and, thus,  satisfies  the unitary property $|{S}_{21}(\omega)|^2+|{S}_{11}(\omega)|^2=1$. Otherwise, interactions may transfer the input signal into another frequency channel or the fluorescence broadens the signal into a 
continuum of frequencies which should be analyzed using correlations. 
Note that the spectral function  of  the voltage fluctuations is 
another quantity also used for analyzing the transmission \cite{GU20171}.

\subsection{Scattered wave operators}

As quantities of interest for the transmission and reflection, we use  the  Heisenberg equations for the  wave operators. Using the definition (\ref{Apm}) and applying the Fourier transform to Eqs.(\ref{He5}) and (\ref{He6}), we find
\begin{eqnarray}\label{eqA}
[\omega+i\eta \mp ck]\hat A^{sc}_{k,\omega,\pm}=
\mp \frac{C'\sqrt{c}}{2C}\int_{-\infty}^\infty\frac{dk'}{2\pi }\left(\frac{1}{k'^+} +\frac{e^{i(k-k')L}}{k'^-}\right)\hat{\dot \alpha}_{k',\omega} -
\frac{ic}{2}\sum_{j=1}^N
e^{ik l_j}\kappa_j \hat\sigma^x_{\omega,j} \,  .
\end{eqnarray}
Multiplying by the propagator both sides, we determine after a contour integration over $k'$ the non trivial scattered waves:
\begin{eqnarray}\label{Asc+}
\hat A^{sc}_{\omega,+}(0^+)=
i \frac{C'}{2C}\int_{-\infty}^\infty\frac{dk'}{2\pi \sqrt{c} }
\left(\frac{1}{k'^+} +\frac{e^{i(\omega/c-k')L}}{k'^-}\right)\hat{\dot \alpha}_{k',\omega} -
\sum_{j=1}^N
e^{i\omega l_j/c}\frac{\kappa_j}{2} \hat\sigma^x_{\omega,j}
\\\label{Asc-}
\hat A^{sc}_{\omega,-}(-L^-)=-
i \frac{C'}{2C}\int_{-\infty}^\infty\frac{dk'}{2\pi \sqrt{c} }
\left(\frac{e^{i\omega L/c}}{k'^+} +\frac{e^{-ik' L}}{k'^-}\right)
\hat{\dot \alpha}_{k',\omega} -
\sum_{j=1}^N
e^{i\omega (L-l_j)/c}\frac{\kappa_j}{2}\hat\sigma^x_{\omega,j}
\,  .
\end{eqnarray}
The superscripts in $0^+$ and $L^-$  are here to indicate that the scattered wave functions are defined outside the cavity and ensure the convergence of the integration. They will be omitted subsequently. 
Taking also into account that 
$\hat{\dot \alpha}^{sc}_{k,\omega}=-(\hat A^{sc}_{k,\omega,+}+ \hat A^{sc}_{k,\omega,-})/\sqrt{c}$, we deduce from Eq.(\ref{eqA}) an integral  equation for $\hat{\dot \alpha}^{sc}_{k,\omega} $ only.  
We integrate the latter as:
\begin{eqnarray}
\hat {\cal I}_\pm &=& \int_{-\infty}^\infty\frac{dk}{2\pi }
\left(\frac{1}{k^+} \pm \frac{ e^{-ik L}}{k^-}\right)\hat {\dot \alpha}^{sc}_{k,\omega} \, ,
\end{eqnarray}
to deduce:
\begin{eqnarray}
\hat {\cal I}_\pm 
&=&\int_{-\infty}^\infty\frac{dk}{2\pi}\frac{1/k^+ \pm e^{-ikL}/k^-}{(\omega + i\eta)^2-c^2k^2}
\biggl[\int_{-\infty}^\infty\frac{dk'}{2\pi}\frac{kC'}{C}\left(\frac{1}{k'^+}+
\frac{e^{i(k-k')L}}{k'^-}\right)
\hat {\dot \alpha}_{k',\omega}
+ \frac{i\omega}{\sqrt{c}} \sum_{j=1}^N
e^{ikl_j} \kappa_j \hat\sigma^x_{\omega,j}\biggr] 
\,  .
\nonumber \\
\end{eqnarray}
Using  the integral results for any real parameter $a \geq 0$:
\begin{eqnarray}
\int_{-\infty}^\infty\frac{dk}{2\pi}
\frac{\cos(ka)}{(\omega + i\eta)^2-c^2k^2}
&=&\frac{\exp(ia\omega/c)}{2i(\omega +i\eta)c} \, ,
\quad\quad  \quad 
\int_{-\infty}^\infty\frac{dk}{2\pi (k\pm i\varepsilon)}
\frac{\exp(\pm ik a)}{(\omega + i\eta)^2-c^2k^2}
= \pm \frac{\exp(i a\omega/c)}{2i(\omega +i\eta)^2}  \, , 
\end{eqnarray}
we obtain an algebraic linear equation for $\hat {\cal I}_\pm$. Combining its solution with the input component, we find the result:  
\begin{eqnarray}\label{dota}
&&\int_{-\infty}^\infty\frac{dk}{2\pi}\left(\frac{1}{k^+}\pm \frac{e^{-ikL}}{k^-}\right)\hat {\dot \alpha}_{k,\omega}
=\frac{2i\omega\int_{-\infty}^\infty\frac{dk}{2\pi}\left(\frac{1}{k^+}\pm\frac{e^{-ikL}}{k^-}\right)\hat {\dot \alpha}^{in}_{k,\omega}+i\sqrt{c}\sum_{j=1}^N \kappa_j\hat \sigma^x_{\omega,j}
[e^{i\omega l_j/c}\mp e^{i\omega (L-l_j)/c}]}{2i\omega-C'c(1\pm e^{i\omega L/c})/C}  \,  .
\end{eqnarray}
After substitution of a linear combination of Eq.(\ref{dota}) into the left hand side of (\ref{Asc+}) and (\ref{Asc-}), we obtain finally for the scattered wave operators:
\begin{eqnarray}\label{res}
\hat A^{sc}_{\omega,+}(0)&=&\sum_{\pm} 
\frac{C'(1\pm e^{i\omega L/c})}{4C}
\frac{2(1\pm e^{-i\omega L/c})(\hat A^{in}_{\omega,+}(0) \pm e^{i\omega L/c}\hat A^{in}_{\omega,-}(0))
-\sum_{j=1}^N \kappa_j\hat \sigma^x_{\omega,j}
[e^{i\omega l_j/c}\mp e^{i\omega (L-l_j)/c}]}{2i\omega-C'c(1 \pm e^{i\omega L/c})/C}
\nonumber \\
&-&\sum_{j=1}^N
e^{i\omega l_j/c}\frac{\kappa_j}{2c} \hat\sigma^x_{\omega,j}
\\\label{res2}
\hat A^{sc}_{\omega,-}(-L)&=&\sum_{\pm}\pm 
\frac{C'(1\pm e^{i\omega L/c})}{4C}
\frac{2(1\pm e^{-i\omega L/c})
(\hat A^{in}_{\omega,+}(0) \pm e^{i\omega L/c}\hat A^{in}_{\omega,-}(0))
-\sum_{j=1}^N \kappa_j\hat \sigma^x_{\omega,j}
[e^{i\omega l_j/c}\mp e^{i\omega (L-l_j)/c}]}{2i\omega-C'c(1\pm  e^{i\omega L/c})/C}
\nonumber \\ &-&
\sum_{j=1}^N
e^{i\omega (L-l_j)/c}\frac{\kappa_j}{2c} \hat\sigma^x_{\omega,j}  \,  .
\end{eqnarray}

\subsection{Transmission and reflection spectrum of the resonator}

In the absence of any qubits, $\kappa_j=0$. In this case, the transmission and reflection is found using the formula (\ref{Smatrix}):
\begin{eqnarray}\label{cav}
{S}_{21}(\omega)&=&
\frac{(2i\omega C/cC')^2}
{\left(1-2i\omega C/cC'\right)^2-\exp(2iL\omega/c)} 
\\ \label{cav2}
{ S}_{11}(\omega)&=&
=\frac{2i[2\omega C/cC'\cos(L\omega/c)+\sin(L\omega/c)]}
{\left(1-2i\omega C/cC'\right)^2-\exp(2iL\omega/c)} \, ,
\end{eqnarray}
that fulfills $|{ S}_{21}(\omega)|^2+|{ S}_{11}(\omega)|^2=1$. The poles of Eqs.(\ref{cav}) and (\ref{cav2}) are the same and correspond to the resonance frequencies $\omega_n$ and the half widths $\gamma_n$ or damping. These  are given in terms of the Lambert special function $W(k,x)$ as:
\begin{eqnarray}
\omega_n -i \gamma_n/2=i\frac{C'c}{2C}
-i\frac{c}{L}W\left(\frac{1- (-1)^n}{4}+\frac{n}{2}, \frac{C'L}{2C}(-1)^n\exp({C'L}/{2C})\right)  \,  .
\end{eqnarray}
Figure \ref{Sm} depicts these reflection and transmission response functions in the frequency ranges for the first three resonances. Although
the broadening has been taken larger than usual, transmission is close to unity around the resonance while the  reflection is dominant outside the resonance. 
Figure \ref{spectrum} represents the spectrum as a function of 
${C}/{C'L}$. 
For high quality factor ($C\rightarrow 0$), we recover the steady resonance $\omega_{n,0}=n\pi c/L$. For small capacitance $C \leq 0.04 C'L$,  we deduce the correction and a non-zero half width: 
\begin{eqnarray}
\omega_n -\omega_{n,0}\cong -\frac{2C}{C'L}\omega_{n,0} \, , \quad \quad \quad 
\gamma_n \cong \frac{4c}{L}\left(\frac{n\pi C}{LC'}\right)^2 \, ,
\end{eqnarray}
in agreement with the reference \cite{doi:10.1063/1.3010859}. 
The quality factor and the peak intensity are:
\begin{eqnarray}
{\cal Q}_n=\frac{\omega_n}{\gamma_n}\cong \frac{C'^2L^2}{2n\pi C^2} = \frac{C'L}{2\omega_{n,0} Z C^2}
\,  .
\end{eqnarray}
In comparison, in the equivalent circuit model proposed in \cite{doi:10.1063/1.3010859}, ones find 
${\cal Q}_n=\frac{C'L}{2\omega_n R_L C^2}$  where $R_L$ is defined as the charge resistance outside the resonator. Since this resistance is generally adjusted to correspond to the impedance $Z$, we may conclude that both results are equivalent. However, we should point out that the model developed in \cite{doi:10.1063/1.3010859} addresses one resonance only in the limit of weak capacitance $C \ll C'L$.  

On the other hand, our expression does not account for the saturation effect at high quality factor \cite{doi:10.1063/1.3010859}. The latter is likely due to the additional currents at the dielectric interfaces caused by the polarisation or some dissipation effect around the capacitance $C$ due to the finite charge mobility.  

\begin{figure}
\begin{center}
 \includegraphics[width=13cm]{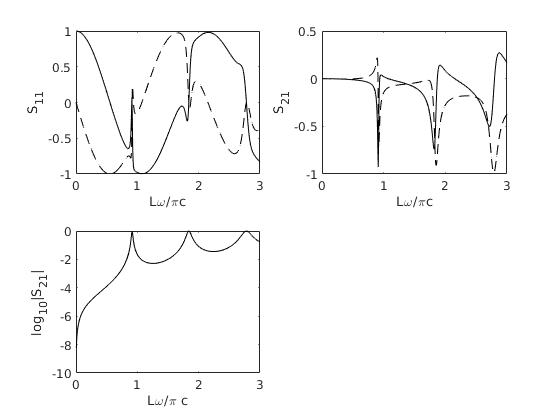}
\end{center}
\caption{\label{Sm} Reflection and Transmission quadrature components for a cavity with $C/(C'L)=20$. The solid curve is for the real part and the dashed curve is for the imaginary part (see Eq.(\ref{cav})). Lower: absolute value of the transmission in logarithm scale.}
\end{figure}

\begin{figure}
\begin{center}
 \includegraphics[width=13cm]{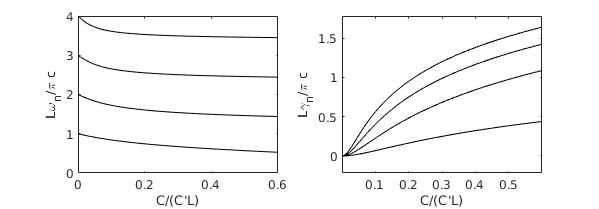}
\end{center}
\caption{\label{spectrum} Representation of the spectrum $\omega_n$ and the damping $\gamma_n$ as a function of the capacitance at the cavity ends and $n=1,2,3,4$. For both graphics, $n$ labels each curve from its lowest to its highest in ascending order.}
\end{figure}

\section{The ``artificial atom'' without a cavity}
\label{G}

\subsection{Quantum stochastic and Master equation derivation}

In this appendix we utilize our framework to reproduce the results for reflection and transmission of a "artificial atom" previously discussed in  \cite{doi:10.1126/science.1181918,Brehm2021}. The interest of the approach here is the use  of a stochastic operator technique \cite{gardiner00} that will be a useful tool  to address the cavity case. 
We focus on the case without cavity ($C \rightarrow\infty$) and derive a master equation using the stochastic formalism.

Using the decomposition into input and scattering component, the Hamiltonian (\ref{Hqbit}) in spatial Fourier components form becomes:
\begin{eqnarray}\label{Hart}
\hat H &=&  
\int_{-\infty}^\infty \frac{dk}{2\pi}
c|k| \hat a^{\dagger in}_{k}(t)\hat a^{in}_{k}(t)+
\int dr_z
\frac{1}{2}\left[
\hat{\dot{\alpha}}^{sc^{ 2}}(r_z,t)- c^2 
{\hat{{\alpha}}^{sc}(r_z,t)\partial_z^2 \hat{{\alpha}}^{sc}(r_z,t)}\right] 
\nonumber \\
&+&
\sum_{j=1}^N\frac{ \omega_j}{2}\hat \sigma^z_j(t)
-\sqrt{ c}\kappa_j\hat \sigma^x_j(t) \left(\int_{-\infty}^\infty \frac{dk}{2\pi}e^{ikl_j}\frac{\hat a^{in}_k(t) +\hat a^{\dagger in}_{-k}(t)}{\sqrt{2c|k|}}
+\hat{{\alpha}}^{sc}(r_z,t)\right) \,  .
\end{eqnarray} 
Here we define the time dependent free operator 
$\hat a_k(t)=e^{-ic|k|t+\eta t}\hat a_k$.
Solving
Eqs.(\ref{sc}) and (\ref{sc2}) for $C\rightarrow  \infty$,
we find the scattered radiation amplitude from the qubits:
\begin{eqnarray}\label{sc1}
\hat \alpha^{sc}_{\omega}(r_z)=
\int_{-\infty}^\infty\frac{dk}{2\pi}
e^{ikr_z}
\hat \alpha^{sc}_{k,\omega}
&= & -
\int_{-\infty}^\infty\frac{dk}{2\pi}
\frac{\sum_{j=1}^N
e^{ik(r_z+l_j)}\sqrt{c}\kappa_j\hat \sigma^x_{\omega,j}}
{(\omega + i\eta)^2-c^2k^2}
=\sum_{j=1}^N \frac{i\exp( i|r_z +l_j|\omega/c)\kappa_{j}\hat \sigma^x_{\omega,j}}{2\omega\sqrt{c}} \,  .
\end{eqnarray}
For a nearly monochromatic input field  involving frequencies close and centered around $\omega >0$, this last expression can be made Markovian in the time domain.  As a result, we obtain the decomposition in positive and negative frequencies for the raising and lowering operators respectively: 
\begin{eqnarray}\label{sc3}
\hat \alpha^{sc}(r_z,t)\cong
\sum_\pm \sum_{j=1}^N \frac{\pm i\exp(\pm i|r_z+ l_j|\omega/c)\kappa_{j}\hat \sigma^\mp_{j}(t)}{2\omega\sqrt{c}} \, ,
\quad \quad \quad
\hat{\dot \alpha}^{sc}(r_z,t)\cong
\sum_\pm \sum_{j=1}^N \frac{\exp(\pm i|r_z+ l_j|\omega/c)\kappa_{j}\hat \sigma^\mp_{j}(t)}{2\sqrt{c}} \,  .
\end{eqnarray}
This procedure amounts to neglecting the Lamb shift effect which appears to have an infrared divergence \cite{gardiner00} that does not concern us here.
Eliminating the scattering component (\ref{sc3}), we  express the Hamiltonian (\ref{Hart}) only in terms of the input field components in the Schr\"odinger picture:
\begin{eqnarray}
\hat H &=&  
\int_{-\infty}^\infty \frac{dk}{2\pi} 
c|k| \hat a^{\dagger in}_{k}\hat a^{in}_{k}+\sum_{j=1}^N\frac{ \omega_j}{2}\hat \sigma^z_j-\kappa_j \int_{-\infty}^\infty \frac{dk}{2\pi}e^{ikl_j}\frac{\hat \sigma^+_j\hat a^{in}_k +\hat a^{\dagger in}_{-k}\hat \sigma^-_j}{\sqrt{2|k|}}
+ \sum_{j,j'=1}^N\frac{\kappa'_{jj'}}{2}\hat \sigma^+_j \hat \sigma^-_{j'} 
\, ,
\nonumber \\
\end{eqnarray}
where
$\kappa'_{jj'}=\kappa_j \kappa_{j'}\sin(|l_j-l_{j'}|\omega /c)/\omega$
are the interaction constant responsible for mutual scattering between the qubits \cite{GU20171,Brehm2021}.
For field frequencies $\omega$ close to the qubit frequencies, we can do  the rotating wave approximation (RWA) by making the unitary transformation 
$\hat a^{in}_k \rightarrow e^{- i\omega t} \hat a^{in}_k$ and  $\hat \sigma^\pm_i \rightarrow e^{\pm i\omega t} \hat \sigma^\pm_i$. The input field in the interaction term can be  rewritten approximately as:
\begin{eqnarray}\label{inp3}
\int_{-\infty}^\infty \frac{dk}{2\pi}e^{ikl_j}
\frac{e^{-ic|k|t} \hat a^{in}_k}{\sqrt{2|k|}}  \cong \sum_ \pm \frac{e^{\pm i \omega l_j/c} }{\sqrt{2\omega}} 
\hat a_{\pm,t} \, ,
\end{eqnarray}
where we define the forward $(+)$ and backward $(-)$ wave time dependent operators 
$\hat a_{\pm,t}=
\int_{-\infty}^\infty \frac{d\omega'}{2\pi}
e^{-i\omega' t} \hat a^{in}_{\pm \omega'/c}$. The latter can be viewed as  stochastic quantum variables expressing the influence of the vacuum noise on the artificial atoms. As long as we operate with frequency field around $\omega$, they are held in a good approximation, the commutation relation $[\hat a_{\pm,t}, \hat a^\dagger_{\pm,t'}]=\delta(t-t')$ and $[\hat a_{\pm,t}, \hat a^\dagger_{\mp,t'}]=0$ and therefore time-dependent backward and forward quantum waves are independent. In the new base of operators using Eq.(\ref{inp3}), we find the time-dependent stochastic Hamiltonian:
\begin{eqnarray}
\hat H'(t)=\sum_{j=1}^N-\frac{ \delta \omega_j}{2}\hat \sigma^z_j
+ \sum_{j,j'=1}^N\frac{\kappa'_{jj'}}{2}\hat \sigma^+_j \hat \sigma^-_{j'} +\sum_{j=1}^N 
\frac{\kappa_j}{\sqrt{2\omega}} 
\left[(e^{i\omega l_j/c} \hat a_{+,t}+ e^{-i\omega l_j/c} \hat a_{-,t})e^{i\omega t}\hat \sigma^+_j+ c.c. \right] \, ,
\end{eqnarray}
where we define the detuning $\delta \omega_j=
\omega-\omega_j$. 
With a forward coherent input field of frequency $\omega$ and no backward input, one has 
$\langle \hat A^{in}_{+}(r_z,t)\rangle =\sqrt{Z} \langle \hat I^{in}(r_z,t)\rangle=\sqrt{Z}I_0\cos[\omega(r_z/c-t)]$. For a well defined frequency  
the forward quantum stochastic field is decomposed into 
a coherent and an quantum radiation field contribution: 
$\hat a_{+,t} = -i \sqrt{2/c\omega}\langle \hat A^{in}_{+}(0,t)\rangle + \delta \hat a_{+,t}$.
Using the Rabi frequency notation
$\Omega_{\pm,j}=\pm \frac{\kappa_j \sqrt{Z}I_0}{i\omega}$, the Hamiltonian becomes:
\begin{eqnarray}
 \hat H'(t)&= &\hat H_S +\hat H_I(t)  \, ,
 \quad \quad \quad
 \hat H_S= \sum_{j=1}^N-\frac{ \delta \omega_j}{2}\hat \sigma^z_j-
\sum_{\pm} \frac{ \Omega_{\pm,\omega,j}}{2}\hat \sigma^\pm_j
+ \sum_{j,j'=1}^N\frac{\kappa'_{jj'}}{2}\hat \sigma^+_j \hat \sigma^-_{j'}
\\
\hat H_I(t)& =& \sum_{j=1}^N 
\frac{\kappa_j}{\sqrt{2\omega}} 
\left[(e^{i\omega l_j/c} \delta \hat a_{+,t}+ e^{-i\omega l_j/c} \delta \hat a_{-,t})e^{i\omega t}\hat \sigma^+_j+ c.c. \right] \,  .
\end{eqnarray}
The last term is responsible for the relaxation of the excited qubit state into the ground state by spontaneous photon emission in vacuum.  
We  eliminate the quantum radiation field component using the master equation for the density matrix $\hat \rho(t)$ developed up to the second order in the interaction:
\begin{eqnarray}
\partial_t \hat \rho(t) +i[\hat H_S,\hat \rho(t)]=
-\int_0^t d\tau {\rm Tr}_B\left\{ 
[\hat H_I(t),[\hat H_I(t-\tau),\hat \rho_B(0) \hat \rho(t-\tau)]]\right\}  \,  .
\end{eqnarray}
In the absence of thermal field noise, we use the vacuum state $\hat \rho_B(0)=|0\rangle_B \langle 0|$ for the radiation. We make the Markovian approximation so that the effective equation is:
\begin{eqnarray}\label{master}
\partial_t \hat \rho(t) +i[\hat H_S,\hat \rho(t)]=
-\int_0^\infty d\tau {\rm Tr}_B\left\{ 
[\hat H_I(t),[\hat H_I(t-\tau),|0\rangle_B \langle 0| \hat \rho(t)]]\right\} \,  .
\end{eqnarray}
Determining the matrix element over the vacuum state in the right hand side of Eq.(\ref{master}) and carry out the integration over $\tau$ (as done in detail for instance in \cite{gardiner00}), we obtain the master equation \cite{GU20171}:
\begin{eqnarray}\label{Master}
&&\partial_t \hat \rho(t) -i[\sum_{j=1}^N\frac{ \delta \omega_j}{2}\hat \sigma^z_j+
\sum_{\pm} \frac{ \Omega_{\pm,\omega,j}}{2}\hat \sigma^\pm_j+ \sum_{j,j'=1}^N\frac{\kappa'_{jj'}}{2}\hat \sigma^+_j \hat \sigma^-_{j'},\hat \rho(t)]
=-\sum_{j,j'=1}^N\frac{\kappa_{jj'}}{2}\left[[\hat \sigma^+_j \hat \sigma^-_{j'},\hat \rho(t)]_+
-\hat \sigma_j^- 2\hat \rho(t) \hat \sigma_{j'}^+\right] \, ,
\nonumber \\
\end{eqnarray}
where  $\kappa_{jj'}=\kappa_j \kappa_{j'}\cos[(l_j-l_{j'})\omega /c]/\omega$ are the parameters responsible for the relaxation.
This equation predicts the possibility of superradiance that has been experimentally analyzed in \cite{Brehm2021}. It however does not predict the dephasing due to the Josephson junctions themselves which needs to be considered phenomenologically.

\subsection{One qubit without a cavity} 

Let us solve the master equation Eq.(\ref{Master}) for the case of one qubit with detuning $\delta \omega_1$, $l_1=0$ and decay rate $\Gamma_1 \cong \kappa_1^2/\omega$ and show the equivalence of the approach with \cite{doi:10.1126/science.1181918}. 
Using (\ref{Asc+}) and (\ref{Asc-}) for $C\rightarrow \infty$, and defining the average for any operator as $\langle \hat O\rangle={\rm Tr}(\hat \rho(t)  \hat O)$,
we assume the steady state  solution:
\begin{eqnarray}
\langle \hat \sigma_1^\pm(t)\rangle = 
\sigma^\pm_\omega \, , \quad\quad \quad 
\langle \hat \sigma_1^z(t)\rangle =
\sigma^z_\omega \, ,
\quad \quad \quad
\langle \hat A^{in}_+(0,t)\rangle
=  \frac{i\omega \Omega_-}{2 \kappa_1} e^{- i\omega t}+
c.c. \, ,
\quad \quad \quad
\langle \hat A^{sc}_{\pm}(0,t)\rangle
=-\frac{\kappa_1}{2} \sum_{\pm}
\sigma^\pm_\omega e^{\pm i \omega t} \,  .
\nonumber \\
\end{eqnarray}
Taking these various averages in Eq.(\ref{Master}), we obtain for the steady state equations:
\begin{eqnarray}
i(\Omega_{+}\sigma^-_{\omega}-\Omega_{-}\sigma^+_{\omega})+\Gamma_1 (1+ \sigma^z_{\omega})=0 \, ,
\quad \quad \quad 
( \delta \omega_1\mp i\Gamma'_1 )\sigma^\pm_{\omega}
=\frac{\sigma^z_{\omega}}{2}\Omega_{\pm} \, ,
\end{eqnarray}
where for completeness we have added phenomenologically the dephasing rate $\Gamma_{1,\phi}$ in 
$\Gamma'_1 =\Gamma_1/2 + \Gamma_{1,\phi}$. 
We find the solution:
\begin{eqnarray}
\sigma^z_{\omega}&=&-\frac{(\delta \omega_1^2 +\Gamma_1^{'2})\Gamma_1}{(\delta \omega^2_1 + \Gamma_1^{'2})\Gamma_1 +|\Omega_{\pm}|^2\Gamma'_1} \, ,
\quad \quad \quad
\sigma^\pm_{\omega}=-\frac{1}{2}
\frac{(\delta \omega_1 \pm i\Gamma'_1)\Gamma_1 \Omega_\pm}{(\delta \omega^2_1 + \Gamma_1^{'2})\Gamma_1 +|\Omega_{\pm}|^2\Gamma'_1} \,  .
\end{eqnarray}
For transmission and reflection, we obtain :
\begin{eqnarray}
{S}_{11} =\frac{\langle \hat A^{sc}_{\omega,-}(0)\rangle}{\langle \hat A^{in}_{\omega,+}(0)\rangle}
=-\frac{\Gamma_1}{2\Gamma'_1}\frac{1+i\delta \omega_1/\Gamma'_1}{1+(\delta \omega^2_1/\Gamma'_1)^2 + |\Omega_{\pm}|^2/(\Gamma_1 \Gamma'_1)} \, ,
\quad \quad \quad 
{S}_{21} 
=\frac{\langle \hat A_{\omega,+}(0)\rangle}{\langle \hat A^{in}_{\omega,+}(0)\rangle }
= 1+ { S}_{11}  \,  .
\end{eqnarray}
The unitary relation $
|{ S}_{11}|^2=1 -|{ S}_{21}|^2$ is satisfied when $\Omega_\pm=0$ and $\Gamma_{1,\phi}=0$, otherwise it is  violated  by quantum wave fluctuations generated from fluorescence. In comparison with \cite{doi:10.1126/science.1181918}, the flux parameter $\phi_p$ defined from the relation
$\hbar |\Omega_{\pm}|=\phi_p I_0$ satisfied the identical relation $\Gamma_1=\omega \phi_p^2/(\hbar Z)$. 

The hereafter master equation approach developed here does not take into account the relaxation to radiation modes  other than the TEM mode in the waveguide. Using the experiment parameter value in \cite{doi:10.1126/science.1181918}, we can predict the relaxation rate for this mode and compare with the measured one $\Gamma$.
From the input power is given by ${\cal P}=Z I_0^2/2$, 
we can indeed find the relation $\Gamma_1=|\Omega_{\pm}|^2\hbar \omega_1/(2 P)$.  We estimate $\Gamma_1/ \Gamma \sim 0.15$ within the order of magnitude of unity but with a discrepancy suggesting that other relaxation mechanisms should be taken into account for the specific qubit device used. 

\section{The quantum master equation in a cavity}\label{H}

\subsection{Derivation}

The qubit model derived in Appendix \ref{E} is still to complex to be solved exactly as it involves a set of continuous modes interacting both with a cavity and with the qubits. 
Moreover, for the purpose of single photon detection, we must focus on a frequency range close to a cavity mode frequency which is itself close to the qubit frequency.  

In this section, we use the method  of the previous section to derive a master equation, with the difference that we need to isolate the cavity mode from the other external fields mode starting from 
the expression Eq.(\ref{Hqbit}). The advantage of using a cavity is the enhancement of the coupling of the qubits with the cavity mode field known as  
the Purcell effect.

We start by separating the electromagnetic field between the photon cavity mode and the other modes according to  a discrete and a continuous field :
\begin{eqnarray}
\hat \alpha (r_z,t)&=& 
\hat \alpha_d (r_z,t)+\hat \alpha_c (r_z,t)  \, ,
\quad \quad \quad \quad
\hat {\dot \alpha} (r_z,t)=
\hat {\dot \alpha}_d (r_z,t)+\hat {\dot \alpha}_c (r_z,t) 
\,  .
\end{eqnarray}
The discrete field operates on the interval $r_z \in [-L,0]$ can be decomposed into 
discrete cavity modes of frequency $\omega_{n,0}=\pi nc/L$ as:
\begin{eqnarray}\label{dec}
\hat \alpha_d (r_z,t)=\sum_{n=0}^\infty
\sqrt{\frac{2-\delta_{n,0}}{L}}\cos\left(\frac{\pi n r_z}{L}\right)
\frac{\hat a_n(t) + \hat a^\dagger_n(t)}{\sqrt{2\omega_{n,0}}}
\, , \quad \quad
\dot {\hat \alpha}_d (r_z,t)=\sum_{n=0}^\infty
\sqrt{\frac{2-\delta_{n,0}}{L}}\cos\left(\frac{\pi n r_z}{L}\right)
\sqrt{\omega_{n,0}}
\frac{\hat a_n(t) - \hat a^\dagger_n(t)}{\sqrt{2}i} 
\,  .
\nonumber \\
\end{eqnarray}
Using the projector formalism to the full quantum operator, we obtain an explicit expression for each field element of the decomposition:
\begin{eqnarray}\label{proj}
{\cal P}\hat \alpha (r_z,t)&=& \hat \alpha_d (r_z,t)=
\sum_{n=0}^\infty \frac{2-\delta_{n,0}}{L}\cos\left(\frac{n\pi r_z}{L}\right)\int_{-L}^0dr'_z\cos\left(\frac{n\pi r'_z}{L}\right)\hat \alpha (r_z,t)
\, , \quad \quad \quad 
{\cal Q}=1 -{\cal P}
\\
{\cal Q}\hat \alpha (r_z,t)&=& \hat \alpha_c (r_z,t) 
\, , \quad \quad \quad 
{\cal P}\hat {\dot \alpha} (r_z,t)= \hat  {\dot \alpha}_d (r_z,t)   \, , \quad \quad \quad
{\cal Q}\hat {\dot \alpha} (r_z,t)= \hat {\dot \alpha}_c (r_z,t)
\,  .
\end{eqnarray}
Inverting the relations Eqs.(\ref{dec}) and combining with Eq.(\ref{proj}), the annihilation operator of the cavity mode is expressed  to the full quantum field operator through:
\begin{eqnarray}
\hat a_n(t)&=& 
\sqrt{\frac{2-\delta_{n,0}}{L}}\int_{-L}^0dr'_z\cos\left(\frac{n\pi r'_z}{L}\right)\frac{i\hat{ \dot{\alpha}} (r_z,t)+\omega_{n,0} \hat \alpha (r_z,t)}
{\sqrt{2\omega_{n,0}}} \,  .
\end{eqnarray}
Using the notation $\hat a_{n,\omega}$ in the time Fourier transform, 
this relation combined together with the Heisenberg equations (\ref{sc}, \ref{sc2}) allows to deduce an equation for this discrete set of cavity operators.
We find for $n\not=0$:
\begin{eqnarray}
(\omega+i\eta-\omega_{n,0}) \hat a_{n,\omega} &= &
\frac{1}{\sqrt{L\omega_{n,0}}}\left(c^2
\int_{-\infty}^\infty\frac{dk}{2\pi i}k
(1-(-1)^n e^{-ikL})\hat \alpha_{k,\omega}
-\sqrt{c}\sum_{j=1}^N \kappa_j \cos(\omega_{n,0}l_j/c)\hat \sigma_{\omega,j}^x\right)  \,  .
\end{eqnarray}
To work out the right hand side of this last equation, we use again the Heisenberg equations to derive the equality:  
\begin{eqnarray}
\int_{-\infty}^\infty\frac{dk}{2\pi i}k
(1-(-1)^n e^{-ikL})\hat \alpha_{k,\omega}
=\frac{i\omega}{c^2}
\int_{-\infty}^\infty\frac{dk}{2\pi i}
\left(\frac{1}{k^+} -\frac{(-1)^n e^{-ik L}}{k^-}\right)
\hat  {\dot \alpha}_{k,\omega} \,  .
\end{eqnarray}
By combining this last relation with Eq.(\ref{dota}),
we find the exact result: 
\begin{eqnarray}
\hat a_{n,\omega}&=&
\sqrt{c}\sum_j \kappa_j \frac{\cos(\frac{\omega l_j}{c})-\cos(\frac{\omega_{n,0} l_j}{c})}{\omega -\omega_{n,0}}
\hat \sigma^x_{\omega,j}+
\frac{1}{\omega-\omega_{n,0}}
\frac{1}{\sqrt{L\omega_{n,0}}}
\frac{1-(-1)^n e^{-i\omega L/c}}
{2i(\omega+i0)- C'c(1-(-1)^n e^{ikL})/C}
\nonumber \\
&\times&
\biggl[-2i\omega(\hat A_{\omega,-}^{in}(0)+(-1)^n\hat A_{\omega,+}^{in}(-L))
- \sum_{j=1}^N \left(\cos(\omega l_j/c)(i\omega-\frac{C'c}{C})+\omega\sin(\omega l_j/c)\right)\sqrt{c}\kappa_j\hat \sigma^x_{\omega,j}
\,  .
\nonumber \\
\end{eqnarray}
An inspection of this formula shows that the poles is located 
at $\omega_n-i\frac{\gamma_n}{2}$ with $\omega_n= \omega_{n,0}+\Delta \omega_n$ and that the pole at $\omega_{n,0}$ is apparent.
Close to a resonance for $n\not=0$, the formula for the operator is approximated as:
\begin{eqnarray}
\hat a_{n,\omega} &\cong&
\frac{i\sqrt{\frac{\gamma_n}{4c\omega_n}}(\hat A_{\omega,-}^{in}(0)+(-1)^n\hat A_{\omega,+}^{in}(-L)) - \sum_{j=1}^N \sqrt{1/\pi n}\cos(\omega_n l_j/c)\sqrt{c}\kappa_j\hat \sigma^-_{\omega,j} }
{\omega-(\omega_n-i\gamma_n/2)}
\\ \label{denom}
&\cong&
\frac{-\sqrt{\gamma_n/2}(\hat a_{\omega/c}+\hat a_{-\omega/c}) - \sum_{j=1}^N \sqrt{1/\pi n}\cos(n\pi l_j/Lc)\sqrt{c}\kappa_j\hat \sigma^-_{\omega,j} }
{\omega-(\omega_n-i\gamma_n/2)} \, ,
\end{eqnarray}
where we used the relation (\ref{Ainpr}) approximated for frequencies close to $\omega_{n}$. 
Similarly to the appendix (\ref{G}), we define the time dependent stochastic operators 
$\hat a_{\pm,t}=\int_{-\infty}^\infty \exp(-i\omega t) \hat a_{\pm \omega/c} d\omega/2\pi$ valid for frequencies around the resonance frequency $\omega_n$ and identified as the vacuum noise operators of ingoing wave from the left $+$ and right $-$ of the waveguide.
Using this definition and reversing the denominator  into the left hand side, the Eq.(\ref{denom}) can be rewritten  back in  the time domain to become
a quantum stochastic Langevin-like equation: 
\begin{eqnarray}\label{lang}
[i\partial_t-(\omega_{n}-i\frac{\gamma_n}{2})] \hat a_n(t) &= &
-\sqrt{\gamma_n/2}\sum_\pm \hat a_{\pm,t} -
\sum_{j=1}^N \frac{\kappa_j}{\sqrt{\pi n}}
\cos(n\pi l_j/Lc)\hat \sigma_j^-
\,  .
\end{eqnarray}
Similarly to Appendix \ref{G}, the stochastic  field is decomposed into its coherent and incoherent part 
$\hat a_{\pm,t} = -i \sqrt{2/\omega}\langle \hat A^{in}_{\pm}(0,t)\rangle + \delta \hat a_{+,t}$ where we use the notation $ A^{in}_{\pm}(0,t)=
\langle \hat A^{in}_{\pm}(0,t)\rangle$.
We neglect other noise channels (TM or TE or  spherical wave) as topologically less favorable for effective interaction. 
Using the analogy  based on stochastic formalism, we obtain eventually the equivalent Hamiltonian:
\begin{eqnarray}
\hat H(t)
&=& \hat  H_S(t) + \hat H_{I}(t)=\hat H_c+\hat H_{ext}(t)+ \hat H_{I}(t)
\label{Heq} \\
\hat H_c&=&\sum_{j=1}^N
-\frac{ \delta \omega_i}{2}\hat \sigma^z_i
-\sum_{n=0}^\infty \frac{\kappa_j}{\sqrt{\pi n}}\cos(\pi l_j/L)(\hat a^\dagger_n \hat \sigma^-_j+\hat a_n \hat \sigma^+_j)
+ (\omega_{n} -\omega)\hat a_n^\dagger \hat a_n
 \\ \label{Hext}
\hat H_{ext}(t)&=&i\sum_\pm \sqrt{\frac{\gamma_n }{\omega}}(A^{in*}_{\pm}(0,t)\hat a_n e^{-i\omega t}- A^{in}_{\pm}(0,t)\hat a^\dagger_n e^{i \omega  t} )
\\ \label{HI2}
\hat H_{I}(t)&=&-\sqrt{\frac{\gamma_n }{2}}\sum_\pm
(\delta \hat a^\dagger_{\pm,t}\hat a_n e^{-i\omega t}+ \hat a^\dagger_n e^{i\omega t} \delta \hat a_{\pm,t})
\,  .
\end{eqnarray}
It contains the Jaynes-Cumming contribution $\hat H_c$ with the addition of the external field term $\hat H_{ext}(t)$ and the interaction with the vacuum field $\hat H_{I}(t)$.
Note that these Eqs.(\ref{Heq}-\ref{HI2}) are formulated in the RWA 
picture with the substitution $\hat a_n \rightarrow e^{- i\omega t} \hat a_n$ and $\hat \sigma^\pm_j \rightarrow e^{\pm i\omega t} \hat \sigma^\pm_j$. 
To prove the equivalence of this last Hamiltonian with the Langevin equation (\ref{lang}), we derive the Heisenberg equations using an Ito-like procedure to obtain the damping term \cite{gardiner00}. For a finite time interval, 
the Heisenberg equation for $\hat a_n(t)$ is up to the second order:
\begin{eqnarray}
\frac{\hat a_n(t+\Delta t)-\hat a_n(t)}{\Delta t} =\frac{1}{i\Delta t}
\int_0^{\Delta t}dt'[\hat H(t'),\hat a_n(t)]
-\frac{1}{2\Delta t } 
\int_0^{\Delta t}dt' \int_0^{t'}dt'' [\hat H(t'),[\hat H(t''),\hat a_n(t)]] +{\cal O}(\Delta t^2)  
\,  .
\end{eqnarray}
It appears that the second term expansion has a finite contribution in the limit $\Delta t \rightarrow 0$. We notice  the nontrivial contribution:
\begin{eqnarray}
\frac{1}{2\Delta t } \int_0^{\Delta t}dt' \int_0^{t'}dt'' [\hat H_I(t'),[\hat H_I(t''),\hat a_n(t)]]
\rightarrow \frac{\gamma_n}{\Delta t}
\int_0^{\Delta t}dt' \int_0^{t'}dt'' \delta(t''-t')\hat a_n(t') =\frac{\gamma_n}{2\Delta t}
\int_0^{\Delta t}dt' \hat a_n(t)=
\frac{\gamma_n}{2}\hat a_n(t) \, ,
\nonumber \\
\end{eqnarray}
which corresponds to the damping term of Eq.(\ref{lang}). The Heisenberg equations for $\hat \sigma^\pm_j(t)$ and $\hat \sigma^z_j(t)$ are also recovered using (\ref{Heq}) when the quantum radiation field is approximated by its cavity mode component.

For the particular case $n=1$, the cavity mode is at frequency 
$\omega_{c}=\omega_1 \cong \pi c/L$ and damping $\gamma_c=\gamma_1$.
The associated system Hamiltonian is:
\begin{eqnarray}\label{Hs}
\hat H_S(t)=\sum_{j=1}^N
-\frac{ \delta \omega_j}{2}\hat \sigma^z_j
-g_j(\hat a^\dagger_1 \hat \sigma^-_j+\hat a_1 \hat \sigma^+_j)
+ (\omega_c -\omega)\hat a_1^\dagger \hat a_1
+i\sum_\pm \sqrt{\frac{\gamma_c }{\omega}}(A^{in*}_{\pm}(0,t)\hat a_1 e^{-i\omega t}- A^{in}_{\pm}(0,t)\hat a^\dagger_1 e^{i \omega  t} )  \, ,
\nonumber \\
\end{eqnarray}
where we renormalized the coupling constant into $g_j=\kappa_j \cos(\pi l_j/L) /\sqrt{\pi}$. From this expression, we note that  
the transmon should be placed ideally close to a cavity end for a maximum coupling to the field. 
Again we repeat the procedure of the previous section to derive a master equation with the elimination of the radiation field but using instead the interaction term (\ref{HI2}) for $n=1$. 
Adding the dephasing $\Gamma_{\phi,j}$ due to Josephson junction and the decay rate $\Gamma_j$ to a transverse channel (due to other cavity modes), we obtain:
\begin{eqnarray}\label{mast2}
&&\partial_t \hat \rho(t) +i[\hat H_S,\hat \rho(t)]
\nonumber \\&=&
-\frac{\gamma_c}{2} \left[\{\hat a^\dagger_1 \hat a_1, \hat \rho(t)\} 
-2\hat a_1  \hat \rho(t) \hat a^\dagger_1 \right]- \sum_{j=1}^N(\frac{\Gamma_j}{2}+\Gamma_{\phi,j})\{\hat \sigma^+_j \hat \sigma^-_j,\hat \rho(t)\}
-\Gamma_j\hat \sigma_j^- \hat \rho(t) \hat \sigma_j^+
  \,  .
\end{eqnarray}
This simulation is different from the one used for the artificial atom. Now in addition to the relaxation times for decay $T_{1,j}=1/\Gamma_j$ and for pure dephasing $T_{2,j}=1/\Gamma_{\phi,j}$, there is a lifetime for 
the photon within the cavity $T_{ph}=1/\gamma_c$. 

Unlike the model in the absence of cavity Eq.(\ref{mast2}), 
the qubits interact more strongly only with one field mode. This Purcell effect generates the  more complex dressed Jaynes-Cumming Hamiltonian spectrum with   anticrossing curves when the qubit frequency is changed by an external magnetic field \cite{fink2008climbing}. The presence of a cavity increases the decoherence time of the whole system as a result of the relaxation mainly due to the cavity field mode \cite{Brehm2021}.

\subsection{Reflection and Transmission for one cavity mode}

The master equation (\ref{mast2}) describes 
only the dynamics of qubit due to an external field. 
However, we need to specify another relation with the scattered field generated from the qubit evolution.
Restricting the frequency domain close to the first cavity mode, the operator relation (\ref{res}) and (\ref{res2}) is approximated  for a forward wavevector  incident field only as:
\begin{eqnarray}
\hat A_{\omega,+}(0)
=-\sum_\pm \frac{(\gamma_c/2)\hat A^{in}_{\omega,+}(0)
\pm i\sqrt{ \gamma_c \omega_c}/2\sum_{j=1}^N g_j  \hat \sigma^\mp_{\omega,j}
}{i(\omega -(\pm \omega_c - i\gamma_c/2))}
= \hat A^{sc}_{\omega,-}(-L)-\hat A^{in}_{\omega,+}(0)
\,  .
\end{eqnarray}
Carry out the quantum average of the scattered operators, the transmission is:
\begin{eqnarray}
S_{21}=\frac{\langle \hat A_{\omega,+}(0)\rangle}{
\langle \hat A^{in}_{\omega,+}(0)\rangle}
= \sum_{\pm}
\frac{-1}{\omega -(\pm \omega_c - i\gamma_c/2)}
\left[i\gamma_c/2 \pm
\frac{ \sqrt{\gamma_c \omega_c}}{2}\sum_{j=1}^N \frac{g_j\langle \hat \sigma_{\omega,j}^\mp \rangle}{\langle  \hat A^{in}_{\omega,+}(0) \rangle} \right] \, ,
\end{eqnarray}
and the reflection is: 
\begin{eqnarray}
S_{11} =
\frac{\langle \hat A^{sc}_{\omega,-}(-L)\rangle}{
\langle \hat A^{in}_{\omega,+}(0)\rangle}
=1-\sum_{\pm}
\frac{1}{\omega -(\pm \omega_c - i\gamma_c/2)}
\left[i\gamma_c/2 \pm
\frac{ \sqrt{\gamma_c \omega_c}}{2}\sum_{j=1}^N  \frac{g_j\langle \hat \sigma_{\omega,j}^\mp \rangle}{\langle  \hat A^{in}_{\omega,+}(0) \rangle} \right]  \,  .
\end{eqnarray}

\subsection{The Schrieffer-Wolff transformation for many qubits}

For the purpose of a quantum non-demolition measurement, the qubits frequencies must depart strongly from the cavity frequency $|\omega_i-\omega_c| \gg g_i$ but still remains small in comparison to the cavity frequency $|\omega_i-\omega_c| \ll \omega_c$. In that situation, by means of the  Schrieffer-Wolff, we can  rewrite the Hamiltonian in a diagonal form for interaction between the dressed qubit and cavity states.   
We start from the  Hamiltonian (\ref{Hs}) in absence of external field:
\begin{eqnarray}
\hat H_c=\hat H_0 + \hat V&=&
(\omega_c -\omega)\hat a_1^\dagger \hat a_1 +
\sum_{j=1}^N
-\frac{\delta \omega_j}{2}\hat \sigma^z_j
-g_j(\hat a^\dagger_1 \hat \sigma^-_j+\hat a_1 \hat \sigma^+_j)
\,  .
\end{eqnarray}
The Schrieffer-Wolff transformation consists in finding a unitary transformation $e^{\hat S}$ such that: $[\hat H_0, \hat S]=\hat V$ so that the new Hamiltonian $\hat H'_c=e^{\hat S}\hat H_c e^{-\hat S}=\hat H_0 +[\hat S, \hat V]/2$ is in the second order in $\hat V$ or $g_j$. We find the operator:
\begin{eqnarray}
\hat S = -\sum_{j=1}^N\frac{-g_j}{\omega_{j}-\omega_c}(\hat \sigma^+_j\hat a_1-\hat a^\dagger_1\hat \sigma^-_j  )
\,  .
\end{eqnarray}
Up to the second order, the qubit and field operators become under this transformation:
\begin{eqnarray}
\hat \sigma'^x_j\cong e^{\hat S}\hat \sigma^x_j e^{-\hat S}=\hat \sigma^x_j
-\frac{g_j}{\omega_{j}-\omega_c}
(\hat a^\dagger_1 +\hat a_1)\hat \sigma^z_j
\, , \quad \quad \quad 
\hat a'_1 \cong e^{\hat S}\hat a_1 e^{-\hat S}=\hat a_1 -\sum_{j=1}^N\frac{g_j}{\omega_{j}-\omega_c}\hat \sigma^-_j
\,  .
\end{eqnarray}
Inserting this result into the transformed Hamiltonian, we obtain finally:
\begin{eqnarray}
\hat H'_c&\cong&\sum_{j=1}^N
-\frac{\delta \omega_j}{2}\hat \sigma^z_j
+
\left(\sum_{j,j'=1}^N\frac{g_j g_{j'}
(\hat \sigma^+_j \hat \sigma^-_{j'}+\hat \sigma^+_{j'} \hat \sigma^-_{j})}{2(\omega_{j}-\omega_c)}+
\sum_{j=1}^N \chi_j\hat a^\dagger_1 \hat a_1 \hat \sigma^z_j \right)
+ (\omega_c -\omega)\hat a^{\dagger}_1\hat a_1 \, ,
\nonumber \\
\end{eqnarray}
where  
$\chi_j=g^2_j/(\omega_j-\omega_c)$ is the non-demolition interaction coupling responsible for the Stark shift of the qubit. We note also an additional term of interaction between the qubits that we shall neglect as its amplitude  contribution is small in comparison to the detuning energy for a frequency close to cavity frequency $\omega=\omega_c$.
The reverse transformation are:
\begin{eqnarray}\label{reverse}
\hat \sigma^-_j=\hat \sigma'^-_j+\frac{g_j}{\omega_{j}-\omega_c}\hat a'_1\hat \sigma'^z_j
\, , \quad \quad \quad 
\hat a_1 =\hat a'_1 +\sum_{j=1}^N\frac{g_j}{\omega_{j}-\omega_c}\hat \sigma'^-_j  \,  .
\end{eqnarray}

\section{Transmission for various 
profiles of the signal photon}\label{I}

\subsection{Signal field and probe field }

In this appendix, we perform the explicit calculation of the linear transmission of an off-resonance probe field of frequency 
$\omega_p$ for various profile of the signal photon field of frequency $\omega$, namely the coherent state, the incoherent state and the thermal state. For each of these cases, we provide analytical formulas that can be directly compared to experiments.  

\subsection{Model 1: Coherent state detection}

When the signal to be detected has the profile of a coherent state, the external field is expressed in terms of the coherent unknown mode and a probe mode:
\begin{eqnarray}\label{inpcoh}
\langle \hat A_{+}^{in}(0,t) \rangle=i\sqrt{\frac{\omega}{\gamma_c}}\left(\frac{\Omega}{2}e^{-i\omega t}+
\frac{\Omega_p}{2}e^{-i\omega_p t} \right)
+c.c.  \,  .
\end{eqnarray}
The external field term (\ref{Hext}) containing the unknown field and the probe field becomes under the Schrieffer-Wolff transformation:
\begin{eqnarray}
&&\hat H'_{ext}(t)=e^{\hat S}\hat H_{ext}(t)e^{-\hat S}
\nonumber \\
&=&
\frac{\Omega}{2}\left(\hat a_1 + \hat a^\dagger_1+\sum_{j=1}^N\frac{g_j}{\omega_{j}-\omega_c}\hat \sigma^x_j\right) + \frac{\Omega^*_{p}}{2}e^{i(\omega_p-\omega) t}\left(\hat a_1 - \sum_{j=1}^N\frac{g_j}{\omega_{j}-\omega_c}\hat \sigma^-_j\right) + \frac{\Omega_{p}}{2}e^{-i(\omega_p-\omega) t}\left(\hat a^\dagger_1 +\sum_{j=1}^N\frac{g_j}{\omega_{j}-\omega_c}\hat \sigma^+_j\right) \,  .
\nonumber \\
\end{eqnarray}
Similarly, defining $\hat \rho'(t)=e^{\hat S}\hat \rho(t) e^{-\hat S}$, the master equation (\ref{mast2}) becomes: 
\begin{eqnarray}\label{mast3}
&&\partial_t \hat \rho'(t) +i[\hat H'_c+\hat H'_{ext}(t),\hat \rho'(t)]= 
\nonumber \\&& -\frac{\gamma_c}{2} \left[\left\{\hat a^\dagger_1 \hat a_1 + \sum_{j,j'=1}^N\frac{g_j g_{j'}}{(\omega_{j}-\omega_c)(\omega_{j'}-\omega_{c})}
\hat \sigma^+_j \hat \sigma^-_{j'} 
, \hat \rho'(t)\right\} 
-2\hat a_1  \hat \rho'(t) \hat a^\dagger_1 
- 2\sum_{j=1}^N\frac{g_j}{\omega_{j}-\omega_c}\hat \sigma^-_{j} \hat \rho'(t) \sum_{j'=1}^N\frac{g_{j'}}{\omega_{j'}-\omega_{c}}\hat \sigma^+_{j'}\right]
\nonumber \\
&& -\sum_{j=1}^N(\frac{\Gamma_j}{2}+\Gamma_{\phi,j})\left \{\hat \sigma^+_j \hat \sigma^-_j +\frac{g^2_j}{(\omega_{j}-\omega_c)^2}
\hat a^\dagger_1 \hat a_1 ,\hat \rho'(t)\right\}
-\Gamma_j\left(\hat \sigma_j^- \hat \rho'(t) \hat \sigma_j^+
+ \frac{g^2_j}{(\omega_{j}-\omega_c)^2}
\hat a_1\hat \sigma^z_j\hat \rho'(t) 
\hat a^\dagger_1 \hat \sigma^z_j
\right) \,  .
\end{eqnarray}
In this equation, we shall neglect the smaller  higher order terms in $g_j^2$. 
We define $\beta$ as the amplitude of the coherent field to be detected inside the cavity. It distinguishes from the quantum part of the field operator. Using the Glauber-Sudershan transformation $\hat D$, we rewrite this field as: 
\begin{eqnarray}\label{beta}
\hat a_1 \rightarrow  \hat D^\dagger \hat a_1 \hat D=\hat a_1+\beta =\hat a_1 +
\frac{\Omega/2}{\omega -\omega_c^* +i\gamma_c/2} \, ,
\end{eqnarray}
where $\omega_{c}^*=\omega_c-\sum_{j=1}^N \chi_j$ is the renormalized cavity frequency due to the presence of the qubits when they are in the ground state. Using $\hat \rho''(t)=\hat D^\dagger \hat \rho'(t) \hat D$, we reduce the master equation (\ref{mast3}) into:
\begin{eqnarray}\label{mast4}
\partial_t \hat \rho''(t) &+&i[\hat H^{(2)}(t),\hat \rho''(t)]
=-\frac{\gamma_c}{2} \left[\{\hat a^\dagger_1 \hat a_1 , \hat \rho''(t)\} 
-2\hat a_1  \hat \rho''(t) \hat a^\dagger_1 \right]- \sum_{j=1}^N(\frac{\Gamma_j}{2}+\Gamma_{\phi,j})\{\hat \sigma^+_j \hat \sigma^-_j,\hat \rho''(t)\}
-\Gamma_j\hat \sigma_j^- \hat \rho''(t) \hat \sigma_j^+
\nonumber \\
\hat H^{(2)}(t)&=&\sum_{j=1}^N
-\frac{\delta \omega_j}{2}\hat \sigma^z_j
+
\sum_{j=1}^N \chi_j
(\hat a^\dagger_1 +\beta^*)(\hat a_1 +\beta)(\hat \sigma^z_j+1)
+(\omega_c^* -\omega)\hat a^{\dagger}_1\hat a_1 +\frac{\Omega}{2}\sum_{j=1}^N\frac{g_j}{\omega_{j}-\omega_c}\hat \sigma^x_j
\nonumber \\
&+& \frac{\Omega^*_{p}}{2}e^{i(\omega_p-\omega) t}\left(\hat a_1 +\sum_{j=1}^N\frac{g_j}{\omega_{j}-\omega_c}\hat \sigma^-_j \right) + \frac{\Omega_{p}}{2}e^{-i(\omega_p-\omega) t}\left(\hat a^\dagger_1 +\sum_{j=1}^N\frac{g_j}{\omega_{j}-\omega_c}
\hat \sigma^+_j\right)  \,  .
\end{eqnarray}
The term proportional to $\hat \sigma^x_j$ can be neglected for sufficiently strong detuning $\delta \omega_j$.
Without the probe field, in this dressed picture, the vacuum field together with qubit in their ground state correspond to the solution of this master equation:
\begin{eqnarray}
\hat \rho_{0}=  |0\rangle \langle 0|\prod_{j=1}^N
\frac{1-\hat \sigma^z_j}{2} \,  .
\end{eqnarray}
A non-zero probe field $\Omega_p$ introduces a perturbation of the form:
\begin{eqnarray}\label{rho1}
\hat \rho_1(t)=\hat \rho''(t)- \hat \rho_0 = e^{-i(\omega_p-\omega) t}\sum_{j=1}^N\left(
\frac{1-\hat \sigma^z_j}{2}|\Psi_{j0}\rangle \langle 0| +
\hat \sigma^+_j |\Psi_j\rangle \langle 0|\right)\prod_{j'\not=j=1}^N
\frac{1-\hat \sigma^z_{j'}}{2} +c.c.  \,  .
\end{eqnarray}
The insertion of the latter in Eq.(\ref{mast4}) imposes that the unknown wave function states solve the non-Hermitian state equations:
\begin{eqnarray}
\left[ (\omega_p-\omega)
- (\omega_c^* -\omega-i\gamma_1/2)\hat a^{\dagger}_1\hat a_1 \right]|\Psi_{j0}\rangle
&=&
\frac{\Omega_{p}}{2}\hat a^\dagger_1 
|0\rangle \,  ,
\end{eqnarray}
\begin{eqnarray}\label{psij}
\left[ (\omega_p-\omega_j)
-
2\chi_j(\hat a^\dagger_1 +\beta^*)(\hat a_1 +\beta)
- (\omega_c^* -\omega-i\gamma_1/2)\hat a^{\dagger}_1\hat a_1 + i(\Gamma_j/2+\Gamma_{\phi,j})\right]|\Psi_j\rangle
&=&
\frac{\Omega_{p}}{2}\frac{g_j}{\omega_{j}-\omega_c}
|0\rangle  \,  .
\end{eqnarray}
The former wave function $|\Psi_{j0}\rangle$ has a simple solution:
\begin{eqnarray}\label{psi0}
|\Psi_{j0}\rangle&=&
\frac{\Omega_{p}/2}{ (\omega_p-\omega_c^* +i\gamma_1/2)}
\hat a^\dagger_1 
|0\rangle \,  .
\end{eqnarray}
It corresponds to the free part propagation of  probe field transmission and therefore insensitive to the presence of the signal photon field. The latter wave function however is sensitive to 
the qubit dynamics whose the resonance frequency has a Stark shift depending on the coherent signal field. Inverting formally the operator in the left hand side of 
(\ref{psij}), we obtain a formal solution for $|\Psi_j\rangle$ 
which, after substitution into (\ref{rho1}), gives explicit expressions  
for the expectation  of the lowering and raising qubit operators. 
Defining the prime representation, we obtain
\begin{eqnarray}\label{formsig}
g_j\langle \hat \sigma^-_{\omega_p,j} \rangle'&=&
g_j{\rm Tr}(\hat \rho_1(t) \hat \sigma^-_j )
e^{i(\omega_p-\omega) t}
\nonumber  \\ 
&=&
\frac{\Omega_{p}}{2}\chi_j\langle 0|
\left[ (\omega_p- \omega_j)
-
2\chi_j(\hat a^\dagger_1 +\beta^*)(\hat a_1 +\beta)
- (\omega_c^* -\omega-i\gamma_c/2)\hat a^{\dagger}_1\hat a_1+i(\Gamma_j/2+\Gamma_{\phi,j})\right]^{-1}|0\rangle \, ,
\end{eqnarray}
and similarly $\langle \hat \sigma_{\omega,j}^- \rangle'=\langle \hat \sigma_{j,-\omega}^+ \rangle'$. 
The vacuum matrix element of the non-Hermitian propagator 
is determined explicitly. 
Defining the parameters:
\begin{eqnarray}\label{var}
w_0= (\omega_p -\omega_j)
-2\chi_j
|\beta|^2+i(\Gamma_j/2+\Gamma_{\phi,j}) \, , \quad \quad \quad
w=  \omega_c^*+ 2\chi_j-\omega-i\gamma_c/2 \, , \quad \quad \quad 
b=2\chi_j\beta \, ,
\end{eqnarray}
and using the creation-annihilation operator algebra, 
we  calculate successively:
\begin{eqnarray}
&&\langle 0|
\frac{1}{w_0 -w \hat a^\dagger_1\hat a_1 
-b\hat a^\dagger_1 -b^*\hat a_1 }|0\rangle
=-i\int_0^\infty dt
\langle 0|
\exp[i(w_0 -w \hat a^\dagger_1\hat a_1 
-b\hat a^\dagger_1 -b^*\hat a_1 )t]|0\rangle
\nonumber \\
&=&-i\int_0^\infty dt
\langle 0|
e^{i[w_0 -w (\hat a^\dagger_1+b^*/w)(\hat a_1 
+b/w) +|b|^2/w ]t}|0\rangle
=-i\int_0^\infty dt
\langle b/w|
e^{i[w_0 -w \hat a^\dagger_1 \hat a_1 +|b|^2/w ]t}|b/w\rangle
\nonumber \\
&=&-i\int_0^\infty dt
\langle b/w|
e^{i[w_0 +|b|^2/w ]t}|e^{-iwt}b/w\rangle
=-i\int_0^\infty dt\, e^{i[w_0 +|b|^2/w ]t} e^{-(1-\exp(-iwt))|b|^2/w^2}
\nonumber \\ \label{hyperg}
&=&e^{-|b|^2/w^2}\sum_{n=0}^\infty
\frac{(|b|^2/w^2)^n}{n!}\frac{(-1)}{nw-w_0-|b|^2/w}
=\frac{e^{-|b|^2/w^2}}{w_0+|b|^2/w}
\,_1 F_1 (-w_0/w-|b|^2/w^2; 1-w_0/w-|b|^2/w^2;|b|^2/w^2) \, ,
\nonumber\\
\end{eqnarray}
where $_1 F_1 (a;b;z)$ is the hypergeometric function.
Using this result with the appropriate parameter substitution into Eq.(\ref{formsig}),
we obtain  a formula with a sideband  spectrum:
\begin{eqnarray}\label{sigcoh}
g_j\langle \hat \sigma^-_{\omega_p,j} \rangle'&=&
\frac{\Omega_{p}}{2}\chi_j
e^{-W}\sum_{n=0}^\infty
\frac{W^n}{n!}
\nonumber \\
&\times&
\frac{1}{ (\omega_p -\omega_j)
-2\chi_j
(|\beta|^2+n)-n (\omega_c^* -\omega-i\gamma_c/2)+
i(\Gamma_j/2+\Gamma_{\phi,j})
+
\frac{4\chi_j^2 |\beta|^2
}{ (\omega_c^*+2\chi_j -\omega-i\gamma_c/2)
}}
\\
&\stackrel{\beta =0}{=}&
\frac{\Omega_{p}}{2}\chi_j
\frac{1}{\omega_p -\omega_j+
i(\Gamma_j/2+\Gamma_{\phi,j})
}  \, ,
\nonumber 
\\
W&=&
\frac{4\chi_j^2|\beta|^2}{ (\omega_c^*+2\chi_j -\omega-i\gamma_c/2)^2}  \,  .
\end{eqnarray}
The pole of each term in the infinite sum allows to  identify 
the resonance frequencies and their width revealing a comb structure (as for example in Fig.\ref{fig1}). Alternatively we can use also the hypergeometric function in (\ref{hyperg}) for a more direct computation. 
For the case when $\chi_j \gg \gamma_c$ and matching frequency $\omega=\omega_c$, we find the more explicit formula:
\begin{eqnarray}\label{weakgam}
g_j\langle \hat \sigma^-_{\omega_p,j} \rangle'&=&
\frac{\Omega_{p}}{2}\chi_j
e^{-|\beta|^2}\sum_{n=0}^\infty
\frac{|\beta|^{2n}}{n!}
\frac{1}{ (\omega_p -\omega_j)
-2\chi_jn +i(n+|\beta|^2)\gamma_c/2+
i(\Gamma_j/2+\Gamma_{\phi,j})} \, .
\end{eqnarray}
For high quality factor, the  width  is mostly controlled by the  the radiative and dephasing times $T_1$ and $T_2$. 
For the opposite case $\chi_j \ll \gamma_c$, we find more simply:
\begin{eqnarray}
g_j\langle \hat \sigma^-_{\omega_p,j} \rangle'&=&
\frac{\Omega_{p}}{2}
\frac{\chi_j}{ (\omega_p -\omega_j)
-2\chi_j |\beta|^2 +
i(\Gamma_j/2+\Gamma_{\phi,j})}  \,  .
\end{eqnarray}
For a lower quality factor,
we obtain a frequency  shift 
revealing the presence of the  photon field of intensity $|\beta|^2$ to be detected.

Determining from Eq.(\ref{psi0}) $\langle \hat a_{\omega_p,1} \rangle=\Omega_p/[2(\omega_p -\omega_c^* +i\gamma_c/2)]$  and returning to the original representation Eq.(\ref{reverse}), we determine after averaging finally the qubit dynamics:
\begin{eqnarray}
g_j\langle \hat \sigma^-_j\rangle(t)= e^{-i\omega_p t}
\left(\langle \hat \sigma^-_{\omega_p,j} \rangle'+
\frac{\chi_j \Omega_p/2}{( \omega_p -\omega_c^* +i\gamma_c/2)}\right)
+
e^{-i\omega t}\chi_j \beta  \,  .
\end{eqnarray}
Using the properties $\langle \hat \sigma_{j,\omega}^- \rangle=\langle \hat \sigma_{-\omega,j}^+ \rangle$, we determine two transmissions as a result of the input signal frequency $\omega$ and the input probe frequency $\omega_p$, :
\begin{eqnarray}
{S}_{21}|_\omega&=&
\sum_{\pm}
\frac{-1}{\omega -(\pm \omega_c - i\gamma_c/2)}
\left[i\gamma_c/2 \pm
\frac{\sqrt{ \gamma_c \omega_c}}{2}\sum_{j=1}^N   \frac{g_j\langle \hat \sigma_{\omega,j}^\mp \rangle}{\langle  \hat A^{in}_{\omega,+}(0) \rangle} \right]
\\
&=&
\sum_{\pm}
\frac{-i\gamma_c/2}{\omega -(\pm \omega_c - i\gamma_c/2)}
\left[1 \pm \sum_{j=1}^N \frac{\mp \chi_j}{(\pm \omega -\omega_c^* \pm i\gamma_c/2)} \right]
\nonumber \\
&=&\sum_{\pm}
\frac{-i\gamma_c/2}{\omega -(\pm \omega_c^*  - i\gamma_c/2)}
\,  ,
\end{eqnarray}
\begin{eqnarray}\label{S21p}
S_{21}^{coh}={ S}_{21}|_{\omega_p}&=&
\sum_{\pm}
\frac{-i\gamma_c/2}{\omega_p -(\pm \omega_c - i\gamma_c/2)}
\left[1 \mp\sum_{j=1}^N \frac{g_j\langle \hat \sigma_{\omega_p,j}^\mp \rangle'}{\Omega_p} -\frac{\chi_j}{ \pm \omega_p -\omega_c^* \pm i\gamma_c/2}\right]
\\
&=&\sum_{\pm}
\frac{-i\gamma_c/2}{\omega_p -(\pm \omega_c^*- i\gamma_c/2)}\mp 
\sum_{j=1}^N \frac{  i\gamma_c}{\omega_p -(\pm \omega_c - i\gamma_c/2)}\frac{g_j\langle \hat \sigma_{\omega_p,j}^\mp \rangle'}{\Omega_p}
\,  .
\end{eqnarray}
The latter is the  probe transmission used for the cavity photon detection  off cavity resonance  while the former corresponds to the signal transmission much larger on resonance.
Therefore, we need to use a higher intensity probe in order to detect the small off-resonance response. 

Finally, we connect the coherent cavity field $\beta$ to the forward photon flux  given by  ${{\cal J}}=  c\langle {\hat {n}^{in}}\rangle/{\cal L}$ for a waveguide of large length ${\cal L}$. The input field in the forward $\langle {\hat {n}^{in}}\rangle$ is  related to the input coherent field $\alpha^{in}$ associated to the operator $\hat a_{\omega/c}/\sqrt{\cal L}$. 
Using the relation in (\ref{inpcoh}) and (\ref{beta}), the field amplitude inside the cavity and its average photon number 
$\overline{n}= |\beta|^2$ are therefore expressed respectively in terms of the input field and the photon flux as:
\begin{eqnarray}\label{incav}
\beta= \frac{\sqrt{\gamma_c/2\cal L}\alpha^{in} }{\omega-\omega_c^* +i\gamma_c/2} \, ,
\quad \quad \quad 
\overline{n}=\frac{\gamma_c{{\cal J}}/2}{(\omega-\omega_c^*)^2 +\gamma_c^2/4} \,  .
\end{eqnarray}
For matching frequency, the cavity photon number becomes $\overline{n}=2{{\cal J}}/\gamma_c$ which is for high quality factor  much higher than the photon  density outside the cavity.

\subsection{Model 2: Incoherent input radiation}

The incoherent light  is a radiation field characterized by an absence of absolute phase \cite{gardiner00}.
It has the same distribution of black body radiation with the difference that it is assumed here to be monochromatic of frequency $\omega$ with an infinite coherence time. 
Given an input field mode $\hat a_{\omega/c}$ in the  forward direction and the associated 
average photon number $\langle {\hat {n}^{in}}\rangle= \langle \hat a^\dagger_{\omega/c}\hat a_{\omega/c}\rangle $,
its density matrix has the form:
\begin{eqnarray}\label{rhoinc}
{\hat {\rho }}_{B}=\sum_{n=0}^\infty
\frac{\langle {\hat {n}^{in}}\rangle^n}
{(\langle {\hat {n}^{in}}\rangle+1)^{n+1}}
|n \rangle \langle n|  \,  .
\end{eqnarray}
To treat this non-classical field, 
we use the insights so far gained for the coherent case by means of  the P representation for the variable $\alpha^{in}$ associated to the operator $\hat a_{\omega/c}/\sqrt{\cal L}$. As in \cite{gardiner00}, we construct a weight function $P(\alpha^{in} )$  with the property that the density matrix  is diagonal in the basis of input coherent field states   $\{|\alpha^{in}\rangle\}$. More explicitly,
\begin{eqnarray}\label{corr}
{\hat {\rho }}_{B}=\int P(\alpha^{in} )|{\alpha^{in} }\rangle \langle {\alpha^{in} }|\,d^{2}\alpha^{in} ,\qquad d^{2}\alpha^{in} \equiv d\,{\rm {Re}}(\alpha^{in} )\,d\,{\rm {Im}}(\alpha^{in} ) \, .
\end{eqnarray}
The P representation of the fully incoherent light corresponding to (\ref{rhoinc}) has the form:
\begin{eqnarray}
P(\alpha^{in})={\frac {1}{\pi \langle {\hat {n}^{in}}\rangle }}e^{{-|\alpha^{in} |^{2}/\langle {\hat {n}^{in}}\rangle }} \,  .
\end{eqnarray}
As a consequence, the transmission becomes  phase independent and using the correspondence (\ref{corr})  is written as: 
\begin{eqnarray}
{  S}^{inc}_{21}=\int P(\alpha^{in} ) {S}^{coh}_{21}({\alpha^{in}}) \,d^{2}\alpha^{in} \,  .
\end{eqnarray}
Using the relations  (\ref{incav}), this integral is expressed in terms of the photon number inside the cavity:
\begin{eqnarray}
{ S}^{inc}_{21}=\int \frac {1}{\pi \overline{n} }e^{-|\beta |^{2}/\overline{n} } S^{coh}_{21}(\beta) \,d^{2}\beta 
\, \quad \quad \quad 
\langle \hat \sigma^-_{\omega_p,j} \rangle'^{inc}=\int \frac {1}{\pi \overline{n} }e^{-|\beta |^{2}/\overline{n} } \langle \hat \sigma^-_{\omega_p,j} \rangle' d^{2}\beta \, .
\end{eqnarray}
In the simplest case of high quality factor
obeying $\gamma_c \ll (1+\overline{n})(\Gamma_j/2+\Gamma_{\phi,j})/\overline{n}$ and matching frequency, we can use the expression (\ref{weakgam}).
We find after integration over $\beta$ a change in the width:
\begin{eqnarray}
g_j\langle \hat \sigma^-_{\omega_p,j} \rangle'^{inc}&=&
\frac{\Omega_{p}}{2}\chi_j
\sum_{n=0}^\infty \frac{\overline{n}^n/(\overline{n}+1)^{n+1}}{\omega_p -(\omega_j+2\chi_jn-i(n\gamma_c/2 +\Gamma_j/2+\Gamma_{\phi,j}))} \, ,
\end{eqnarray}
with the spectrum given by  
\begin{eqnarray}
\omega_p^{(n)} =\omega_j
+2\chi_jn -in\gamma_c/2-
i(\Gamma_j/2+\Gamma_{\phi,j})  \,  .
\end{eqnarray}
For the general case,  the spectral line is  rewritten in terms of the exponential integral functions 
$E_n(x)=\int_1^\infty dt e^{-xt}/t^n$.
Using the relations:
\begin{eqnarray}
\int \frac {1}{\pi}e^{-c|\beta |^{2} } \frac{|\beta|^{2n}}{n!}
\frac{1}{a+b|\beta|^2}\,d^{2}\beta
=\frac{e^{a c/b}}{bc^n}E_{n+1}\left( \frac{ac}{b}\right) \, ,
\end{eqnarray}
and applying it to Eq.(\ref{sigcoh}) with the variable $w$ defined in (\ref{var}), we obtain:
\begin{eqnarray}\label{siginc}
g_j\langle \hat \sigma^-_{\omega,j} \rangle'^{inc}&=&
\frac{\Omega_{p}}{2}\chi_j
\sum_{n=0}^\infty \frac{1}{\overline{n}}
\frac{e^{x_n}E_{n+1}\left( x_n\right)}
{(2\chi_j -\frac{4\chi_j^2}{w})\left(1/\overline{n}+\frac{4\chi_j^2}{w^2}\right)^n} \, ,
\\
x_n&=&\frac{i(1/\overline{n}+\frac{4\chi_j^2}{w^2})}{(2\chi_j -\frac{4\chi_j^2}{w})}
\left[\frac{\Gamma_j}{2}+\Gamma_{\phi,j}+n\gamma_c/2 -i(\omega_p - \omega_j-2(\chi_j+\omega_c^* -\omega)n )\right]
\,  .
\end{eqnarray}
For the purpose of practical computation, we can use alternatively the Kummer function $e^y E_n(y)=y^{n-1}U(n,n,y)$.
At the singularity for $x_n \sim 0$, we have $E_{1}\left( x_1\right) \sim -\gamma -\ln(x_n)$ and  $E_{n+1}\left( x_n\right) \sim 1/n$ for $n >0$. 
Asymptotically, $E_{n+1}\left( x_n\right) \sim \exp(-x_n)/x_n$. In the limit of strong $\gamma_c \gg \chi_j$, the first term in the sum is dominant and Eq.(\ref{siginc}) becomes:
\begin{eqnarray}
g_j\langle \hat \sigma^-_{\omega,j} \rangle'^{inc}
=\frac{\Omega_{p}}{4\overline{n}}
\exp[ [(\omega_p - \omega_j) +i(\frac{\Gamma_j}{2}+\Gamma_{\phi,j})]/(2\overline{n}\chi_j)]
E_1\left([(\omega_p - \omega_j) +i(\frac{\Gamma_j}{2}+\Gamma_{\phi,j})]/(2\overline{n}\chi_j)\right) \,  .
\end{eqnarray}

\subsection{Model 3: Incoherent state detection with  coherence time  (thermal bath states)}

In the previous subsection, we assume a pure input monochromatic beam 
which means that the frequency broadening is much smaller than the cavity broadening. In this subsection, we  assume the opposite case where the broadening  is much larger so that the coherence time $\tau_c$ fulfills the condition  $\gamma_c\tau_c \ll 1$. 
In that situation, the amplitude average is still zero and 
the averaged photon number is distributed continuously in the momentum space so that the radiation signal can be viewed as a thermal bath for the cavity system .  
As a consequence, the treatment of such photon signal requires to change the density matrix $\hat \rho_B(0)$ from the vacuum state into a thermal-like state. It is in fact sufficient 
to define this incoherent state by the stationary average:
\begin{eqnarray}
\langle \hat a^\dagger_{k,in}\hat a_{k',in}\rangle
= 2\pi 1^+(k)\delta(k-k')\frac{ 2{\cal J}/ \tau_c}{ (\omega-\omega_k)^2 +1/\tau_c^2} \, , \quad
\quad\quad 
\langle \hat a_{k,in}\hat a^\dagger_{k',in}\rangle
= 2\pi\delta(k-k')
\left((1+1^+(k)\frac{ 2{\cal J}/\tau_c}{(\omega-\omega_k)^2 +1/\tau_c^2}\right) \, ,
\nonumber \\
\end{eqnarray}
where ${\cal J}$ is the photon flux coming from the left side of the cavity. The distribution has a Lorentzian form centered around the input frequency $\omega$ but with a frequency 
broadening large enough as to be indistinguishable from a thermal state.

With these different assumptions, the master equation (\ref{mast4}) has to be modified by replacing 
the vacuum from the Lorentzian state and setting $\Omega=0$ and choosing the reference frequency for detuning in the RWA as $\omega=\omega_c^*$.
This procedure leads to the following non-Markovian master equation:
\begin{eqnarray}
\partial_t \hat \rho'(t) &+&i[\hat H^{(3)}(t),\hat \rho'(t)]
=-\frac{\gamma_c}{2} \int_0^\infty dt' 
{\cal J} 
\cos[(\omega-\omega_c^*)t']
e^{-t'/\tau_c}
\left[\{\hat a_1 \hat a^\dagger_1, \hat \rho'(t-t')\} 
-2\hat a^\dagger_1   \hat \rho'(t-t') \hat a_1 \right]+
\nonumber \\
(2\delta(t')&+&{\cal J} 
\cos[(\omega-\omega_c^*)t']
e^{-t'/\tau_c}) 
\left[\{\hat a^\dagger_1 \hat a_1 , \hat \rho(t-t')\} 
-2\hat a_1  \hat \rho'(t-t') \hat a^\dagger_1 \right]
- \sum_{j=1}^N(\frac{\Gamma_j}{2}+\Gamma_{\phi,j})\{\hat \sigma^+_j \hat \sigma^-_j,\hat \rho'(t)\}
-\Gamma_j\hat \sigma_j^- \hat \rho'(t) \hat \sigma_j^+
\nonumber \\
\hat H^{(3)}(t)&=&\sum_{j=1}^N
\frac{ (\omega_j-\omega_c^*)}{2}\hat \sigma^z_j
+
\sum_{j=1}^N \chi_j
\hat a^\dagger_1 \hat a_1 (\hat \sigma^z_j +1) 
+
\sum_{j=1}^N\left[\frac{\Omega^*_{p}}{2}\left(\hat a_1-\frac{g_j}{\omega_{c}-\omega_j}\hat \sigma^-_j\right)
e^{i(\omega_p -\omega_c^*) t}+ c.c.\right] \,  .
\end{eqnarray}
In absence of the probe field ($\Omega_p=0$), the stationary solution has a form where the qubits are in their ground state and the radiation has a thermal distribution: 
\begin{eqnarray}
\hat \rho_0
=
\hat \rho_T \prod_{j=1}^n \frac{1 - \hat \sigma_j^z}{2} \, ,
\quad\quad \quad
\hat \rho_T
=
\frac{\exp(-\hbar \omega_c^* \hat a^\dagger_1 \hat a_1 /k_B T_e)}{\overline{n}+1} \,  .
\end{eqnarray}
The effective temperature $T_e$ is defined  from the effective Bose Einstein factor for average population inside the cavity $\overline{n}$ which is itself related to the Lorentzian distribution parameter: 
\begin{eqnarray}
\overline{n}= \frac{1}{\exp(\hbar \omega_c^*/k_B T_e)-1}= \frac{{\cal J}/\tau_c}{ (\omega-\omega_c^*)^2 +1/\tau_c^2}\stackrel{\omega= \omega^*_c}{=}
\tau_c {\cal J} \,  .
\end{eqnarray}
Therefore, because of the larger broadband, the rate of signal photon passing through the cavity is considerably reduced by a factor $\tau_c\gamma_c/2$ in comparison to the monochromatic beams defined in subsection 1 and 2. 
The probe field introduces a perturbation:
\begin{eqnarray}
\hat \rho'(t)= \hat \rho_0 +\delta \hat \rho(t) 
= \hat \rho_0 + \left[ e^{-i(\omega_p-\omega_c^*) t}
\left(\sum_{j=1}^N \frac{1-\hat \sigma^z_{j}}{2}\hat \rho_{j0} +\hat \sigma^+_j  \hat \rho_j \right)\prod_{j'\not=j=1}^N
\frac{1-\hat \sigma^z_{j'}}{2} +c.c.\right] \, ,
\end{eqnarray}
whose matrix elements $\hat \rho_{j0}$ and
$\hat \rho_{j}$  solve respectively:
\begin{eqnarray}
&&-i(\omega_p-\omega)\hat \rho_{j0} 
+\frac{i\Omega_{p}}{2}[\hat a^{\dagger}_1,\hat \rho_T]
\nonumber \\
&&
=-\frac{\gamma_c}{2}\overline{n}_p
\left[\{\hat a_1 \hat a^\dagger_1, \hat \rho_{j0}\} 
-2\hat a^\dagger_1   \hat \rho_{j0} \hat a_1 \right]
-\frac{\gamma_c}{2} (1+\overline{n}_p)
\left[\{\hat a^\dagger_1 \hat a_1, \hat \rho_{j0}\} 
-2\hat a_1   \hat \rho_{j0} \hat a^\dagger_1 \right] \,  ,
\end{eqnarray}
\begin{eqnarray}\label{rho0}
&&-i[\omega_p-\omega_j+i(\frac{\Gamma_j}{2}+\Gamma_{\phi,j})] \hat \rho_j 
+\hat a^{\dagger}_1\hat a_1 \hat \rho_j+\frac{i\Omega_{p}}{2}\frac{g_j\hat \rho_T}{\omega_{j}-\omega_c}
\nonumber \\ \label{rhoj}
&&
=-\frac{\gamma_c}{2}\overline{n}_p
\left[\{\hat a_1 \hat a^\dagger_1, \hat \rho_j\} 
-2\hat a^\dagger_1   \hat \rho_j \hat a_1 \right]
-\frac{\gamma_c}{2} (1+\overline{n}_p) 
\left[\{\hat a^\dagger_1 \hat a_1, \hat \rho_j\} 
-2\hat a_1   \hat \rho_j \hat a^\dagger_1 \right] \,  .
\end{eqnarray}
In both equations, the effective Bose enhancement factor is defined  as
$\overline{n}_p=\overline{n}|_{\tau_c=\tau_p}$
where $\tau_p= \tau_c/(1+i\tau_c(\omega_p-\omega))$ and is much less sensitive to the signal for large detuning between the signal and probe frequency. On the contrary, for relatively long coherence time satisfying $\tau_c(\omega_p-\omega) \ll 1$, the master equation  corresponding to a markovian process, this enhancement is maximal $\overline{n}_p \simeq \overline{n}$.

The first equation (\ref{rho0}) is similar to the case (\ref{psi0}) developed in subsection 1 and provides an identical  shift in the photon transmission spectrum. We need to take only the first order average in the field to obtain the same response $\langle \hat a_{\omega_p,1}\rangle$. The second equation (\ref{rhoj}) is 
solved using the following steps. First, we use  
the Fock representation $f_j(n)=\langle n|\hat \rho_j|n\rangle$, this second equation is rewritten as:
\begin{eqnarray}
[i(\omega_p-\omega_j)-(\frac{\Gamma_j}{2}+\Gamma_{\phi,j}) -i2n\chi_j ]f_j(n)
&=&\gamma_c (1+\overline{n}_p)(nf_j(n)-(n+1)f_j(n+1))
+\gamma_c \overline{n}_p((n+1)f_j(n)-nf_j(n-1)
)
\nonumber \\&+&
\frac{i\Omega_{p}}{2}\frac{g_je^{-\hbar\omega^*_c n/k_bT_e}}{(\omega_{j}-\omega_c)(1+\overline{n})}
\,  .
\nonumber \\
\end{eqnarray}
Second, we define from this representation the generating function:
$h_j(x)=e^{v_jx}\sum_{n=0}^\infty x^n f_j(n)/n!$. This function  obeys the differential equation:
\begin{eqnarray}
&&\left[\gamma_c (1+\overline{n}_p)(x\partial_x^2+\partial_x)
+[2\gamma_c(v_j-1)(\overline{n}_p+1)+\gamma_c-2i\chi_j]x\partial_x
+
i[\omega_p-\omega_j+i(\frac{\Gamma_j}{2}+\Gamma_{\phi,j})]+\gamma_c(v_j(\overline{n}_p+1)-\overline{n}_p)\right] h_j(x)
\nonumber \\
&&\quad\quad \quad=
\frac{i\Omega_{p}}{2}\frac{g_j e^{[\overline{n}/(1+\overline{n})-v_j]x}}{(1+\overline{n})(\omega_{j}-\omega_c)} 
\end{eqnarray}
where we choose $v_j=1+(i\chi_j-\gamma_c/2-
S_j)/[\gamma_c(1+\overline{n}_p)]$
with
$S_j=\sqrt{\gamma_c^2/4+\gamma_c(2\overline{n}_p+1)i\chi_j-\chi^2_j}$. Third, we solve the differential equation using the Laguerre polynomials expansion: $h_j(x)=\sum_{k=0}^\infty
c_n L_n\left(2S_jx/\gamma_c(\overline{n}_p+1)\right))$. 
These polynomials are eigenfunctions of the differential 
operator and leads to a quantification condition of the spectrum for $\omega_p$  
according to
\begin{eqnarray}
\omega_p^{(n)}=\omega_j -i(\frac{\Gamma_j}{2}+\Gamma_{\phi,j})
-i(2n+1)S_j- \chi_j+i\gamma_c/2 \,  .
\end{eqnarray}
Fourth, we determine the coefficient $c_n$ using the variation of constants and then determine $\langle \hat \sigma^-_{\omega,j} \rangle'={\rm Tr}(\hat \rho_j)$. After a straightforward calculation, we find the general expression: 
\begin{eqnarray}\label{sigth}
g_j\langle \hat \sigma^-_{\omega,j} \rangle'&=&
\frac{\Omega_{p}\chi_j}{2}
\sum_{n=0}^\infty \frac{2S_j\gamma_c}{\omega_p -\omega_p^{(n)}}
\frac{[\gamma_c/2-S_j-i\chi_j]^n}{[\gamma_c/2+S_j-i\chi_j]^{n+1}}
\frac{[\gamma_c(v_j-\overline{n}/(1+\overline{n}))(1+\overline{n}_p)]^n}{[
\gamma_c(v_j-\overline{n}/(1+\overline{n}))(1+\overline{n}_p)+2S_j]^{n+1}} \,  .
\end{eqnarray}
Similarly to previous models, the summation is done over propagators with poles that represent the resonance of the spectrum. For practical computation of the infinite sum, we may use the
related hypergeometric function defined as 
$\Phi(x,1,a)=\, _2F_1 (1,a;1+a;x)/a=\sum_{n=0}^\infty x^n /(n+a)$.
For the case of vanishing coherence time, we recover the thermal case with $\overline{N}=\overline{N}_p$. We find 
successively in the limit of high quality factor and no signal photon:
\begin{eqnarray}
g_j\langle \hat \sigma^-_{\omega,j} \rangle'
&=&
\frac{\Omega_{p}}{2}\chi_j
\sum_{n=0}^\infty \frac{2S_j\gamma_c }{\omega_p -\omega_p^{(n)}}
\frac{[(S_j-\gamma_c/2)^2+\chi^2_j]^n}{[(S_j+\gamma_c/2)^2+\chi^2_j]^{n+1}}
\\
&\stackrel{\gamma_c \ll \chi_j}{=}&
\sum_{n=0}^\infty \frac{ (\Omega_{p}/2)\chi_j \overline{n}^n/(\overline{n}+1)^{n+1}}{\omega_p -(\omega_j+2\chi_jn-i(\gamma_c [(2\overline{n}+1)n+\overline{}]+\frac{\Gamma_j}{2}+\Gamma_{\phi,j}))}
\\
&\stackrel{\overline{n}=0}{=}&
\frac{\Omega_{p}}{2}\chi_j
\frac{ 1}{\omega_p -\omega_j+ i(\frac{\Gamma_j}{2}+\Gamma_{\phi,j})}  \,  .
\end{eqnarray}

\end{widetext}

\end{document}